\documentclass[useAMS,usenatbib]{mn2e}
\usepackage{times}
\usepackage{rotating} 
\usepackage{amsmath}
\usepackage{amssymb}
\usepackage{longtable}
\usepackage{hyperref}
\usepackage{epsfig}
\usepackage{natbib}

\newcommand{\tel}{$T_\mathrm{e}$} 		                    % electron temperature
\newcommand{\nel}{$N_\mathrm{e}$} 	                         %electron density
\newcommand{\hi}{H~{\sc i}}

\newcommand{\hei}{He~{\sc i}}
\newcommand{\heii}{He~{\sc ii}}

\newcommand{\ci}{C~{\sc i}}
\newcommand{\cii}{C~{\sc ii}}
\newcommand{\ciii}{C~{\sc iii}}

\newcommand{\crii}{Cr~{\sc ii}}

\newcommand{\clii}{Cl~{\sc ii}}
\newcommand{\cliii}{Cl~{\sc iii}}
\newcommand{\cliv}{Cl~{\sc iv}}

\newcommand{\nii}{N~{\sc ii}}
\newcommand{\niii}{N~{\sc iii}}

\newcommand{\oi}{O~{\sc i}}
\newcommand{\oii}{O~{\sc ii}}
\newcommand{\oiii}{O~{\sc iii}}

\newcommand{\neii}{Ne~{\sc ii}}
\newcommand{\neiii}{Ne~{\sc iii}}
\newcommand{\neiv}{Ne~{\sc iv}}

\newcommand{\mnii}{Mn~{\sc ii}}

\newcommand{\mnv}{Mn~{\sc v}}

\newcommand{\mgi}{Mg~{\sc i}}

\newcommand{\ariii}{Ar~{\sc iii}}
\newcommand{\ariv}{Ar~{\sc iv}}

\newcommand{\sii}{S~{\sc ii}}
\newcommand{\siii}{S~{\sc iii}}

\newcommand{\fei}{Fe~{\sc i}}
\newcommand{\feii}{Fe~{\sc ii}}
\newcommand{\feiii}{Fe~{\sc iii}}
\newcommand{\feiv}{Fe~{\sc iv}}

\newcommand{\apj}{ApJ}

\newcommand{\apjss}{ApJS}
\newcommand{\ana}{A\&A}
\newcommand{\anass}{A\&AS}
\newcommand{\piaus}{Proc. IAU Symp.}

\newcommand{\mn}{MNRAS}

\newcommand{\pna}{NGC~6153}
\newcommand{\pnb}{M\,1-42}
\newcommand{\pnc}{Hf\,2-2}

\title[VLT deep spectroscopy of Galactic PNe]{Very Large 
Telescope Deep Echelle Spectroscopy of Galactic Planetary Nebulae 
NGC\,6153, M\,1-42 and Hf\,2-2\thanks{Based on observations 
collected at the European Southern Observatory, Chile. Proposal ID 
ESO~69.D-0174(A).}}

\author[McNabb, Fang \& Liu]
 {I.~A. McNabb$^{1}$,
  X. Fang$^{2,3,4}$\thanks{E-mail: fangx@hku.hk},
  and 
  X.-W. Liu$^{1,5}$
\\
$^{1}$Kavli Institute for Astronomy and Astrophysics, Peking University, 
Beijing 100871, China\\
$^{2}$Instituto de Astrof\'{i}sica de Andaluc\'{i}a - CSIC, Glorieta de 
la Astronom\'{i}a, s/n. E-18008 Granada, Spain\\
$^{3}$Laboratory for Space Research, Faculty of Science, University of 
Hong Kong, Pokfulam Road, Hong Kong, China\\ 
$^{4}$Department of Physics, University of Hong Kong, Pokfulam Road, 
Hong Kong, China\\ 
$^{5}$Department of Astronomy, School of Physics, Peking University, 
Beijing 100871, China}

\begin{document}

\date{Received XXXX, accepted XXXX, Published online XXX}

\maketitle

\label{firstpage}

\begin{abstract}
We present deep spectroscopy of three Galactic planetary nebulae 
(PNe) with large abundance discrepancy factors (ADFs): NGC\,6153, 
M\,1-42 and Hf\,2-2. 
The spectra were obtained with VLT/UVES and cover the whole optical 
range (3040--11\,000\,{\AA}) with a spectral resolution of $\sim$20\,000. 
For all three PNe, several hundred emission lines were detected 
and identified, with more than 70 per cent of them as permitted lines. 
Most of these permitted lines are excited by recombination. 
Numerous weak optical recombination lines (ORLs) of O~{\sc ii}, C~{\sc ii}, 
N~{\sc ii} and Ne~{\sc ii} were detected in the spectra and accurate 
fluxes measured.  Line flux tables were compiled 
and ready for use by the community of nebular astrophysics.  These ORLs 
were critically analyzed using the effective recombination coefficients 
recently calculated for the optical recombination spectrum of N~{\sc ii} 
and O~{\sc ii} under the physical conditions of photoionized gaseous 
nebulae.  Plasma diagnostics based on the heavy element ORLs were carried 
out using the new atomic data.  Elemental abundances 
derived from the ORLs were systematically higher than those derived from 
the collisionally excited lines (CELs) by a factor of $\sim$11, 22 and 
80 for NGC\,6153, M\,1-42 and Hf\,2-2, respectively.  The electron 
temperatures derived from the heavy element ORLs are 
systematically lower than those derived from the CELs.  These ORL versus 
CEL abundance and temperature discrepancies, previously observed in the 
three PNe through deep spectroscopy with medium to low spectral 
resolution, are thus confirmed by our analysis of the deep echelle 
spectra using the new atomic data. 
\end{abstract}


\section{Introduction}

Planetary nebulae (PNe) evolve from the low- and intermediate-mass 
($\leq$8--10~$M_{\odot}$) stars, and are common in the universe. 
They are characterized by rich, bright mission line 
spectrum that act as a vital tool to acquire knowledge of their 
chemical composition.  This chemical information sheds lights on the 
basic astrophysics regarding the formation and evolution of the 
progenitor stars. 
Since the $\alpha$-elements are in principle not enriched through 
nucleosynthesis inside the low- and intermediate-mass stars, the 
abundances of O, Ne, S, Ar and Cl of a PN represent the chemistry 
of the interstellar medium (ISM) at the time when its progenitor star 
formed.  The bright and narrow emission lines of PNe make them 
not only observable at very distant ($\sim$100~Mpc; 
\citealt{ger05,ger07}) galaxies but also excellent tracers of 
kinematics \citep[e.g.,][]{Lon15a,Lon15b}.

The traditional method for the analysis of PN spectra and deriving 
ionic abundances is mainly based on the strong collisionally excited 
lines (CELs; also usually called the forbidden lines) of 
heavy elements.  However, emissivities of the CELs relative to that 
of H$\beta$ are very sensitive to the electron temperature under 
typical physical conditions of the photoionized 
gaseous nebulae, and therefore the derived ionic abundances could be 
greatly underestimated if temperature fluctuations exist.  These 
electron temperatures are derived using the traditional method based 
on the CEL ratios \citep{OF06}.  Another method of nebular 
analysis utilizes the intensity ratio of an optical recombination 
line (ORL) of either helium or a heavy element with the H$\beta$ (see 
the reviews \citealt{L03,L06a,L06b}).  Unlike the CELs, the relative 
emissivities of ORLs change weakly with both temperature and density 
because they only have a power-law dependence on the temperature and 
negligible dependence on the density \citep[e.g.,][]{L06a,L06b}.  This 
method is thus much less affected by temperature fluctuations, and the 
abundances derived from ORLs are expected to be more reliable. 
However, accurate measurements of the heavy element ORLs are extremely 
demanding due to their weakness ($\sim$10$^{-4}$--10$^{-3}$ of the 
H$\beta$ flux, or even fainter), while the typical fluxes of the CELs 
are more than 10$^{3}$--10$^{4}$ times those of the heavy element ORLs 
(e.g., the [O~{\sc iii}] $\lambda$5007 line is usually 10 times the 
H$\beta$ flux).

Modern technologies in observations have enabled detection of 
many heavy element ORLs that were previously too weak to be detected. 
The recombination lines of \oii, \cii, \nii, and \neii, can now be 
measured with high accuracy in the deep spectra of PNe.  Some of these 
recombination lines are important nebular diagnostics, such as the 
density-sensitive ORL ratios \oii~$\lambda$4649/$\lambda$4662 and 
\nii~$\lambda$5679/$\lambda$5666 and the temperature-sensitive ratios 
\oii~$\lambda$4649/$\lambda$4089 and \nii~$\lambda$5679/$\lambda$4041, 
which have recently been applied to nebular analysis (e.g., 
\citealt{MFL13}; \citealt{FL13}). 
Accompanied with the steady improvement of observational 
technologies, atomic data calculations, especially for the 
effective recombination coefficients for nebular lines of the 
heavy element ions, have achieved significant progress over the past 
decades.  Calculations of the accurate atomic data can help the 
astrophysical community to better understand the 
physics in gaseous nebulae as well as the excitation mechanisms of 
emission, and this effort is no doubt an organic part of nebular 
astrophysics \citep{GJF03}. 
Plasma diagnostics of Galactic PNe and H~{\sc ii} regions based 
on the heavy element ORLs have been carried out with the availability 
of new atomic data \citep[e.g.,][]{MFL13}.  Among the calculations for 
the relatively abundant heavy elements in the nebulae, the most 
comprehensive treatment of the recombination and radiative-collisional 
processes has been for the nebular lines of N~{\sc ii} 
\citep{FSL11,FSL13} and O~{\sc ii} (P.~J. Storey, unpublished; private 
communications).  These new effective recombination coefficients 
enabled plasma diagnostics using the ORLs of heavy elements 
\citep{MFL13}.

Nebular analyses based on the CELs and ORLs/continua have revealed 
one of the key problems in nebular astrophysics:  the ionic and 
elemental abundances of carbon, nitrogen, oxygen and neon relative to 
hydrogen derived from ORLs are systematically higher than those derived 
from the much brighter CELs.  This phenomenon, known as the ``abundance 
discrepancy'', was originally noticed in the \emph{IUE} observations of 
the C$^{2+}$/H$^{+}$ abundance ratio derived from the \ciii$]$ 
$\lambda\lambda$1907,\,1909 forbidden/semi-forbidden lines and the 
ground-based spectroscopic observations of the same ionic abundance 
derived from the C~{\sc ii} $\lambda$4267 line \citep[e.g.,][and 
references therein]{B82,B91}.  Another discrepancy 
in nebular studies that has been proved to be related with the 
``abundance discrepancy'' is that the nebular electron temperature 
derived from the Balmer jump (at 3646\,{\AA}) is always lower than 
those derived from the CEL ratios \citep[e.g.,][]{P71,LD93}.

The ``abundance discrepancy'' problem has been investigated in 
some particular Galactic PNe: 
e.g., NGC\,7009 \citep{LSB95,Luo01}, NGC\,6153 \citep{LSB00}, 
M\,1-42 and M\,2-36 \citep{LLB01}, NGC\,6543 \citep{WL04}, Hf\,2-2 
\citep{LBZ06}, and NGC\,6720 \citep{GD01}.  Spectroscopic surveys of 
Galactic PNe \citep[e.g.,][]{TBL03,TBL04,LLB04a,LLB04b,RG05,WLB05,WL07} 
have also been carried out, which allows investigation of the 
distribution of the abundance discrepancy among the PNe 
\citep[][]{Liu12a}. 
The abundance discrepancy factor (ADF), defined as the ratio of the 
ORL and CEL abundances of the same ion, lies in the range 1.6--3.2 for 
most Galactic PNe \citep{L06b}.  There are many PNe with ADFs 
$\sim$5--20.  The highest ADF has been discovered in Hf\,2-2 
\citep[ADF$\sim$70;][]{LBZ06}.  Different from PNe, where the ADF 
values span a wide range, ADFs of H~{\sc ii} regions are mostly 
between 1 and 2 \citep[e.g.,][]{GE07}.

Over decades of research in PNe and H~{\sc ii} regions, several 
mechanisms have been suggested to explain the abundance and 
temperature discrepancies (e.g., \citealt{P67,R89,VC94,LSB00}; see 
also reviews by \citealt{L06a,L06b} and \citealt{pp06}). 
The discrepancies have been traditionally interpreted in terms of 
large temperature fluctuations by \citet{P67}, who first revealed the 
temperature discrepancy through spectroscopic observations of H~{\sc 
ii} regions.  The temperature fluctuation model well explains the 
observed discrepancies in H~{\sc ii} regions, but has difficulties 
in the interpretation of high ADFs ($>$5) in PNe, which needs such 
large temperature fluctuations that are difficult to explain 
physically.  The bi-abundance nebular model \citep{LSB00} suggests 
that PNe (probably also in H~{\sc ii} regions) contain a previously 
unknown, metal-rich component.  This model well explains 
the problem.  In this model, the faint optical 
recombination lines of heavy elements are mainly emitted from the 
`cold' ($\leq$1000~K), metal-rich (probably H-deficient) component, 
where the cooling through the infrared fine-structure lines of the 
heavy elements is very efficient, while the stronger CELs come 
predominantly from the hot ($\sim$10$^{4}$~K) ionized gas with 
nearly Solar chemical composition. 
Deep recombination line surveys and spectroscopic analyses 
of Galactic PNe have provided strong evidence for the existence of 
the `cold', metal-rich component (see the review \citealt{Liu12a}), 
although the origin of this component is still unclear.

Abundance calculations and temperature determinations for PNe 
so far have all been based on the theory that free electrons have 
Maxwell-Boltzmann energy distributions \citep{spitzer48}, a widely 
accepted assumption in nebular astrophysics.  Recently, \citet{NDS12} 
proposed that energies of the free electrons in gaseous nebulae could 
deviate from the classical Maxwell-Boltzmann distribution and suggested 
that the ``$\kappa$-distribution'' of electron energy that has long 
been observed in the solar system plasmas, can explain the temperature 
and abundance discrepancies.  The effects of this mechanism on the 
nebular abundance and temperature determinations have been under 
investigation \citep[e.g.,][]{DSN13,NDS13}.  Direct measurements of 
the electron energy distribution from observations is important to 
assess whether the Maxwell-Boltzmann distribution or the 
$\kappa$-distribution is more suitable for the free electrons in PNe. 
This purpose has been attempted by \citet{ss13}, who studied the 
C~{\sc ii} dielectronic recombination lines in several PNe.  However, 
the uncertainties are too large to make any conclusion.  Through 
studies of the H~{\sc i} free-bound emission near the Balmer jump of 
four PNe with the highest ADFs, \citet{ZLZ14} found that a nebular 
model with two Maxwell-Boltzmann components, along with the 
$\kappa$-distributed electron energies, can equally fit the observed 
H~{\sc i} recombination continuum.  A Model fitting of the nebular 
continuum of Hf\,2-2 (ADF$\sim$70) near the Balmer jump also favors 
the two components with Maxwell-Boltzmann distribution of electron 
energies over the $\kappa$-distributed electron energies \citep{ss14}. 
Atomic data calculations incorporating the $\kappa$-distributed 
electron energies have also been carried out for the H~{\sc i} 
recombination lines \citep{ss15a} and the [O~{\sc iii}] collisionally 
excited lines \citep{ss15b}.

Galactic PNe NGC\,6153, M\,1-42 and Hf\,2-2 are among those with the 
highest ADFs.  The heavy-element ORLs observed in the three objects 
have previously been analyzed using deep spectra with medium spectral 
resolution \citep{LSB00,LLB01,LBZ06}.  These spectra were mostly 
obtained at the ESO 1.52~m telescope.  The deep spectroscopy revealed 
that ADF$\sim$10, 20 and 70 for NGC\,6153, M\,1-42 and Hf\,2-2, 
respectively.  The effective recombination coefficients of the heavy 
elements (C~{\sc ii}, N~{\sc ii}, O~{\sc ii} and Ne~{\sc ii}) used in 
these analyses are relatively outdated, although the results are still 
of high confidence.  In this paper, we present deep echelle spectra 
of NGC\,6153, M\,1-42 and Hf\,2-2 obtained at the ESO 8.2~m Very Large 
Telescope.  The Ultraviolet and Visual Echelle Spectrograph used in 
the observations provide both broad wavelength coverage 
($\sim$3040--11\,000\,{\AA}) and high spectral resolution 
($R\sim$20\,000).  These high-quality spectra will enable accurate 
flux measurements for the faint heavy element ORLs that are critical 
for nebular plasma diagnostics.  With the availability of new 
effective recombination coefficients for the N~{\sc ii} and O~{\sc ii} 
recombination lines, it is worthwhile to reappraise the properties of 
the three archetypal PNe and investigate the ORL versus CEL abundance 
discrepancy in details.

The contents of this paper are as follows:  Observations and data 
reduction are described in Section~2.  The reduced 1D spectra are 
presented in Section~3.  Emission line table and line identifications 
are presented in Section~4.  We carry out plasma diagnostics in 
Section~5, and present the abundance analysis in Section~6. 
We discuss the results in Section~7, and summarize conclusions in 
Section~8.

\begin{figure*}
\centering
\epsfig{file=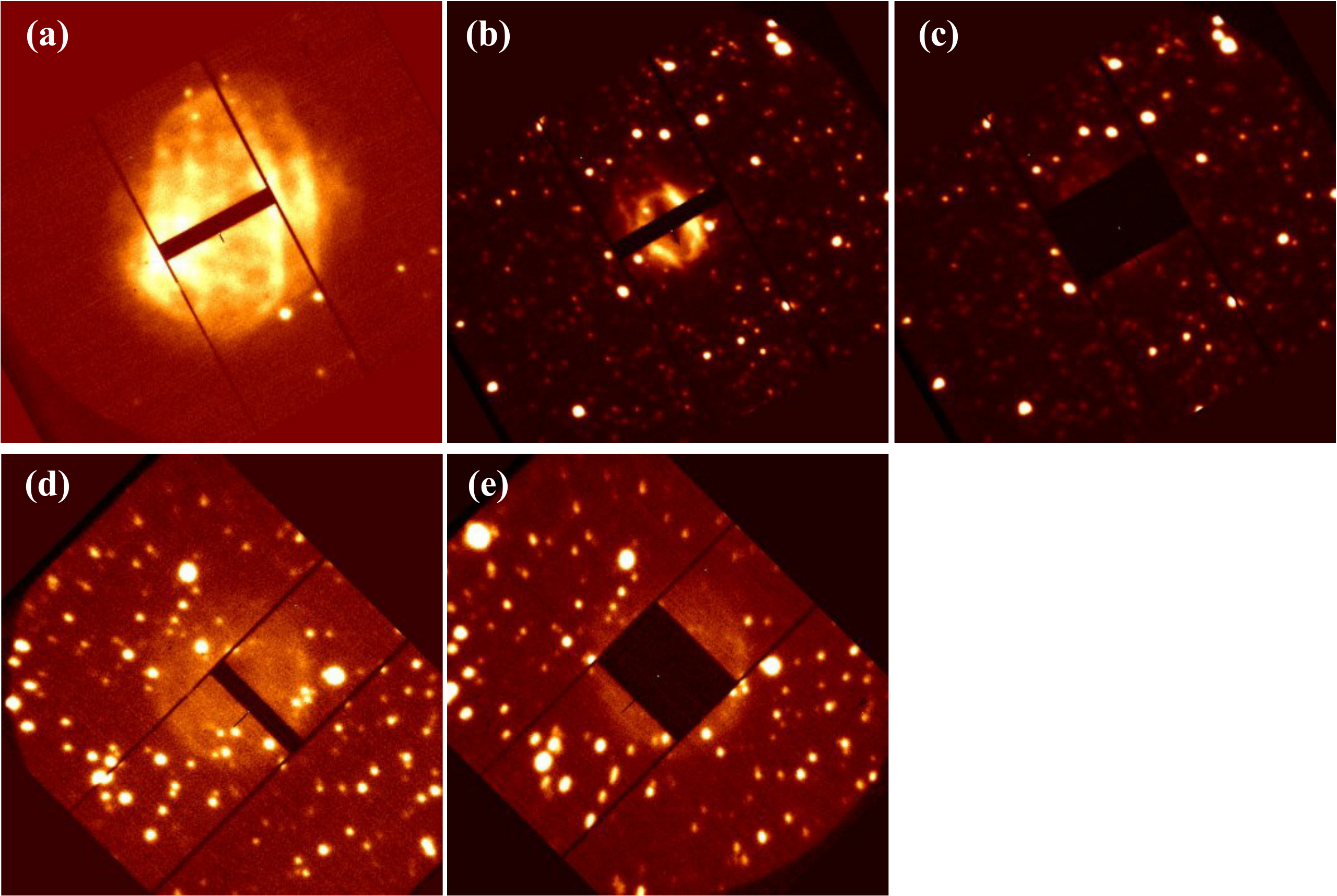,width=1.0\textwidth,angle=0}
\caption{Slit views of NGC\,6153 (panel a), M\,1-42 (panels b and c), 
and H\,2-2 (panels d and e).  The overlaid slit widths are 2~arcsec 
for NGC\,6153, and 2~arcsec and 10~arcsec for both M\,1-42 and Hf\,2-2. 
The field of view of each panel is roughly 40~arcsec$\times$40~arcsec. 
North is up and east to the left.}
\label{slitviews}
\end{figure*}

\section{Observations and Data Reduction}

The spectra were obtained on June 8, 2002, using the ESO Ultraviolet 
and Visual Echelle Spectrograph \citep[UVES;][]{DCD00} on the UT2 
Kueyen Telescope of the 8.2\,m Very Large Telescope (VLT) at Cerro 
Paranal, Chile (prop.~ID: 69.D-0174(A), P.I.: J. Danziger).  UVES is 
a high-efficiency echelle spectrograph that covers a broad wavelength 
range, from the atmospheric cut-off at 3000\,{\AA} to about 
11\,000\,{\AA}.  The light beam entering the telescope is split into 
two arms (from the UV to Blue, and Visual to Red) with a resolving 
power of about 40\,000 when a 1~arcsec slit is 
used\footnote{http://www.eso.org/sci/facilities/paranal/instruments/uves/index.html} 
\citep{DDK00}.  The dichroics DIC1 and DIC2 were used.  Cross dispersers 
CD\#1 and CD\#2 were adopted for DIC1.  Combined with the slit viewers, 
these two cross dispersers covered the wavelength ranges 303--388\,nm in 
the blue arm and 476--684\,nm in the red.  For DIC2, cross dispersers 
CD\#3 and CD\#4 were used, covering the wavelength ranges 373--499\,nm 
in the blue arm and 660--1060\,nm in the red.  A slit length of 10~arcsec 
was used in the blue arm and 13~arcsec for the red.  The observations 
and instrument setups are summarized in Table \ref{tab:observations}.

\begin{table*}
\centering
\setlength{\tabcolsep}{0.015in}
\caption{Journal of observations.}
\begin{tabular}{lcccccc}
\hline
\hline
Object & \multicolumn{1}{c}{~~~Exp.~Time~~~} & \multicolumn{1}{c}{~~~Dichroic~~~} & \multicolumn{1}{c}{~~~~~~Cross~Disperser~~~~~~} & \multicolumn{1}{c}{~~~Wave.~Range~~~} & \multicolumn{1}{c}{~~~Slit~Length~~~} & \multicolumn{1}{c}{Pixel~Scale} \\
~ & \multicolumn{1}{c}{(min)} & ~ & ~ & \multicolumn{1}{c}{(nm)} & \multicolumn{1}{c}{(arcsec)} & (~arcsec/pixel) \\
\hline
NGC\,6153  & 20 & DIC1 & CD\#1 & 303--388 & 10 & 0.25 \\
                 ~  & ~   & DIC2 & CD\#3 & 373--499 & 10 & 0.18 \\
                 ~  & ~   & DIC1  & CD\#2 & 476--684 & 13 & 0.25 \\
                 ~  & ~   & DIC2 & CD\#4 & 660--1060 & 13 & 0.17 \\
~  & 2 & DIC1 & CD\#1 & 303--388 & 10 & 0.25 \\
                 ~  & ~   & DIC2  & CD\#3 & 373--499 & 10 & 0.18 \\
                 ~  & ~   & DIC1  & CD\#2 & 476--684 & 13 & 0.25 \\
                 ~  & ~   & DIC2 & CD\#4 & 660--1060 & 13 & 0.17 \\
M\,1-42 & 30 & DIC1 & CD\#1 & 303--388 & 10 & 0.25 \\
                 ~  & ~   & DIC2  & CD\#3 & 373--499 & 10 & 0.18 \\
                 ~  & ~   & DIC1  & CD\#2 & 476--684 & 13 & 0.25 \\
                 ~  & ~   & DIC2 & CD\#4 & 660--1060 & 13 & 0.17 \\
~  & 15 & DIC1 & CD\#1 & 303--388 & 10 & 0.25 \\
                 ~  & ~   & DIC2 & CD\#3  & 373--499 & 10 & 0.18 \\
                 ~  & ~   & DIC1  & CD\#2 & 476--684 & 13 & 0.25 \\
                 ~  & ~   & DIC2 & CD\#4 & 660--1060 & 13 & 0.17 \\
~  & 1 & DIC1 & CD\#1 & 303--388 & 10 & 0.25 \\
                 ~  & ~   & DIC2  & CD\#3 & 373--499 & 10 & 0.18 \\
                 ~  & ~   & DIC1  & CD\#2 & 476--684 & 13 & 0.25 \\
                 ~  & ~   & DIC2 & CD\#4 & 660--1060 & 13 & 0.17 \\
Hf\,2-2 & 30 & DIC1 & CD\#1 & 303--388 & 10 & 0.25 \\
                 ~  & ~   & DIC2  & CD\#3 & 373--499 & 10 & 0.18 \\
                 ~  & ~   & DIC1  & CD\#2 & 476--684 & 13 & 0.25 \\
                 ~  & ~   & DIC2 & CD\#4 & 660--1060 & 13 & 0.17 \\
~ & 15 & DIC1 & CD\#1 & 303--388 & 10 & 0.25 \\
                 ~  & ~   & DIC2  & CD\#3 & 373--499 & 10 & 0.18 \\
                 ~  & ~   & DIC1  & CD\#2 & 476--684 & 13 & 0.25 \\
                 ~  & ~   & DIC2 & CD\#4 & 660--1060 & 13 & 0.17 \\
\hline
\end{tabular}
\end{table*}
\label{tab:observations}

Multiple exposures, typically long and short ones, were utilized for 
observations with each instrument setup.  A 2~arcsec wide slit was used 
for the observations of NGC\,6153.  The long exposure is 20~min and the 
short exposure is 2~min.  The slit has a position angle of 118$^{\circ}$ 
and covers the central part of the PN.  Fig.~\ref{slitviews} (panel a) 
shows the slit view for NGC\,6153.  For M\,1-42, there are three 
exposures: a 30~min exposure with a slit width of 2~arcsec, a 15~min 
exposure with a slit width of 10~arcsec, and a 1~min exposure with a 
slit width of 2~arcsec.  The slit has a position angle of 120$^{\circ}$ 
covering the central part of the PN.  Fig.~\ref{slitviews} (panels b 
and c) shows the slit views for M\,1-42.  For Hf\,2-2, there are two 
exposures: a 30~min exposure with a slit width of 2~arcsec and 15~min 
exposure with a slit width of 10~arcsec.  The slit has a position angle 
of 45$^{\circ}$, spanning from the central star to the bright 
condensation in the southwest.  Fig.~\ref{slitviews} (panels d and 
e) shows the slit views of Hf\,2-2.  The full width at half-maximum 
(FWHM) of the emission lines is $\sim$0.15--0.6\,{\AA} from the 
blue to the red part.

Spectra were reduced using the ESO software 
{\sc midas}\footnote{{\sc midas} is developed and distributed by the 
European Southern Observatory.}.  The spectra were first de-biased, 
corrected for the flat field, and cosmic-ray removed.  Wavelength 
calibration was then performed using exposures of arc lamps.  We 
then flux-calibrated the spectra using the spectroscopic observations 
of the spectrophotometric standard stars CD\,$-$32$^{\circ}$\,9927, 
Feige~56, LTT~3864 and LTT~9293 \citep{HWS92,HPW94}. 
It is difficult to find a clear sky region in the 2D echelle spectra, 
therefore sky lines were removed manually.  Due to the high spectral 
resolution, most of the lines do not suffer much from blending.

\begin{figure*}
\centering
\epsfig{file=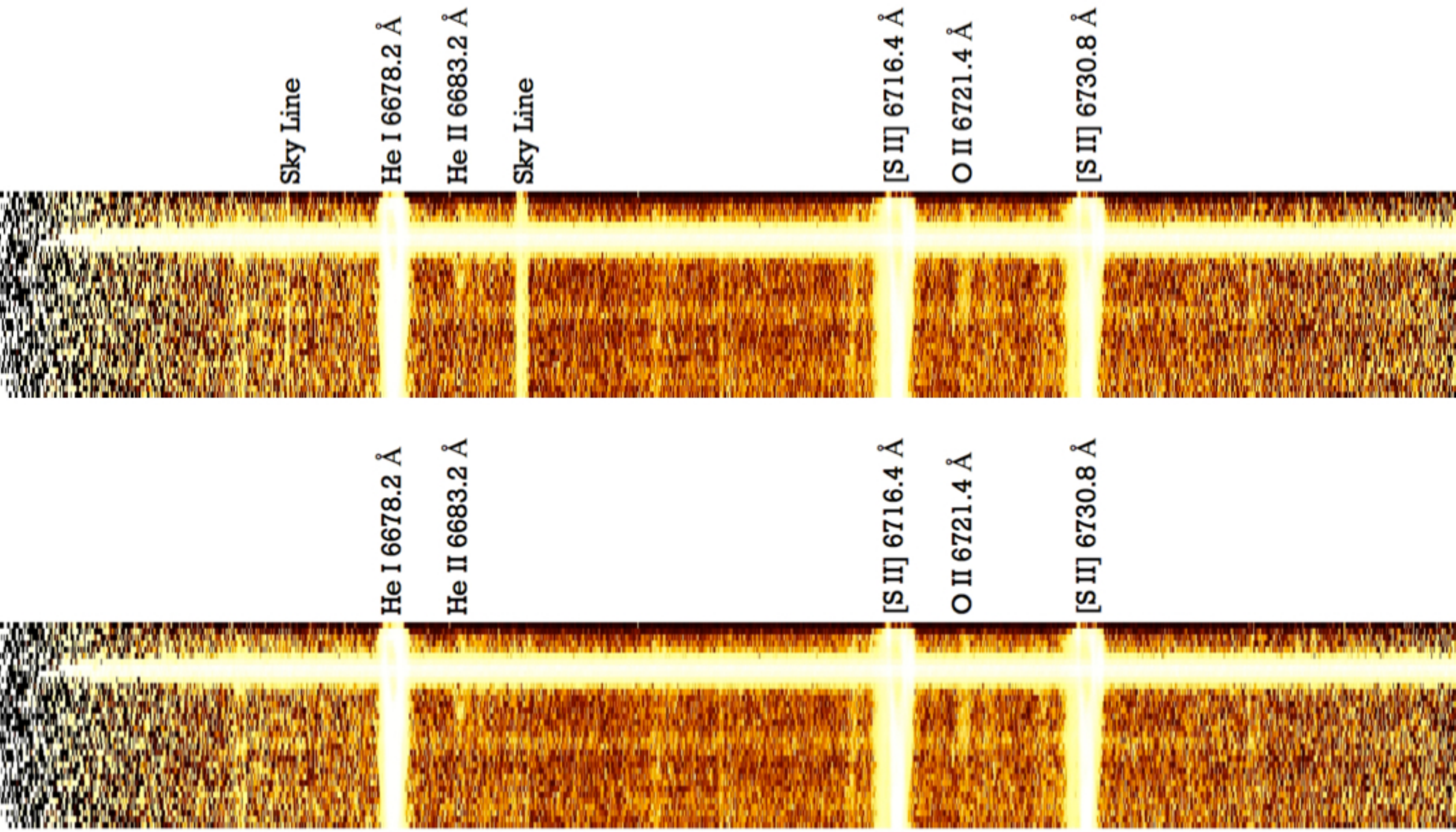,width=0.9\textwidth,angle=0}
\caption{UVES 2D spectrum of Hf\,2-2 (30~min exposure) from 6649 
to 6760\,{\AA}.  The top panel shows the spectrum before sky 
subtraction and the bottom panel shows the same section after sky 
subtraction.  Emission lines are identified and labeled, including 
two sky lines.} 
\label{hf22sky}
\end{figure*}

Considerable effort was devoted to sky subtraction.  There were numerous 
skylines in the two-dimensional (2D) spectra, especially in the 
wavelength region beyond 5000\,{\AA}.  Some skylines are blended with 
the nebular emission lines, complicating the data reduction.  Sky 
subtraction on the 2D frames is 
not straightforward because no region on the slit could be defined as 
the sky background, as shown in Fig.~\ref{hf22sky}, where the [\sii] 
$\lambda\lambda$6716,\,6731 nebular lines and the \oii\ $\lambda$6721 
recombination line are located.  While emission lines in the blue 
region of the VLT UVES spectrum (mainly from 3000 to 5000\,{\AA}) 
suffer less 
from the contamination by skylines, the red part (from 5000 to 
11\,000\,{\AA}) are littered with the skylines.  In order to remove the 
them, we first need to measure their widths and determine the central 
wavelengths.  The weak and rich emission lines within the blue region 
also made skyline-locating fairly difficult, especially when the nebular 
expansion affects the profiles of the emission lines and a skyline is 
totally embedded in the emission feature.  Despite the difficulties in 
skyline removal, we have developed the {\sc idl} routines to solve this 
problem.

Skylines that are not blended with any nebular emission lines were simply 
removed and replaced with local background continuum.  Skylines that are 
completely blended with nebular lines are difficult to remove.  These 
skylines were removed by fitting the unaffected region of the emission 
line with a polynomial and extrapolated to the whole slit, assuming the 
emission line has a symmetric profile.  Any residual skylines after the 
subtraction only occur at either end of the CCD frame, where the detector 
efficiencies are low.  For those sky lines utterly embedded in the 
relatively bright nebular lines, subtraction is impossible.  Removal of 
these skylines will create artificial uncertainties in nebular line 
fluxes.  Fortunately, these skylines do not affect the analysis of the 
most critical nebular emission lines. 
%We have tried to improve sky subtraction using the archival skyline 
%tables from the ESO Science Archive, but with no success.

Fig.~\ref{hf22sky} shows an example of sky subtraction for the spectrum 
of Hf\,2-2 in the range 6649--6760\,{\AA}.  The upper panel of the 
figure shows the spectrum before sky subtraction and the lower panel 
shows the spectrum after subtraction.  We identified and removed about 
900 skylines in the red region for NGC\,6153, $\sim$800 for M\,1-42 and 
$\sim$1200 for Hf\,2-2.  The skyline subtraction for the long-exposure 
spectra of NGC\,6153, M\,1-42 and Hf\,2-2 are generally satisfactory, 
whereas a few remain in the short-exposure spectra due to the relatively 
low S/N.

\section{The 1D spectra}

\begin{figure}
\begin{center}
\epsfig{file=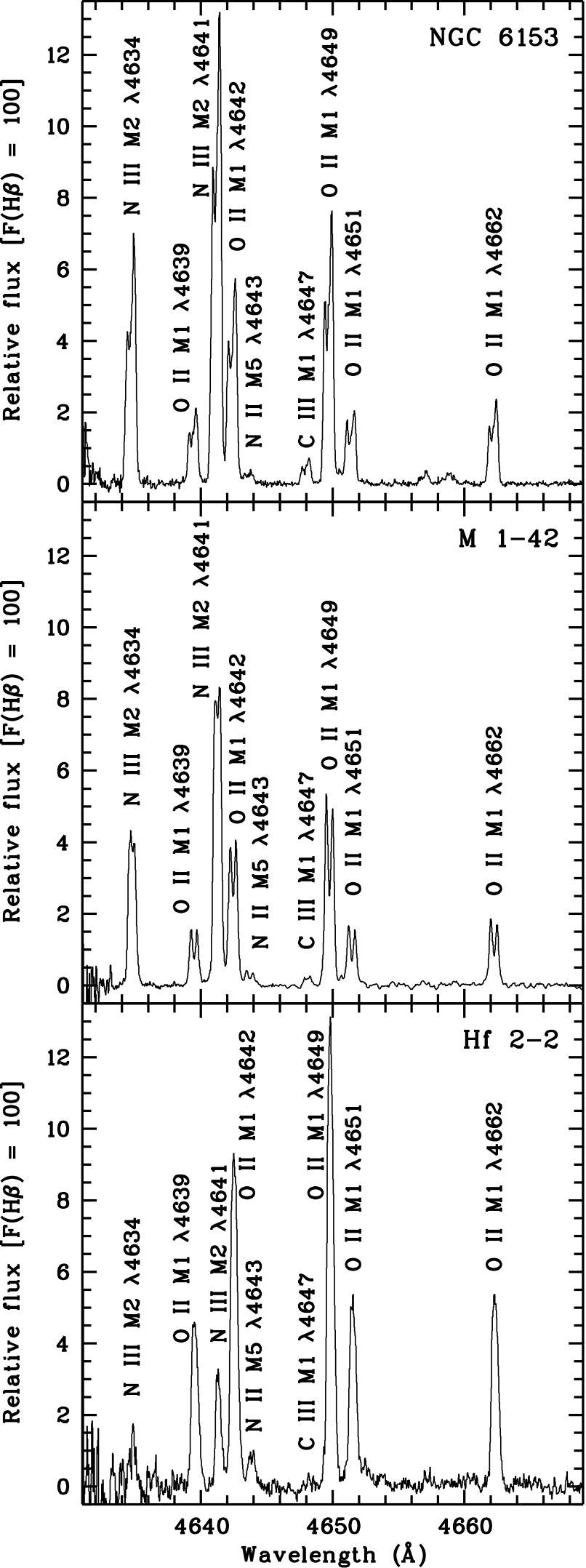,width=8.3cm,angle=0}
\caption{
1D spectra of NGC\,6153 (top), M1\,42 (middle) and Hf\,2-2 (bottom) 
from 4631\,{\AA} to 4669\,{\AA}, showing the ORLs of the O~{\sc ii} M1 
3p\,$^{4}$D$^{\rm o}$ -- 3s\,$^{4}$P multiplet.  Note the double-peak 
line profiles due to nebular expansion.  Continuum has been subtracted 
from all three spectra.  Extinction has not been corrected for.} 
\label{oii_lines}
\end{center}
\end{figure}

The 1D spectra extracted from the cleaned 2D frames are presented in 
Fig.~\ref{1d_spectra}.  Fig.~\ref{oii_lines} shows the ORLs of 
the O~{\sc ii} M1 3p\,$^{4}$D$^{\rm o}$ -- 3s\,$^{4}$P multiplet. 
Line fluxes were then measured by integration over the emission line 
profiles.  Emission of the central stars was excluded in the spectrum 
extraction.  Nebular expansion was clearly seen from the splitting of 
emission line profiles.  Peaks of the two components can not be 
well-fitted with Gaussian profiles, therefore we integrated to get 
the total fluxes of each line.  
The extinction parameter $c({\rm H}\beta)$ was derived from the 
observed H~{\sc i} Balmer line ratios, which were compared with the 
theoretical Case~B ratios of \citet{SH95} assuming $T_\mathrm{e}$ = 
10\,000~K and $N_\mathrm{e}$ = 10$^{4}$~cm$^{-3}$.

For NGC\,6153, we derived a $c({\rm H}\beta)$ value of 1.32, as 
averaged from the $c$(H$\beta$) values derived from several Balmer 
line ratios.  For M\,1-42, our long-exposure (30~min) spectrum yield 
a $c({\rm H}\beta)$ value of 0.699.  The medium-exposure (15~min) 
spectrum yield a $c({\rm H}\beta)$ value of 0.663, and the 
short-exposure (1~min) spectrum yield a $c({\rm H}\beta)$ value of 
0.779.  \citet{LLB01} derived a $c({\rm H}\beta)$ of 0.70 for 
NGC\,6153, which well agrees with our measurements of the 
long-exposure data.  For Hf\,2-2, measurements of the long-exposure 
(30~min) spectrum yield $c({\rm H}\beta)$ = 0.361, while measurements 
of the medium-exposure (15~min) spectrum yield a $c({\rm H}\beta)$ 
value of 0.439.  \citet{LBZ06} derived $c({\rm H}\beta)$ = 0.47 for 
Hf\,2-2.  The small differences between our extinction values and 
the literature probably arise from the different nebular regions 
sampled as well as the measurement errors.  The error contribution 
from flux calibration is expected to be very small.  The line fluxes 
were extinction corrected using 

\begin{equation}
I(\lambda) = 10^{c({\rm H}\beta)f(\lambda)} F(\lambda). 
\end{equation}
\noindent
\noindent
Here $f$($\lambda$) is the standard Galactic extinction curve for 
$R = 3.1$ \citep{CCM89}.  The $c$(H$\beta$) values derived from 
the long-exposure (20 and 30~min) data were accurate and thus 
adopted in the extinction correction.

\begin{table*}
\rotatebox{90}{
\begin{minipage}{\textheight}
\begin{center}
\caption{Emission lines detected in the long-exposure spectra of NGC\,6153,
M\,1-42 and Hf\,2-2.  The observed line fluxes and the extinction-corrected
intensities are normalized such that H$\beta$ = 100.  This is a sample of 
the complete table, which is available online only. } 
\label{lines}
\begin{tabular}{rrrrcrrrcrrrccccrr}
\hline
\hline
~ & \multicolumn{3}{c}{NGC 6153} & ~ & \multicolumn{3}{c}{M 1-42} & ~ & \multicolumn{3}{c}{Hf 2-2}\\
\cline{2-4}
\cline{6-8}
\cline{10-12}
~ & \multicolumn{3}{c}{20 minutes} & ~ & \multicolumn{3}{c}{30 minutes} & ~ & \multicolumn{3}{c}{30 minutes} \\
\cline{2-4}
\cline{6-8}
\cline{10-12}
\multicolumn{1}{c}{$\lambda_{lab}$} & $F$($\lambda$) & $I$($\lambda$) & 
Error &~ & $F$($\lambda$) & $I$($\lambda$) & Error &~ & $F$($\lambda$) & 
$I$($\lambda$) & Error & ID & Mul. & Lower & Upper & $g_{1}$ & $g_{2}$ \\
({\rm \AA}) &  &  & (\%) &~ &  &  & (\%) &~ &  &  & (\%) & & & & & & \\
\hline
 3047.56&    1.86&    9.61&   8.07&~&    1.66&    3.95&   5.44&~&~&~&~&\neii&V8&3p 4P*&3d 4D&  4&  6\\
 3121.64&    1.07&    4.96&  10.19&~&    0.84&    1.89&   4.14&~&~&~&~&\oiii&V12&3p 3S&3d 3P*&  3&  3\\
 3132.79&   12.39&   56.29&   3.07&~&   13.64&   30.38&   1.96&~&    7.46&   11.28&  10.25&\oiii&V12&3p 3S&3d 3P*&  3&  5\\
 3149.09&~&~&~&~&    1.03&    2.26&   1.86&~&~&~&~&\neii&~&3d 2F&5f 2<3>*&  6&  8\\
 3187.74&    2.41&   10.07&   1.95&~&    2.41&    5.14&   1.95&~&    9.19&   13.58&   2.91&\hei&V3&2s 3S&4p 3P*&  3&  9\\
 3203.17&    2.89&   11.81&   2.53&~&    3.46&    7.28&   0.32&~&    4.44&    6.52&   7.61&\heii&3.5a&3d 2D&5f 2F*&  6&  8\\
 3218.19&    1.00&    4.00&   9.22&~&~&~&~&~&    4.96&    7.24&  14.69&\neii&V13&3p 4D*&3d 4F&  8& 10\\
 3229.55&    1.04&    4.11&  24.10&~&~&~&~&~&    5.52&    8.03&   2.50&Si~{\sc ii}&~&3d$^{\prime}$ 2P*&4f$^{\prime}$ 2D&  4&  6\\
 3230.54&~&~&~&~&    1.57&    3.23&  21.15&~&~&~&~&\nii&~&3D&3P*&  7&  5\\
 3244.10&    0.72&    2.79&  12.85&~&    0.73&    1.49&   1.58&~&    3.63&    5.25&   3.53&\neii&V13&3p 4D*&3d 4F&  6&  8\\
 3258.56&~&~&~&~&    0.91&    1.84&  19.45&~&    3.94&    5.67&   0.31&[\mnii]&~&a7S&a3H&  7& 11\\
 3260.85&    0.74&    2.79&  15.01&~&    0.80&    1.62&  13.89&~&~&~&~&\oiii&V8&3p 3D&3d 3F*&  5&  7\\
 3265.32&    0.70&    2.62&  19.77&~&~&~&~&~&~&~&~&\oiii&V8&3p 3D&3d 3F*&  7&  9\\
 3267.20&~&~&~&~&~&~&~&~&    5.24&    7.52&   2.26&\oiii&V8&3p 3D&3d 3F*&  3&  5\\
 3277.35&~&~&~&~&    0.63&    1.26&  11.62&~&~&~&~&\feii]&V1&a4D&z6D*&  8& 10\\
 3289.98&~&~&~&~&    0.68&    1.34&  15.86&~&~&~&~&\oii&V23&3p 4P*&4s 4P&  2&  4\\
 3299.40&    1.22&    4.41&   3.52&~&    1.13&    2.23&   2.35&~&~&~&~&\oiii&V3&3s 3P*&3p 3S&  1&  3\\
 3304.31&~&~&~&~&    0.72&    1.41&  13.24&~&~&~&~&\feiii&~&2S 4s3S&4D 4p3P*&  3&  5\\
 3312.32&    1.95&    6.95&   3.67&~&    2.01&    3.93&   1.32&~&~&~&~&\oiii&V3&3s 3P*&3p 3S&  3&  3\\
 3322.53&~&~&~&~&    0.74&    1.44&  10.23&~&~&~&~&[\feiii]&~&5D&7S&  9&  7\\
 3334.84&    0.93&    3.23&   4.82&~&    1.03&    1.99&   2.74&~&    3.53&    4.96&  10.84&\neii&V2&3s 4P&3p 4D*&  6&  8\\
 3340.76&    2.49&    8.63&   3.64&~&    2.61&    5.04&   1.42&~&    3.75&    5.27&   6.80&\oiii&V3&3s 3P*&3p 3S&  5&  3\\
 3342.44&    0.74&    2.56&  25.96&~&    0.66&    1.27&   2.33&~&    2.93&    4.11&   2.70&[\neiii]&~&2p4 1D&2p4 1S&  5&  1\\
 3355.02&    0.99&    3.37&   3.90&~&    0.92&    1.77&   7.80&~&    4.06&    5.68&   0.68&\neii&V2&3s 4P&3p 4D*&  4&  6\\
 3360.60&    0.66&    2.23&   0.43&~&    0.64&    1.23&  17.26&~&~&~&~&\neii&V2&3s 4P&3p 4D*&  1&  3\\
 3367.22&    1.31&    4.42&   0.79&~&    0.73&    1.38&   8.75&~&~&~&~&\neii&V20&3p 2D*&3d 2F&  6&  8\\
 3390.55&~&~&~&~&    0.73&    1.38&  19.01&~&~&~&~&\neii&V12&3p 4D*&3d 4D&  2&  4\\
 3405.71&    0.84&    2.76&   1.89&~&~&~&~&~&~&~&~&\oiii&V15&3p 3P&3d 3P*&  1&  3\\
 3408.12&    0.82&    2.68&   0.56&~&~&~&~&~&~&~&~&\oiii&V15&3p 3P&3d 3P*&  3&  1\\
 3416.91&~&~&~&~&    0.87&    1.61&   6.21&~&~&~&~&\neii&V21&3p 2D*&3d 2D&  6&  6\\
 3417.69&    0.95&    3.07&   1.59&~&~&~&~&~&    4.04&    5.57&   0.45&\neii&V19&3p 2D*&3d 4F&  6&  8\\
 3428.62&    1.41&    4.52&   3.62&~&    1.25&    2.33&   1.91&~&~&~&~&\oiii&V15&3p 3P&3d 3P*&  3&  5\\
 3430.57&    0.88&    2.82&  12.86&~&~&~&~&~&~&~&~&\oiii&V15&3p 3P&3d 3P*&  5&  3\\
 3444.06&    5.00&   15.89&   3.33&~&    4.72&    8.69&   2.11&~&    4.37&    5.99&   5.85&\oiii&V15&3p 3P&3d 3P*&  5&  5\\
 3447.59&    0.99&    3.13&  10.99&~&    0.90&    1.65&   1.28&~&    3.83&    5.24&   2.21&\hei&V7&2s 1S&6p 1P*&  1&  3\\
 3456.86&    0.85&    2.66&  19.35&~&~&~&~&~&~&~&~&\hei&~&2p 3P*&19d 3D&  9& 15\\
\hline
\end{tabular}
\end{center}
\end{minipage}}
\end{table*}

\section{Line Identifications}

After sky subtraction and extraction of the 1D spectra, we measured 
wavelengths and fluxes of the emission lines, and carried out emission 
line identification, following the procedure of \citet{FL11}. 
Table~\ref{lines} presents the deep spectra of NGC\,6153, M\,1-42 and 
Hf\,2-2, and will be available in its complete form as the online data. 
Generally there are two steps in emission line identification: 
(1) We first identified the emission lines manually, using the empirical 
method based on the available atomic transition database.  Emission 
line lists from the literature \citep{SWB03,ZLL05a,FL11} were also used 
as references.  Thanks to the high-spectral resolution, the majority of 
the important nebular emission lines seldom suffer from line blending. 
For the nebular lines with double peaks (due to nebular expansion), the 
average wavelength of the two components were used in line identification. 
(2) After a complete (also preliminary) line list was constructed, it 
was used as an input file for the {\sc emili}\footnote{{\sc emili} was 
originally developed by \citet{SWB03} and was used to identify weak 
emission lines, particularly the weak recombination lines detected in 
deep, high-dispersion spectra.} code running for 
further identification.

Our line identification was first based on the observed 
wavelengths compared with the laboratory or theoretical wavelengths. 
We also checked the observed relative line fluxes against the 
predicted ones where available.  The atomic transition tables compiled 
by \citet{HH95} as well as the line lists of Galactic PNe IC\,418 
\citep{SBW04}, NGC\,7027 \citep{ZLL05a}, NGC\,7009 \citep{FL11} were 
used in the line identification.  For some features that we were 
unable to assign reasonable identifications, we used the online 
atomic transition database\footnote{Atomic Line List v2.05 by P.~A.~M. 
van Hoof, website: http://www.pa.uky.edu/$\sim$peter/newpage/} as an aid. 
This empirical identification might be wrong for some faint lines. 
Once the {\sc emili} runs were completed, we carefully checked the line 
lists for possible misidentifications.  The correctness of line 
identifications was confirmed by searching for the fine-structure 
lines within a given multiplet of transition.

\begin{figure*}
\centering
\epsfig{file=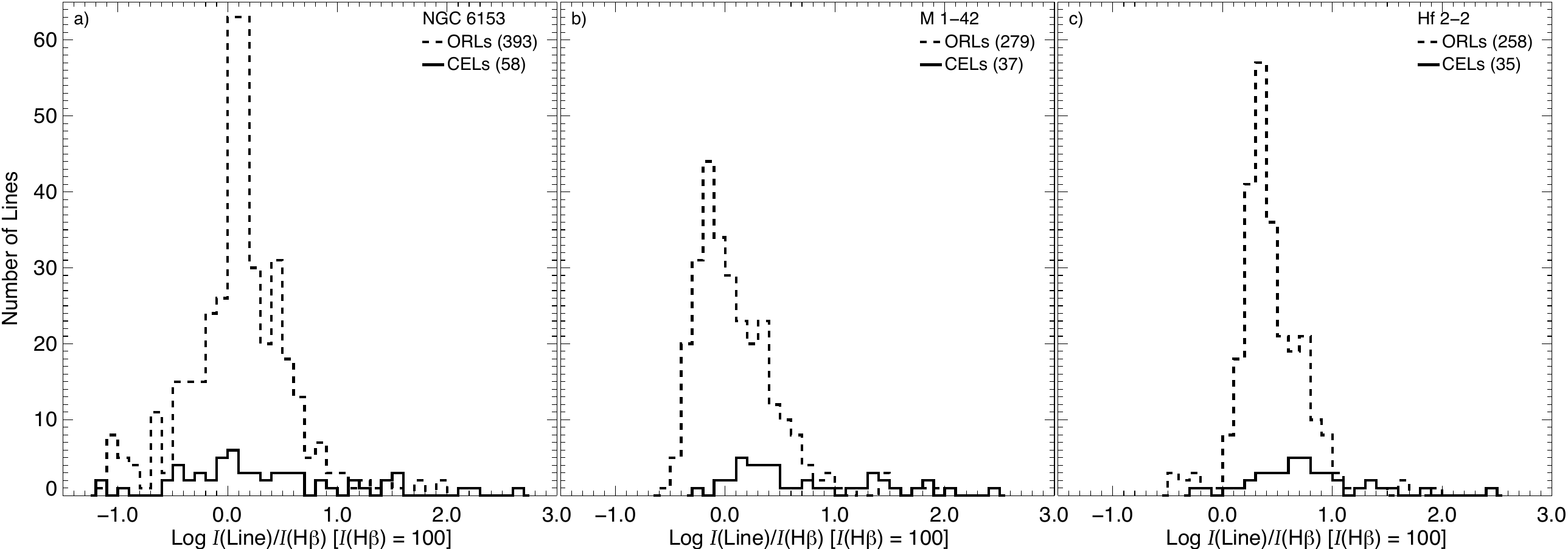,width=1.0\textwidth,angle=0}
\caption{Number distributions of the ORL (dashed lines) and CEL 
(solid lines) intensities for the long-exposure data of NGC\,6153 
(a), M\,1-42 (b), and Hf\,2-2 (c). }
\label{fig:orlceldist}
\end{figure*}

\begin{figure*}
\centering
\epsfig{file=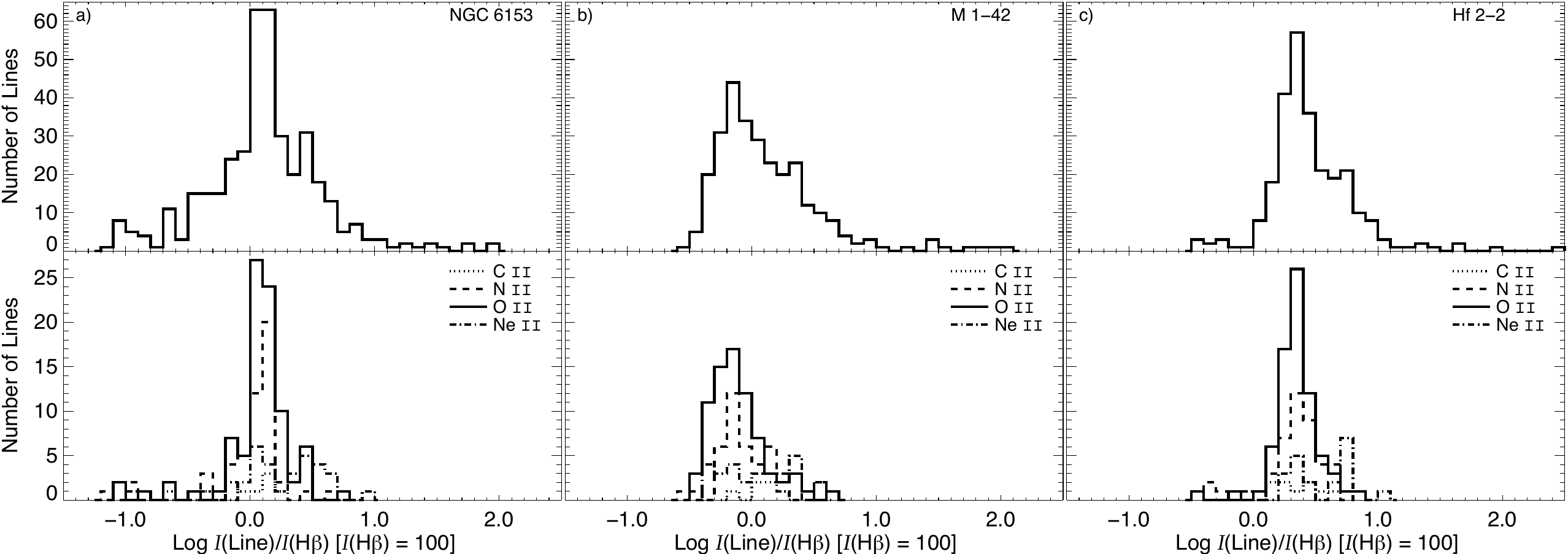,width=1.0\textwidth,angle=0}
\caption{Number distributions of emission line intensities of 
NGC\,6153 (left), M\,1-42 (middle) and Hf\,2-2 (right).  The upper 
panels show the intensity distribution of all emission lines, and 
the lower panels show the distribution of the \cii, \nii, \oii, and 
\neii\ ORLs.} 
\label{fig:linedist}
\end{figure*}

About 470 emission lines were identified in the long-exposure 
(20~min) spectrum of NGC\,6153, whereas 240 lines were identified in 
the short-exposure (2~min) spectrum.  About 330 emission lines were 
identified in the long-exposure (30~min) spectrum of M\,1-42, whereas 
320 lines were identified in the intermediate-exposure (15~min) and 150 
in the short-exposure (1~min) spectrum.  310 emission lines were 
identified in the long-exposure (30~min) spectrum of Hf\,2-2, and 290 
in the medium-exposure (15~min).  The emission line tables of the 
deep spectra of the three PNe are presented in Table~\ref{lines} (the 
complete table is available as the online data). 
Fig.~\ref{fig:orlceldist} shows the intensity distributions of 
the collisionally excited lines and the permitted lines identified 
in NGC\,6153, M\,1-42 and Hf\,2-2.  These permitted lines are emitted 
by the most abundant heavy element ions (C~{\sc ii}, N~{\sc ii}, 
O~{\sc ii} and Ne~{\sc ii}). 
Fig.~\ref{fig:linedist} shows the histograms of the intensities of 
all emission lines identified in the deep spectra of the three PNe. 
Numbers of the ORLs of different ions are also presented.

\begin{table*}
\centering
\setlength{\tabcolsep}{0.015in}
\caption{References for atomic data.}
\begin{tabular}{lccc}
\hline
\hline
Ion & \multicolumn{3}{c}{Collisionally excited lines} \\
\hline
~ & Transition probabilities & & Collision Strengths \\
\cline{2-2}
\cline{4-4}
$[$N~{\sc i}]$$~& \citet{Z82} & & \citet{BB81} \\
$[$N~{\sc ii}$]$~& \citet{NR79} & & \citet{SBH94} \\
$[$N~{\sc iii}$]$~& \citet{FKP93} & & \citet{BP92} \\
$[$O~{\sc ii}$]$~& \citet{Z82} & & \citet{P76} \\
$[$O~{\sc iii}$]$~& \citet{NS81} & & \citet{A83} \\
$[$Ne~{\sc iii}$]$~& \citet{M83} & & \citet{BZ94} \\
$[$S~{\sc ii}$]$~& \citet{MZ82} & & \citet{KAB96} \\
~ & \citet{KHO93} & & ~ \\
$[$S~{\sc iii}$]$~& \citet{MZ82} & & \citet{M83} \\
$[$Cl~{\sc iii}$]$~& \citet{M83} & & \citet{M83} \\
$[$Ar~{\sc iii}$]$~& \citet{PB95} & & \citet{VW92} \\
$[$Ar~{\sc iv}$]$~& \citet{MZ82} & & \citet{BZ89} \\
\\
Ion & \multicolumn{3}{c}{Optical recombination lines} \\
\hline
~ & Effective recomb. coeffs. & & Case \\
\cline{2-2}
\cline{4-4}
\hi~& \citet{SH95} & & B \\
\hei~& \citet{BSS99} & & B: singlets \\
~ & \citet{B72} & & A: triplets \\
\heii~& \citet{SH95} & & B \\
\cii~& \citet{DSK00} & & A \\
\nii~& \citet{FSL13} & & B \\
\niii~& \citet{PPB91} & & A \\
~ & \citet{NS84} & & ~ \\
\oii~& Storey (unpublished) & & B \\
\oiii~& \citet{PPB91} & & A \\
\neii~& \citet{KSD98} & & B: doublets \\
~ & Storey (unpublished) & & Case A: quartets \\
~ & \citet{NS87} & & Dielectronic recombination \\
\hline
\end{tabular}
\label{tab:atomicdata}
\end{table*}

\begin{table*}
\centering
\setlength{\tabcolsep}{0.04in}
\caption{Plasma diagnostics.}
\begin{tabular}{rlrrcrrcrr}
\hline
\hline
ID & Diagnostic & \multicolumn{8}{c}{Results} \\
\hline
~ & ~ & \multicolumn{8}{c}{\tel~(K)} \\
\cline{3-10}
~ & ~ & \multicolumn{2}{c}{NGC\,6153} & ~ & \multicolumn{2}{c}{M\,1-42} & ~ & \multicolumn{2}{c}{Hf\,2-2} \\
\cline{3-4}
\cline{6-7}
\cline{9-10}
~ & ~ & \multicolumn{1}{l}{20 minutes$^{*}$} & \multicolumn{1}{l}{Literature$^{\rm a}$} & 
~ & \multicolumn{1}{l}{30 minutes$^{*}$} & \multicolumn{1}{l}{Literature$^{\rm b}$} & 
~ & \multicolumn{1}{l}{30 minutes$^{*}$} & \multicolumn{1}{l}{Literature$^{\rm c}$} \\
\hline
1 & [N {\sc ii}] ($\lambda$6548+$\lambda$6584)/$\lambda$5754 & 
10550$^{100}_{150}$ & 10200 &~&
10000$^{750}_{1000}$ &  8900 &~&
12500$^{400}_{350}$ & 14500 \\ [0.9ex]
2 & [O {\sc iii}] ($\lambda$4958+$\lambda$5007)/$\lambda$4363 &  
9850$^{350}_{400}$ &  9100 &~&
8900$^{300}_{300}$ &  9200 &~&
9650$^{450}_{400}$ &  8700 \\ [0.9ex]
3 & [S {\sc iii}] ($\lambda$9069+$\lambda$9531)/$\lambda$6312 &  
17050$^{500}_{600}$ &  9100 &~&
14750$^{1000}_{1500}$ &  9200 &~&
11250$^{1000}_{1000}$ &  8700 \\ [0.9ex]
4 & [O {\sc ii}] ($\lambda$7320+$\lambda$7330)/($\lambda$3726 + $\lambda$3729) & 
11800$^{4550}_{1000}$ & 17900$^{\rm d}$ &~&
14000$^{2000}_{2000}$ & 16850 &~&
13200$^{1000}_{1001}$ &     ~ \\ [0.9ex]
5 & [Ar {\sc iii}] ($\lambda$7135+$\lambda$7752)/$\lambda$5192 & 
9350$^{500}_{250}$ &  9200$^{\rm e}$ &~&
18900$^{650}_{600}$ &  7100$^{\rm e}$ &~&
    ~$^{    ~}_{    ~}$ &     ~ \\ [0.9ex]
6 & [S {\sc ii}] ($\lambda$6717+$\lambda$6731)/($\lambda$4069+$\lambda$4076)$^{\rm f}$ & 
11850$^{1500}_{500}$ &     ~ &~&
5500$^{1500}_{1000}$ &     ~ &~&
9250$^{1000}_{1000}$ &     ~ \\ [0.9ex]
~ & Average CEL temperature &  
11750$\pm$2800 & 11600$\pm$4200 &~&
12050$\pm$4800 & 10200$\pm$3400 &~&
11200$\pm$1750 & 11600$\pm$4000 \\ [0.9ex]
7 & [Ne {\sc iii}] ($\lambda$3868+$\lambda$3967)/$\lambda$3342 & 
25400$^{3650}_{3600}$ &  8600$^{\rm g}$ &~&
24450$^{2000}_{1750}$ &  8800$^{\rm g}$ &~&
23000$^{1500}_{1500}$ &     ~ \\ [0.9ex]
~ & He I $\lambda$7281/$\lambda$6678 & 
3100$^{450}_{400}$ &     ~ &~&
3100$^{450}_{400}$ &     ~ &~&
3300$^{550}_{350}$ &     ~\\ [0.9ex] 
~ & He I $\lambda$7281/$\lambda$5876 & 
3800$^{500}_{650}$ &     ~ &~&
3800$^{500}_{650}$ &     ~ &~&
3400$^{250}_{400}$ &      \\ [0.9ex] 
~ & He I $\lambda$6678/$\lambda$4471 & 
3250$^{350}_{450}$ &     ~ &~&
3250$^{350}_{450}$ &     ~ &~&
2700$^{400}_{250}$ &     ~\\ [0.9ex] 
~ & \hei~$\lambda$5876/$\lambda$4471 & 
2900$^{300}_{250}$ &     ~ &~&
2900$^{300}_{250}$ &     ~ &~&
2750$^{350}_{200}$ &     ~\\ [0.9ex] 
~ & \hi~BJ/H11 &  
6250$^{150}_{100}$ &  6100 &~&
3450$^{200}_{250}$ &  3560 &~&
1050$^{250}_{150}$ &   930\\ [0.9ex] 
~ & \hi~PJ/P11 &  
2500$^{500}_{650}$ &     ~ &~&
600$^{150}_{200}$ &     ~ &~&
550$^{300}_{250}$ &     ~\\ [0.9ex]
\hline
~ & ~ & \multicolumn{8}{c}{$\log$~\nel~[cm$^{-3}$]} \\
\cline{3-10}
\cline{3-10}
~ & ~ & \multicolumn{2}{c}{NGC\,6153} & ~ & \multicolumn{2}{c}{M\,1-42} & ~ & \multicolumn{2}{c}{Hf\,2-2} \\
\cline{3-4}
\cline{6-7}
\cline{9-10}
~ & ~ & \multicolumn{1}{l}{20 minutes$^{*}$} & \multicolumn{1}{l}{Literature$^{\rm a}$} & 
~ & \multicolumn{1}{l}{30 minutes$^{*}$} & \multicolumn{1}{l}{Literature$^{\rm b}$} & 
~ & \multicolumn{1}{l}{30 minutes$^{*}$} & \multicolumn{1}{l}{Literature$^{\rm c}$} \\
\hline
8 & [Ar {\sc iv}] $\lambda$4740/$\lambda$4711 &   
3.71$^{0.15}_{0.10}$ & 3.40 &~&
3.82$^{0.05}_{0.05}$ & 2.80 &~&
4.18$^{0.65}_{0.40}$ &    ~ \\ [0.9ex]
9 & [Cl {\sc iii}] $\lambda$5537/$\lambda$5517 &   
3.23$^{0.10}_{0.05}$ & 3.60 &~&
3.20$^{0.05}_{0.15}$ & 3.20 &~&
3.22$^{0.15}_{0.25}$ &    ~ \\ [0.9ex]
10 & [S {\sc ii}] $\lambda$6731/$\lambda$6716$^{\rm g}$ &   
2.60$^{0.25}_{0.40}$ & 3.60 &~&
2.30$^{0.40}_{0.05}$ & 3.10 &~&
2.90$^{0.10}_{0.05}$ & 2.60 \\ [0.9ex]
11 & [O {\sc ii}] $\lambda$3729/$\lambda$3726 &   
3.46$^{0.10}_{0.10}$ &    ~ &~&
3.19$^{0.05}_{0.05}$ & 3.20 &~&
2.99$^{0.15}_{0.10}$ & 3.20 \\ [0.9ex]
~ & Average CEL density &   
3.47$\pm$2.43 & 3.50$\pm$0.10 &~&
3.40$\pm$3.59 & 3.00$\pm$0.20 &~&
3.46$\pm$6.27 & 2.90$\pm$0.30 \\ [0.9ex]
12 & [N {\sc i}] $\lambda$5200/$\lambda$5198$^{\rm h}$ &   
2.88$^{0.20}_{0.05}$ &    ~ &~&
2.87$^{0.10}_{0.05}$ &    ~ &~&
2.70$^{0.30}_{0.10}$ &    ~ \\ [0.9ex]
~ & Balmer Decrement & 
4.0$^{0.75}_{0.50}$ &     3.3 &~&
4.2$^{0.35}_{0.60}$ &     $\leq$4.0 &~&
5.0$^{0.50}_{0.95}$ &     2.5\\ [0.9ex]
~ & Paschen Decrement & 
4.0$^{0.25}_{0.35}$ &     ~ &~&
4.5$^{1.20}_{1.00}$ &     ~ &~&
4.2$^{1.50}_{0.75}$ &     ~\\ [0.9ex]
\hline
\end{tabular}
\begin{description}
\item$^{*}$The super- and subscripts are the errors.
\item[$^{\rm a}$] \citet{LSB00}, observations of the entire nebula.
\item[$^{\rm b}$] \citet{LLB01}.
\item[$^{\rm c}$] \citet{LBZ06}, 2~arcsec~wide-slit spectroscopy.
\item[$^{\rm d}$] From the line ratio ($\lambda$7320+$\lambda$7330)/$\lambda$3727.
\item[$^{\rm e}$] From the line ratio $\lambda$7135/$\lambda$5192.
\item[$^{\rm f}$] The critical densities of the [\sii] $\lambda\lambda$6716 and 6731 lines are only 1400 and 3800 cm$^{-3}$, respectively. Thus, at least the 6716 line could be suppressed by electron collisions, i.e. collisionally de-excited and the derived density could be unreliable.
\item[$^{\rm g}$] From the line ratio 15.5$\mu$m/($\lambda$3868+$\lambda$3967).
\item[$^{\rm h}$] The critical densities of the [N~{\sc i}] $\lambda\lambda$5198 and 5200 lines are 1800 and 780 cm$^{-3}$, respectively.  Thus, these two lines could be collisionally de-excited, i.e. suppressed by electron collisions.
\end{description}
\label{tab:plasmadiag}
\end{table*}

\section{Plasma Diagnostics}

Electron temperatures and densities of NGC\,6153, M\,1-42 and Hf\,2-2 
were deduced using the CEL and ORL ratios of heavy elements detected in 
the spectra of the three PNe.  Electron temperatures can also be derived 
from the hydrogen recombination continuum discontinuities \citep{ZLW04}, 
the Balmer jump at 3646\,{\AA} and the Paschen jump at 8204\,{\AA}. 
Electron densities can be constrained using the high-order Balmer and 
Paschen decrements \citep[e.g.,][]{LSB00}.  The He~{\sc i} recombination 
line ratios can also be used as temperature diagnostics \citep{ZLL05b}. 
Recombination may enhance the most commonly used temperature-sensitive 
[\nii], [\oii] and [\oiii] auroral lines (\citealt{LSB00}). This 
possibility is discussed below, with the amounts of enhancements 
estimated.

%\clearpage
\begin{figure*}
\centering
\begin{tabular}{cc}
\epsfig{file=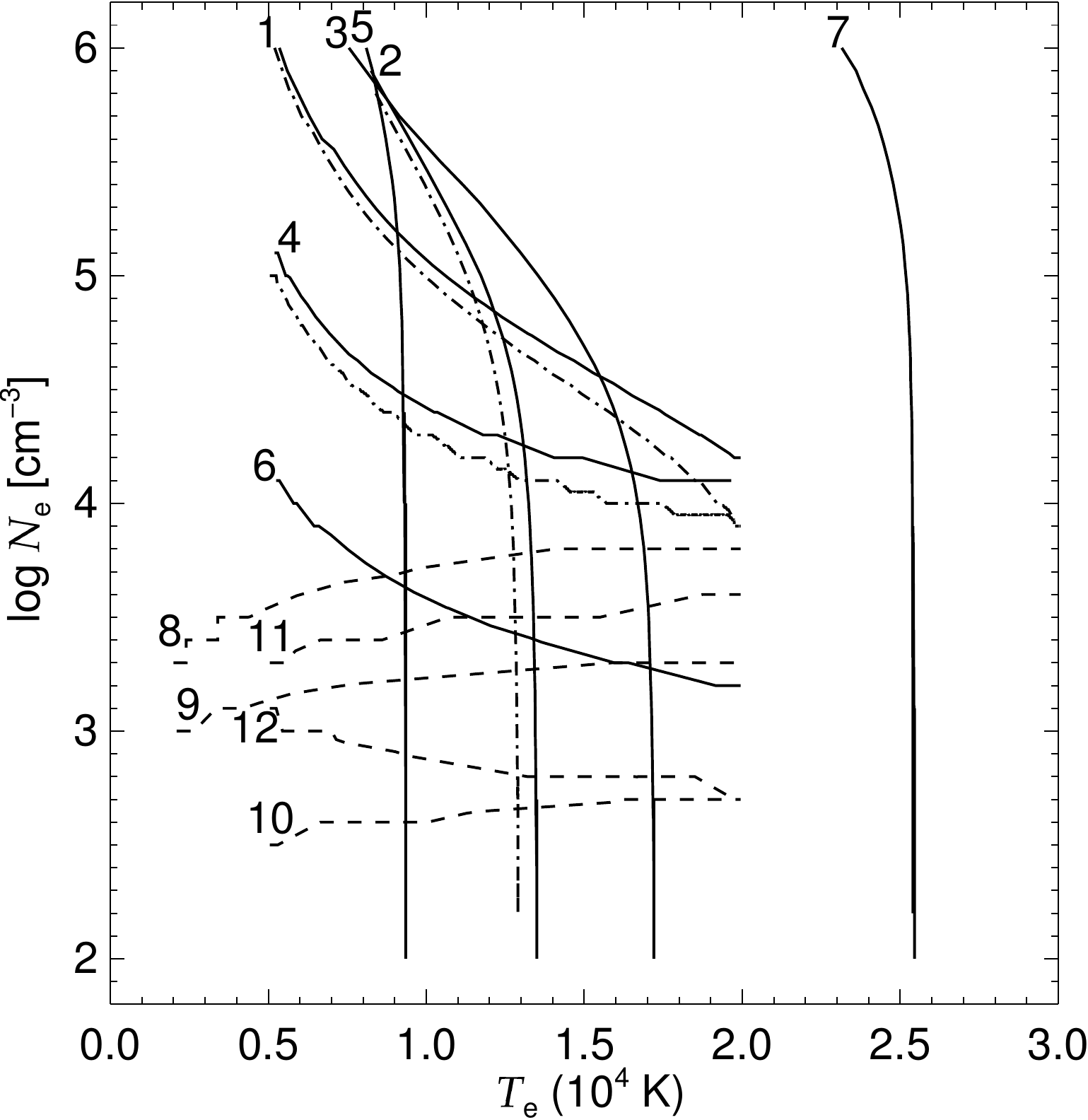,width=0.425\textwidth,angle=0}
\epsfig{file=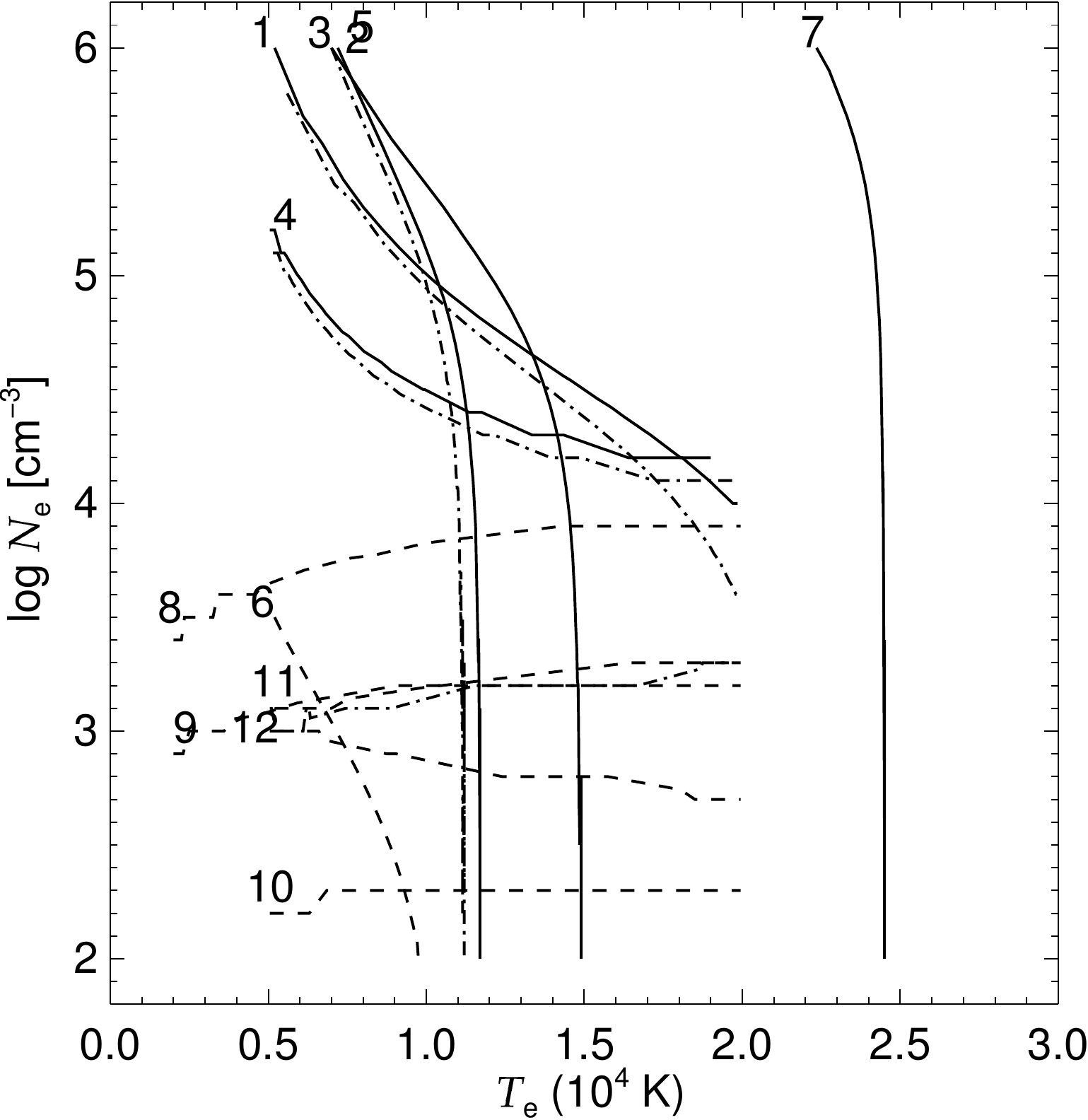,width=0.425\textwidth,angle=0}
\end{tabular}
\vskip0.1truein
\begin{tabular}{c}
\epsfig{file=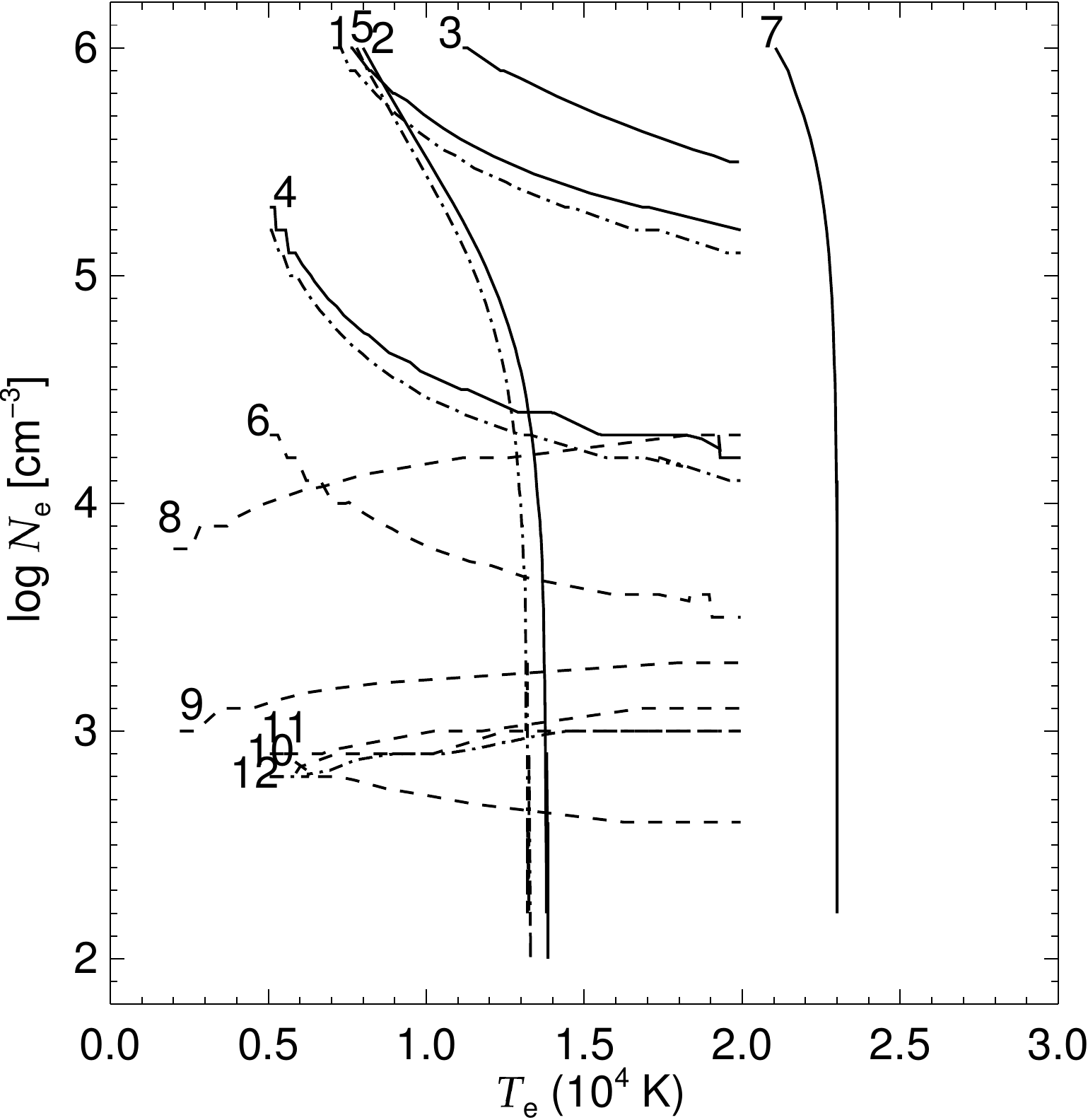,width=0.425\textwidth,angle=0}
\end{tabular}
\caption{Plasma diagnostic diagrams for the long-exposure spectra of 
NGC\,6153 (top-left panel), M\,1-42 (top-right panel) and Hf\,2-2 
(bottom panel).  Each curve is labeled with an ID number given in 
Table~\ref{tab:plasmadiag}.  The solid curves are the temperature 
diagnostics, and the dashed curves the density diagnostics. The 
dash-dot curves represent the temperature-diagnostic curves of 
[\nii], [\oii] and [\oiii] after correction of auroral line fluxes 
for the contribution from recombination excitation, and the solid 
curves represent those without corrections.} 
\label{fig:plasmadiag}
\end{figure*}

\subsection{Plasma diagnostics based on CELs}

CELs are the traditional tools for plasma diagnostics and abundance 
determinations.  Numerous CELs were detected in the echelle spectra of 
NGC\,6153, M\,1-42 and Hf\,2-2.  We derived physical properties (electron 
temperatures and densities) and then derived ionic abundances using the 
fluxes of the CELs measured in the spectra.  This analysis was carried 
out with the aid of the {\sc equib} code, which was originally developed
by \citet{ha81} to solve the equations of statistical equilibrium in 
multi-level atoms in nebular conditions. 
%and yielded level populations and line emissivities for a specified 
%physical condition, appropriate to the zones where the ions are 
%expected to exist. 
The atomic data used by the {\sc equib} code were updated manually. 
References for the atomic data set used for plasma diagnostics based on 
CELs, as well as for ionic abundance determinations, are presented in 
Table~\ref{tab:atomicdata}.  Results of our plasma diagnostics are 
presented in Table~\ref{tab:plasmadiag}.  Discussion of the physical 
conditions yielded by our plasma diagnostics is presented in Section~7.2.

\begin{figure*}
\centering
\begin{tabular}{cc}
\epsfig{file=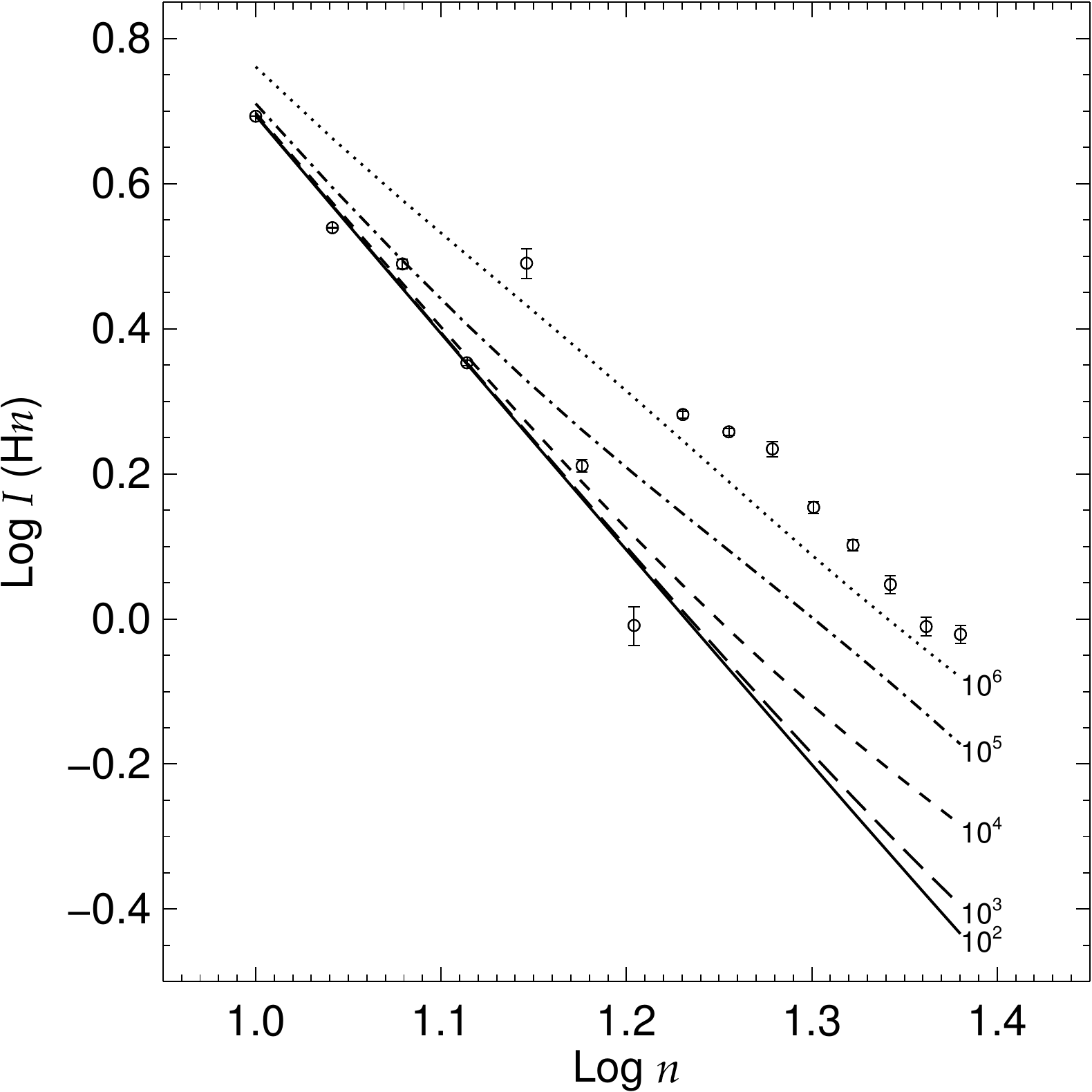,width=0.45\textwidth,angle=0}
\epsfig{file=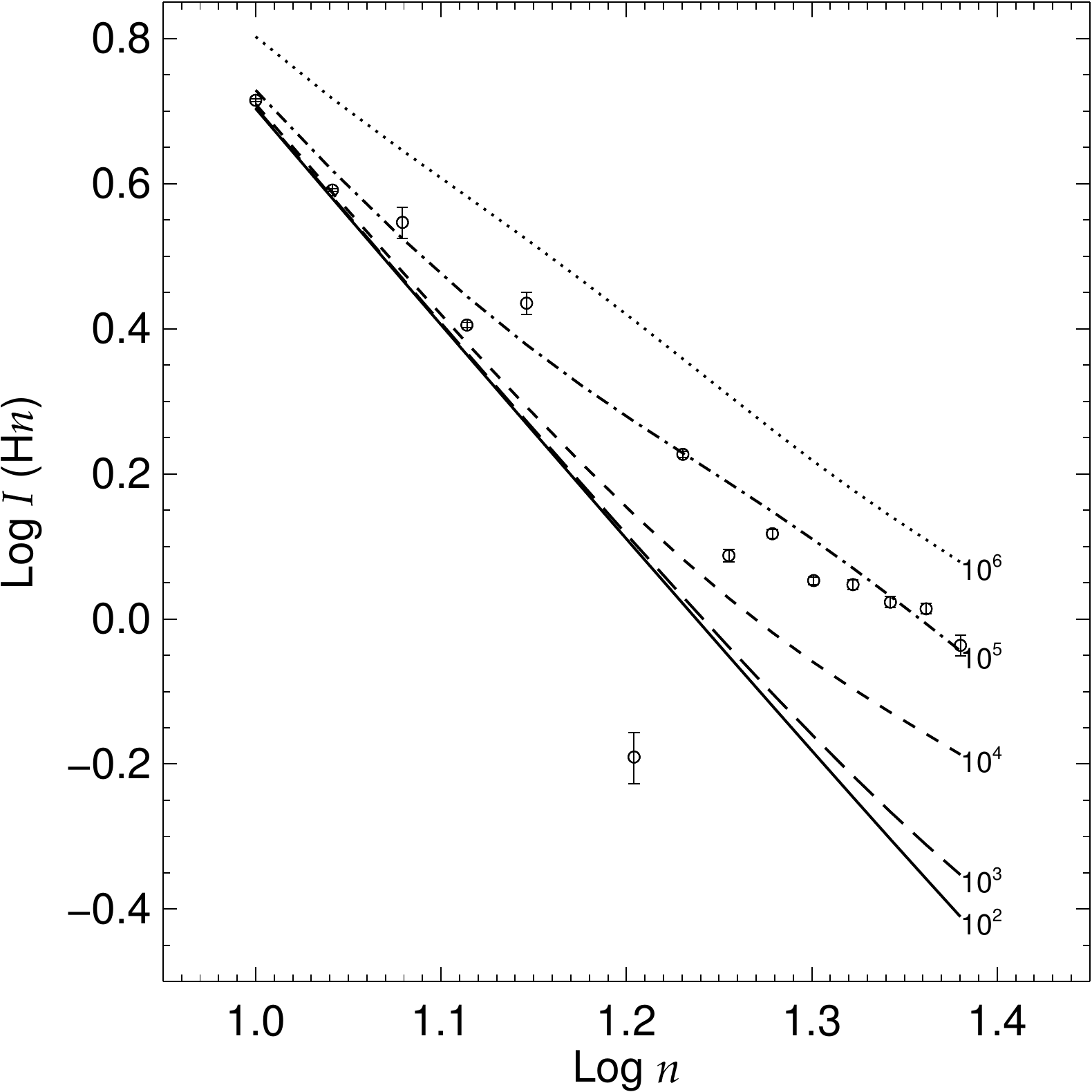,width=0.45\textwidth,angle=0}
\end{tabular}
\vskip0.1truein
\begin{tabular}{c}
\epsfig{file=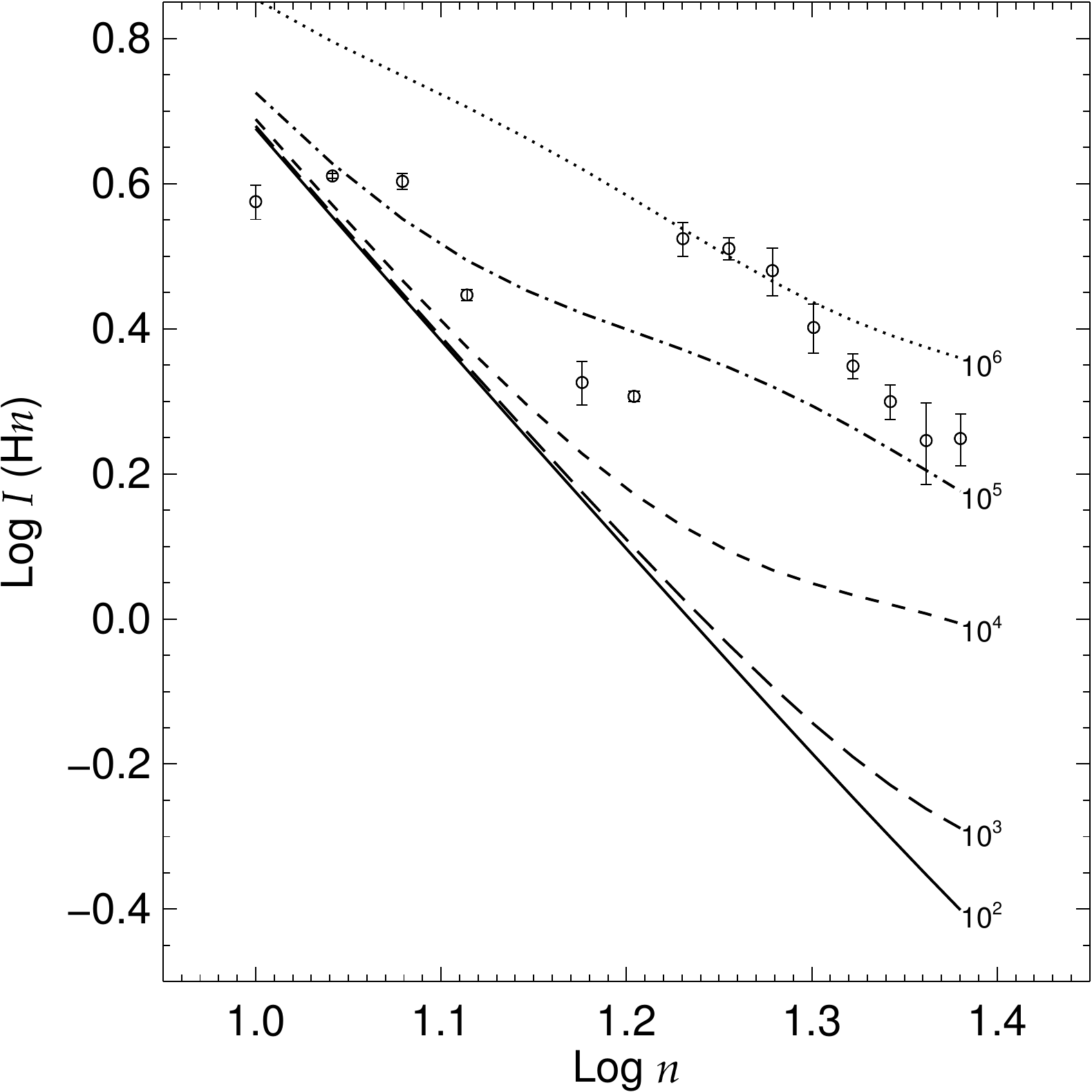,width=0.45\textwidth,angle=0}
\end{tabular}
\caption{Observed intensities (normalized to H$\beta$ = 100) of 
high-$n$ ($n$ = 10, 11, 12, ..., 24) Balmer lines as a function 
of the principal quantum number $n$ for NGC\,6153 (top-left panel), 
M\,1-42 (top-right panel) and Hf\,2-2 (bottom panel).  Some of 
the high-order Balmer lines may be overestimated due to line 
blending.  The curves are the predicted Balmer decrements for a 
number of density cases, from 10$^{2}$ to 10$^{6}$~cm$^{-3}$.  The 
curves are created using the hydrogenic theory of \citet{SH95}. 
The assumed electron temperatures are presented in 
Table~\ref{tab:plasmadiag} for NGC\,6153, M\,1-42 and Hf\,2-2.}
\label{fig:balmerdec}
\end{figure*}

\begin{figure*}
\centering
\begin{tabular}{cc}
\epsfig{file=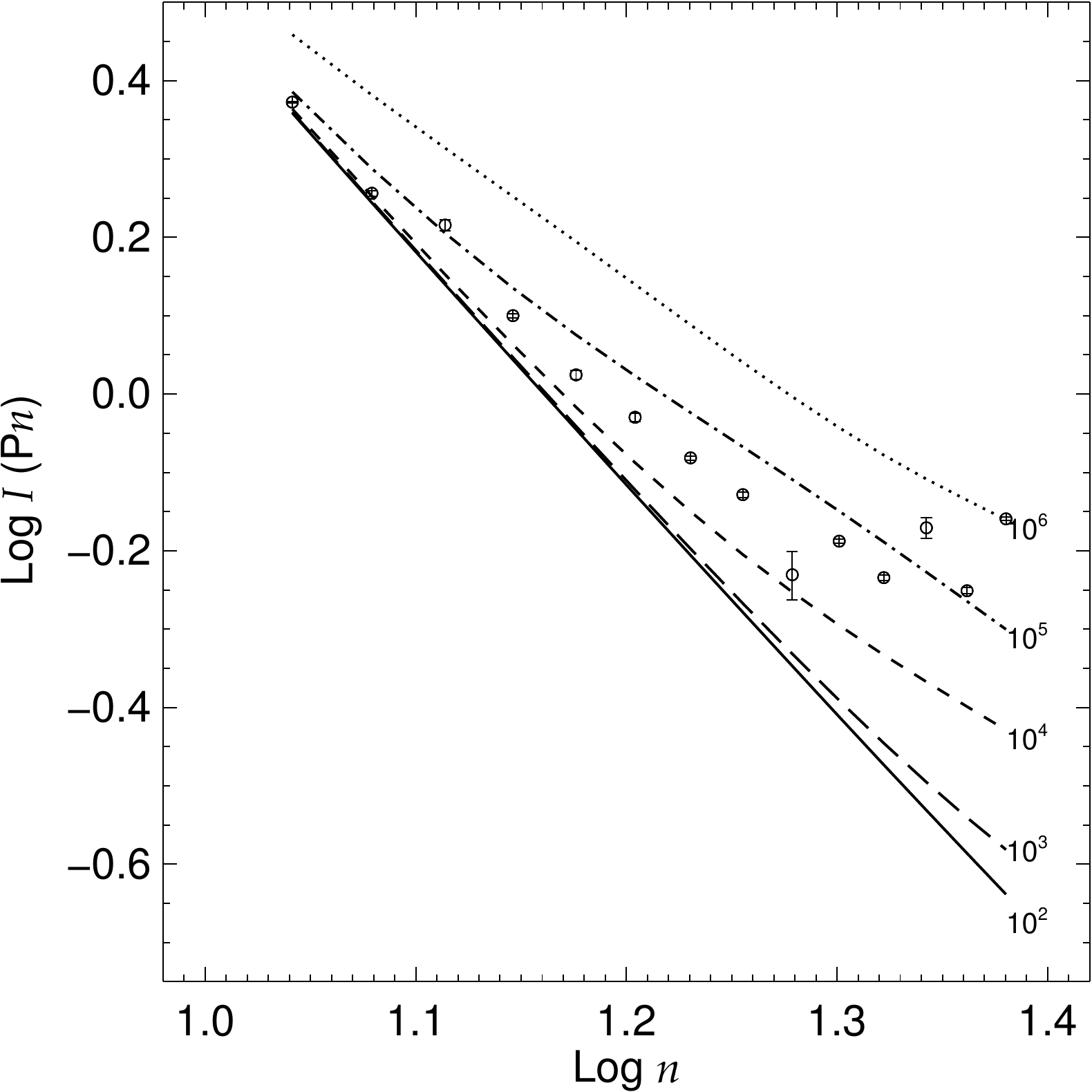,width=0.45\textwidth,angle=0}
\epsfig{file=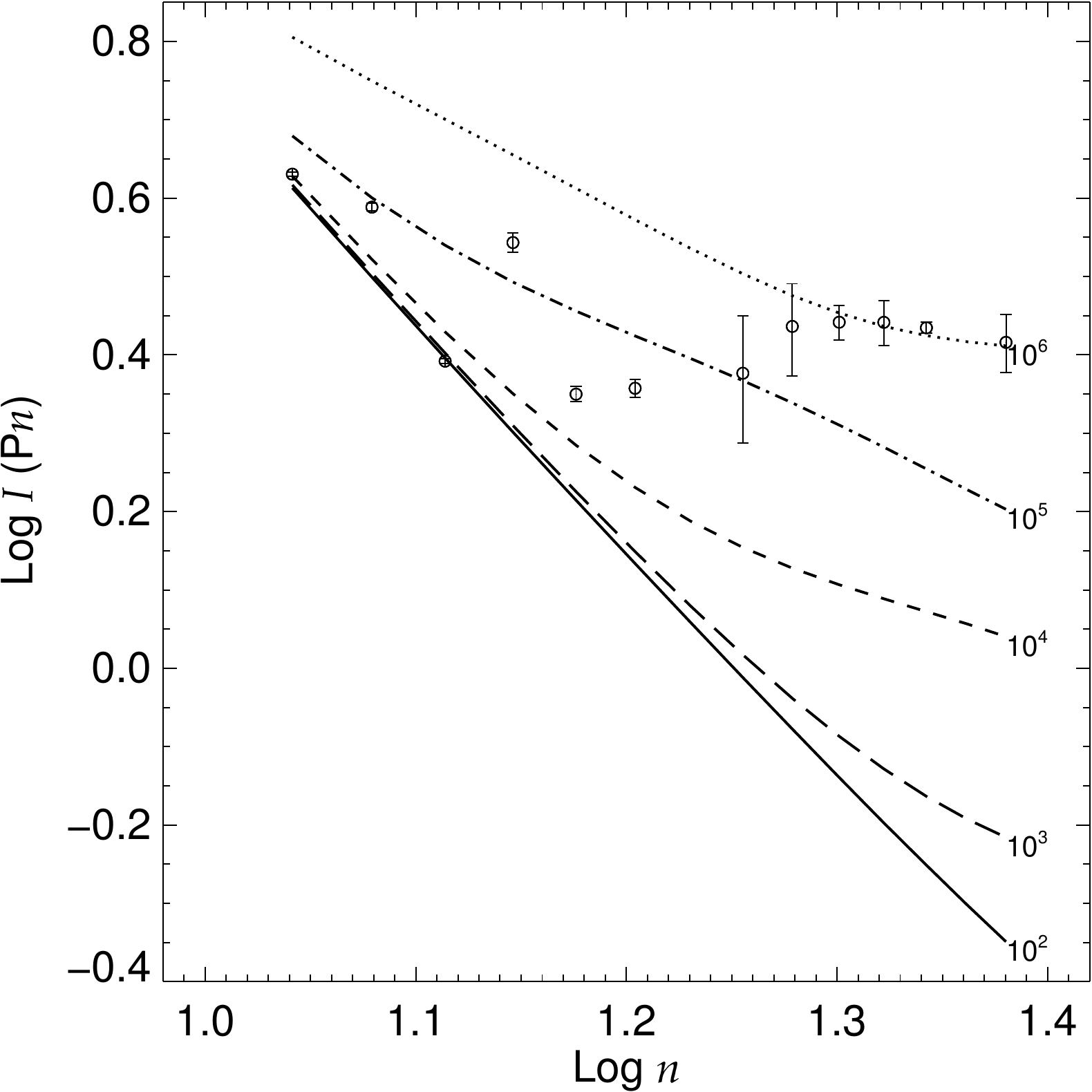,width=0.45\textwidth,angle=0}
\end{tabular}
\vskip0.1truein
\begin{tabular}{c}
\epsfig{file=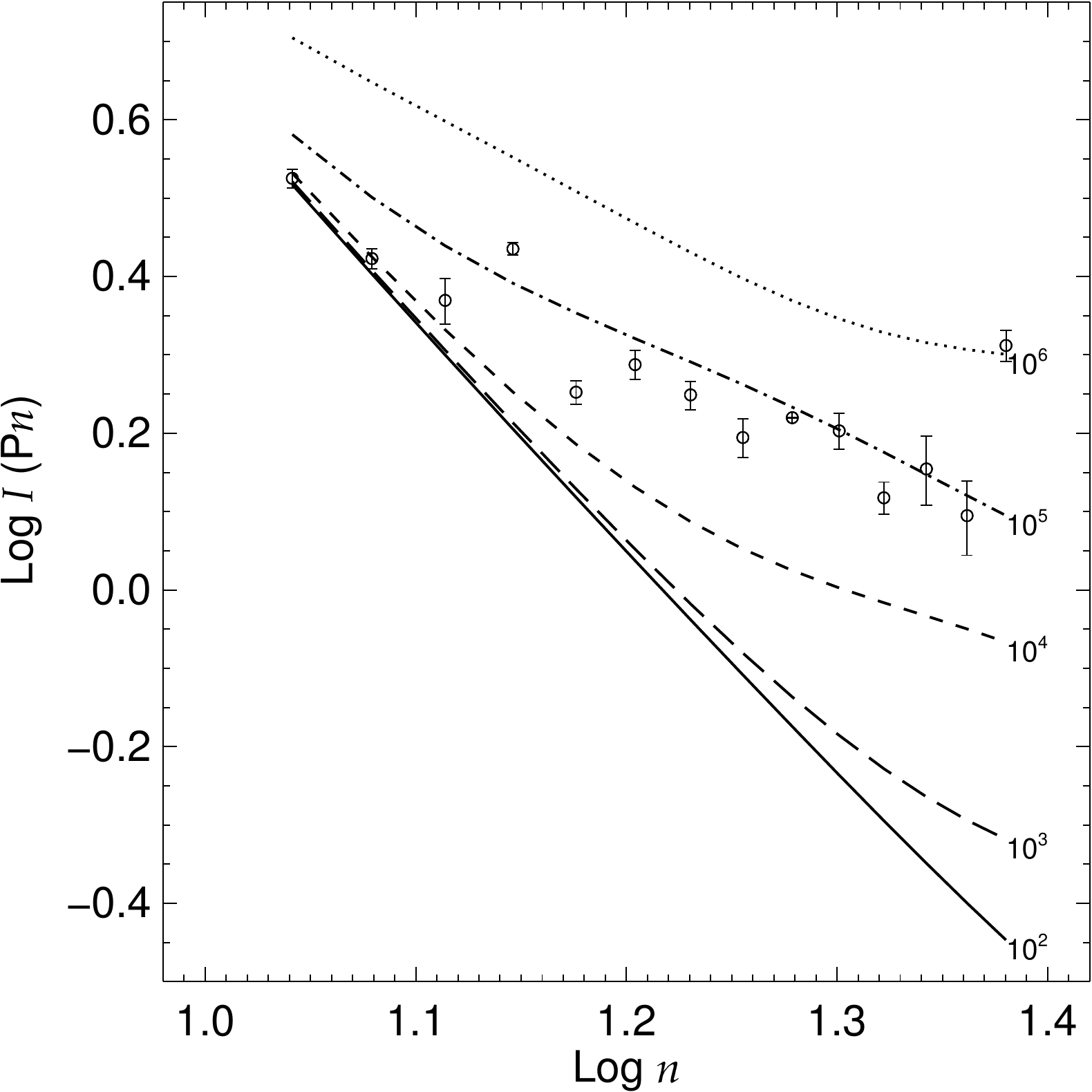,width=0.45\textwidth,angle=0}
\end{tabular}
\caption{Observed intensities (normalized to H$\beta$ = 100) of 
high-$n$ ($n$ = 11, 12, ..., 23) Paschen lines as a function of 
the principal quantum number $n$ for NGC\,6153 (top-left panel), 
M\,1-42 (top-right panel) and Hf\,2-2 (bottom panel).  The intensities 
of P22, P23 and P24 may be overestimated due to line blending.  The 
curves are the predicted Paschen decrements for a number of density 
cases, from 10$^{2}$ to 10$^{6}$~cm$^{-3}$.  The assumed electron 
temperature is presented in Table~\ref{tab:plasmadiag} for NGC\,6153, 
M\,1-42 and Hf\,2-2.}
\label{fig:paschendec}
\end{figure*}

The electron temperatures were derived by assuming a constant electron 
density, which is averaged from the densities derived from the [\oii], 
[\cliii] and [\ariv] line ratios.  The temperatures that are either 
too low or too high were not included in the average calculation. 
The [\oi] ($\lambda$6300 + $\lambda$6363)/$\lambda$5577 temperature 
diagnostic ratio is not used because of possible residuals from the 
sky subtraction.

Fig.~\ref{fig:plasmadiag} shows the plasma diagnostic curves based on 
the CEL ratios measured in the long-exposure spectra of NGC\,6153, 
M\,1-42 and Hf\,2-2.  As previously mentioned, the [\nii], [\oii] and 
[\oiii] auroral lines could be enhanced by the recombination process. 
The diagnostic curves derived for these three ions before and after 
correction for the recombination contribution are represented with 
different line types in Fig.~\ref{fig:plasmadiag} (see the figure 
caption for details).  If a PN has uniform physical condition, the 
electron temperatures and densities derived from different diagnostics 
ratios should be confined within a narrow region in the diagram. 
These curves do not intersect at a single point, indicating the 
temperature and density inhomogeneity in the three PNe 
\citep[e.g.,][]{K00}.

%\clearpage
%\begin{landscape}
%\begin{rotate}{180}
\begin{table*}
\centering
\setlength{\tabcolsep}{0.05in}
\caption{CEL ionic abundances.}
%\vskip-0.1truein
\begin{tabular}{lclrrcrrcrr}
\hline
\hline
Ion & Spectrum & Lines Used & \multicolumn{8}{c}{Abundances ($\times$10$^{-5}$)} \\
\cline{4-11}
~ & ~ & ~ & \multicolumn{2}{c}{NGC\,6153} & ~ & \multicolumn{2}{c}{M\,1-42} & ~ & \multicolumn{2}{c}{Hf\,2-2} \\
\cline{4-5}
\cline{7-8}
\cline{10-11}
~ & ~ & ~ & \multicolumn{1}{c}{20 minutes$^{*}$} & \multicolumn{1}{c}{Literature$^{\rm a}$} & ~ & 
                     \multicolumn{1}{c}{30 minutes$^{*}$} & \multicolumn{1}{c}{Literature$^{\rm b}$} & ~ & 
                     \multicolumn{1}{c}{30 minutes$^{*}$} & \multicolumn{1}{c}{Literature$^{\rm c}$} \\
\hline
N$^{0}$& [N~{\sc i}] & $\lambda\lambda$5198.5200 &  
12.90$^{(9.9)}_{(2.5)}$&   ~ & ~ & 
16.70$^{(4.3)}_{(2.1)}$&   ~ & ~ & 
10.07$^{(13.3)}_{(3.5)}$&   ~ \\ [0.9ex]
N$^{+}$& [N~{\sc ii}] & $\lambda\lambda$5755, 6548, 6583 &   
1.29$^{(8.0)}_{(5.3)}$&    1.24 & ~ & 
5.30$^{(10.5)}_{(6.1)}$&  5.86 & ~ & 
0.87$^{(6.0)}_{(6.2)}$&    0.97 \\ [0.9ex]
O$^{+}$& [O~{\sc ii}] & $\lambda\lambda$3726+29, 7320+30 &   
2.88$^{(3.2)}_{(3.9)}$&    2.62 & ~ & 
4.24$^{(4.3)}_{(5.5)}$&  3.88 & ~ & 
2.74$^{(6.4)}_{(10.1)}$&    2.13 \\ [0.9ex]
O$^{+}$& [O~{\sc ii}] & $\lambda\lambda$3726+29 &   
2.14$^{(9.4)}_{(9.4)}$&    2.62 & ~ & 
3.46$^{(3.2)}_{(3.2)}$&  3.88 & ~ & 
2.62$^{(6.0)}_{(4.0)}$&    2.13 \\ [0.9ex]
O$^{2+}$& [O~{\sc iii}] & $\lambda\lambda$4363, 4959, 5007 &  
38.54$^{(8.3)}_{(6.2)}$&  43.30 & ~ & 
21.59$^{(12.8)}_{(8.7)}$& 24.90 & ~ & 
10.82$^{(14.0)}_{(13.8)}$&  10.50 \\ [0.9ex]
Ne$^{2+}$& [Ne~{\sc iii}] & $\lambda\lambda$3342, 3869, 3968 &  
9.66$^{(13.8)}_{(26.3)}$&  14.40 & ~ & 
8.53$^{(10.5)}_{(9.5)}$&  9.62 & ~ & 
3.74$^{(11.8)}_{(12.3)}$&   3.60 \\ [0.9ex]
S$^{+}$& [S~{\sc ii}] & $\lambda\lambda$4068+76, 6716+31 &   
0.05$^{(1.1)}_{(3.3)}$&    0.05 & ~ & 
0.13$^{(1.9)}_{(0.9)}$&  0.13 & ~ & 
0.05$^{(4.0)}_{(0.8)}$&   0.03 \\ [0.9ex]
S$^{+}$& [S~{\sc ii}] & $\lambda\lambda$6716+31 &   
0.04$^{(6.2)}_{(7.7)}$&    0.05 & ~ & 
0.20$^{(2.9)}_{(0.9)}$&  0.13 & ~ & 
0.05$^{(3.6)}_{(1.4)}$&   0.03 \\ [0.9ex]
S$^{2+}$& [S~{\sc iii}] & $\lambda\lambda$6312, 9069, 9532 &  
0.42$^{(4.4)}_{(4.3)}$&   0.46 & ~ & 
0.62$^{(7.2)}_{(7.5)}$&  0.47 & ~ & 
0.12$^{(2.9)}_{(2.1)}$&    0.12 \\ [0.9ex]
Cl$^{2+}$& [Cl~{\sc iii}] & $\lambda\lambda$5517, 5537 &   
0.01$^{(12.3)}_{(6.1)}$&    0.01 & ~ &
0.02$^{(6.6)}_{(13.7)}$&  0.01 & ~ & 
0.01$^{(18.5)}_{(10.2)}$&    ~ \\ [0.9ex]
Ar$^{2+}$& [Ar~{\sc iii}] & $\lambda\lambda$5182, 7136, 7751 &   
0.23$^{(14.9)}_{(4.9)}$&    0.18 & ~ & 
0.18$^{(3.9)}_{(2.0)}$& 0.19 & ~ & 
~&    0.07 \\ [0.9ex]
Ar$^{3+}$& [Ar~{\sc iv}] & $\lambda\lambda$4711, 4740 &   
0.07$^{(2.6)}_{(1.7)}$&   0.06 & ~ & 
0.05$^{(1.3)}_{(1.3)}$&  0.04 & ~ & 
0.04$^{(2.4)}_{(1.7)}$&    ~ \\ [0.9ex]
\hline
\end{tabular}
\begin{description}
\item[$^{\rm *}$] Values in parentheses are the propagated uncertainties based on observational error (in per cent). The super- and subscripts are the errors added to and subtracted from the observed fluxes, respectively, and then corrected for extinction.
\item[$^{\rm a}$] \citet{LSB00}, the observations of the entire nebula.
\item[$^{\rm b}$] \citet{LLB01}.
\item[$^{\rm c}$] \citet{LBZ06}, 2~arcsec~wide-slit spectroscopy.
\end{description}
\label{tab:celabund}
\end{table*}
%\end{landscape}
%\end{rotate}

\subsection{Enhancement of the [\nii], [\oii] and [\oiii] auroral 
lines by recombination excitation}

In NGC\,6153, M\,1-42 and Hf\,2-2, the nitrogen and oxygen atoms are 
mostly in their doubly-ionized stages, N$^{2+}$ and O$^{2+}$. 
Recombination of the N$^{2+}$ and O$^{2+}$ ions (to N$^{+}$ and O$^{+}$,
respectively) may contribute to the excitation of the weak 
[\nii] $\lambda$5754 auroral line and the [\oii] auroral 
$\lambda$$\lambda$7320,7330 and nebular $\lambda$$\lambda$3726,3729 
lines \citep{LSB00}.  This could result in the overestimated 
electron temperatures derived from the ($\lambda$6548 + 
$\lambda$6584)/$\lambda$5754 and ($\lambda$7320 + 
$\lambda$7330)/$\lambda$3727 line ratios \citep{R86}.  Similarly, 
recombination could also enhance the [\oiii] $\lambda$4363 auroral 
line, thereby leading to an overestimated electron temperature.  We 
estimated the contribution by recombination using the empirical fitting 
formulae of \citet[][Equations~1, 2 and 3 therein]{LSB00}, which were 
derived from the atomic data of N$^{+}$, O$^{+}$ and O$^{2+}$. 
%The three equations are as follows

%\begin{equation}
%\frac{I_{R}(\lambda5754)}{I({\rm H}\beta)} = 3.19 t^{0.30} \times \frac{{\rm N}^{2+}}{{\rm H}^{+}}, (0.5 \leq t \leq 2.0), 
%\end{equation}

%\begin{equation}
%\frac{I_{R}(\lambda7320+\lambda7330)}{I({\rm H}\beta)} = 9.36 t^{0.44} \times \frac{{\rm O}^{2+}}{{\rm H}^{+}}, (0.5 \leq t \leq 1.0), 
%\end{equation}
%\noindent
%and

%\begin{equation}
%\frac{I_{R}(\lambda4363)}{I({\rm H}\beta)} = 12.4 t^{0.59} \times \frac{{\rm O}^{3+}}{{\rm H}^{+}}, 
%\end{equation}
%\noindent
%where $t$ = $T_\mathrm{e}$/10$^{4}$ [K].

We estimated that recombination only contributes less than 1 per cent 
to the total intensity of the [N~{\sc ii}] $\lambda$5754 auroral line. 
Here we used an N$^{2+}$/H$^{+}$ recombination line abundance of 
1.75$\times$10$^{-3}$ for NGC\,6153 \citep{LSB00}, 
2.86$\times$10$^{-3}$ for M\,1-42 \citep{LLB01} and 
2.72$\times$10$^{-3}$ for Hf\,2-2 \citep{LBZ06}.  For the 
O$^{2+}$/H$^{+}$ recombination line abundance, we adopted a 
value of 4.07$\times$10$^{-3}$ for NGC\,6153 \citep{LSB00}, 
5.59$\times$10$^{-3}$ for M\,1-42 \citep{LLB01}, and 
7.48$\times$10$^{-3}$ for Hf\,2-2 \citep{LBZ06}.  For the 
O$^{3+}$/H$^{+}$ recombination line abundance, we adopted 
2.37$\times$10$^{-4}$ for NGC\,6153 \citep{LSB00}, 
3.26$\times$10$^{-4}$ for M\,1-42 \citep{LLB01}, and 
4.35$\times$10$^{-4}$ for Hf\,2-2 \citep{LBZ06}.  Here we adopted 
the average temperatures deduced from CEL diagnostic ratios listed 
in Table \ref{tab:plasmadiag} to calculate the contribution by 
recombination.  The corrected electron temperatures were then used 
to derive an average value, which was adopted in the abundance 
calculations.

%\clearpage
%\begin{landscape}
\begin{table*}
\centering
\setlength{\tabcolsep}{0.05in}
\caption{Ionic and elemental abundances of helium.}
\begin{tabular}{lcrrcrrcrr}
\hline
\hline
He$^{i+}$/H$^{+}$ & Line & \multicolumn{8}{c}{Abundance} \\
\cline{3-10}
~ & (\AA) &  \multicolumn{2}{c}{NGC\,6153} & ~ & 
\multicolumn{2}{c}{M\,1-42} & ~ & 
\multicolumn{2}{c}{Hf\,2-2} \\
\cline{3-4}
\cline{6-7}
\cline{9-10}
~ & ~ & \multicolumn{1}{c}{20 minutes}  & \multicolumn{1}{c}{Literature$^{\rm a}$} & ~ & 
\multicolumn{1}{c}{30 minutes} & \multicolumn{1}{c}{Literature$^{\rm b}$} & ~ & 
\multicolumn{1}{c}{30 minutes} & \multicolumn{1}{c}{Literature$^{\rm c}$} \\
\hline
He$^{+}$/H$^{+}$ & He~{\sc i}~$\lambda$4471 &  
0.130$^{(13.8)}_{(14.0)}$&  0.123  & ~& 
0.113$^{(6.1)}_{(6.3)}$&  0.142 &~& 
0.142$^{(2.4)}_{(2.4)}$&  0.102 \\ [0.9ex]
He$^{+}$/H$^{+}$ & He~{\sc i}~$\lambda$5876 &  
0.109$^{(12.3)}_{(12.4)}$&   0.127 & ~& 
0.148$^{(5.3)}_{(5.3)}$&   0.141 & ~& 
0.116$^{(4.4)}_{(4.4)}$&   0.103\\ [0.9ex]
He$^{+}$/H$^{+}$ & He~{\sc i}~$\lambda$6678 &  
0.122$^{(4.5)}_{(4.6)}$&   ~& ~ &
0.143$^{(4.8)}_{4.6)}$&   0.138 & ~& 
0.095$^{(12.6)}_{(13.8)}$&   0.099\\ [0.9ex]
He$^{+}$/H$^{+}$ & Mean &  
0.120$^{(10.2)}_{(10.3)}$&   0.123 & ~& 
0.135$^{(5.3)}_{(5.4)}$&  0.139 & ~& 
0.117$^{(4.4)}_{(4.6)}$&   0.102\\ [0.9ex]
He$^{+}$/H$^{+}$ & He~{\sc i}~$\lambda$7281 &  
0.104$^{(13.6)}_{(13.6)}$&   0.133& ~ &
0.102$^{(9.1)}_{9.5)}$&   0.095 & ~& 
0.106$^{(7.6)}_{(7.8)}$&   ~\\ [0.9ex]
He$^{2+}$/H$^{+}$ & He~{\sc ii}~$\lambda$4686 &  
0.010$^{(12.0)}_{(12.1)}$&   0.011 & ~& 
0.010$^{(8.7)}_{(8.8)}$&   0.009& ~& 
0.001$^{(4.4)}_{(4.6)}$&   0.002\\ [0.9ex]
He/H & ~ &  
0.110$^{(10.0)}_{(10.2)}$&   0.134 & ~& 
0.124$^{(5.0)}_{(5.1)}$&  0.148 & ~& 
0.116$^{(11.4)}_{(11.7)}$&   0.104 \\ [0.9ex]
\hline
~& ~ & \multicolumn{8}{c}{$\log~N({\rm X})/N({\rm H})$ + 12} \\
\cline{3-10}
He & ~ & 
11.12$^{(4.0)}_{(4.1)}$& 11.13 & ~&
11.16$^{(2.2)}_{(2.2)}$&  10.92 & ~&
11.07$^{(4.3)}_{(4.5)}$&  11.02 \\
\hline
\end{tabular}
\begin{description}
\item[$^{\rm a}$] \citet{LSB00}, observations of the entire nebula.
\item[$^{\rm b}$] \citet{LLB01}.
\item[$^{\rm c}$] \citet{LBZ06}, 2~arcsec wide slit spectroscopy.
\item[$^{\rm *}$] Values in parentheses are the propagated uncertainties based on observational error (in per cent). The super- and subscripts are the errors added to and subtracted from the observed fluxes, respectively, and then corrected for extinction.
\end{description}
\label{tab:heabund}
\end{table*}
%\end{landscape}

\begin{table*}
\centering
\setlength{\tabcolsep}{0.001in}
\caption{Recombination line C$^{2+}$/H$^{+}$ abundances for NGC\,6153, 
M\,1-42 and Hf\,2-2. Intensities are normalized such that H$\beta$ = 100. 
The effective recombination coefficients are from \citet{DSK00}.}
\begin{tabular}{llrrcrrcrrcrrcrrcrr}
\hline
\hline
~ & ~ & \multicolumn{5}{c}{NGC\,6153} & ~ & 
              \multicolumn{5}{c}{M\,1-42} & ~ & 
              \multicolumn{5}{c}{Hf\,2-2} \\
\cline{3-7}
\cline{9-13}
\cline{15-19}
Line & Mult.  & \multicolumn{2}{c}{20 minutes} & ~ & \multicolumn{2}{c}{Literature$^{\rm a}$} &
                        & \multicolumn{2}{c}{30 minutes} & ~ & \multicolumn{2}{c}{Literature$^{\rm b}$} & 
                        & \multicolumn{2}{c}{30 minutes} & ~ & \multicolumn{2}{c}{Literature$^{\rm c}$} \\
\cline{3-4}
\cline{6-7}
\cline{9-10}
\cline{12-13}
\cline{15-16}
\cline{18-19}
(\AA) & ~ & \multicolumn{1}{c}{\underline{$I_{\rm obs}$}} & \multicolumn{1}{c}{C$^{2+}$/H$^{+}$$^{*}$}  & ~ & \multicolumn{1}{c}{\underline{$I_{\rm obs}$}} & \multicolumn{1}{c}{C$^{2+}$/H$^{+}$} 
          & ~ & \multicolumn{1}{c}{\underline{$I_{\rm obs}$}} & \multicolumn{1}{c}{C$^{2+}$/H$^{+}$$^{*}$}  & ~ & \multicolumn{1}{c}{\underline{$I_{\rm obs}$}} & \multicolumn{1}{c}{C$^{2+}$/H$^{+}$} 
          & ~ & \multicolumn{1}{c}{\underline{$I_{\rm obs}$}} & \multicolumn{1}{c}{C$^{2+}$/H$^{+}$$^{*}$}  & ~ & \multicolumn{1}{c}{\underline{$I_{\rm obs}$}} & \multicolumn{1}{c}{C$^{2+}$/H$^{+}$}\\
~ & ~ & $I ({\rm H}\beta)$ & ($\times$10$^{-3}$)  & ~ & $I ({\rm H}\beta)$ & ($\times$10$^{-3}$) 
    & ~ & $I ({\rm H}\beta)$ & ($\times$10$^{-3}$)  & ~ & $I ({\rm H}\beta)$ & ($\times$10$^{-3}$)
    & ~ & $I ({\rm H}\beta)$ & ($\times$10$^{-3}$)  & ~ & $I ({\rm H}\beta)$ & ($\times$10$^{-3}$) \\
\hline
%~ & ~ & \multicolumn{5}{c}{\bf 3--3 transitions} \\
$\lambda$7231.32 & M3 & 
1.30 & 0.74$^{(5.1)}_{(5.2)}$ & ~ & 
0.36 & 0.85 & ~ & 
1.72 & 0.98$^{(1.5)}_{(1.5)}$& ~ & 
0.56 & 1.26 & ~ & 
2.93 & 1.67$^{(3.0)}_{(3.0)}$& ~ & 
~ & ~ \\ [0.9ex]
{\bf M3 3d $^{2}$D -- 3p $^{2}$P$^{{\rm o}}$} & 
~ & ~ &  ~ & ~ & 
~ & ~ &  ~ \\
$\lambda$4267.38 & M6 & 
4.62 & 2.26$^{(2.5)}_{(2.5)}$ & ~ & 
2.40 & 2.25 & ~ & 
3.79 & 3.72$^{(2.0)}_{(2.1)}$& ~ & 
2.55 & 1.98 & ~ & 
10.13 & 3.31$^{(4.0)}_{(4.0)}$& ~ & 
7.60 & 3.68 \\ [0.9ex]
{\bf M6 4f $^{2}$F$^{{\rm o}}$ -- 3d $^{2}$D} & 
~ & ~ &  ~ & ~ & 
~ & ~ &  ~ & ~ & 
~ & ~ &  ~ & ~ & 
~ & ~ &  ~ \\ [0.9ex]
$\lambda$6151.43 & M16.04 & 
0.71 & 1.62$^{(2.8)}_{(2.8)}$ & ~ & 
0.08 & ~ & ~ & 
~ & ~ & ~ & ~ & 
~ & ~ & 
1.92 & 4.37$^{(3.5)}_{(3.6)}$ & ~ & 
0.08 & ~ \\ [0.9ex]
{\bf M16.04 6f $^{2}$F$^{{\rm o}}$ -- 4d $^{2}$D} & 
~ & ~ &  ~ &~ & 
~ & ~ &  ~ \\
$\lambda$9903.46 & M17.02 & 
0.45 & 1.83$^{(3.8)}_{(3.8)}$ & ~ & 
~ & ~ & ~ & 
1.13 & 4.64$^{(2.9)}_{(3.0)}$& ~ & 
~ & ~ & ~ & 
1.39 & 5.71$^{(1.7)}_{(1.8)}$& ~ & 
~ & ~ \\ [0.9ex]
{\bf M17.02 5g $^{2}$G -- 4f $^{2}$F$^{{\rm o}}$} & 
~ & ~ & {\bf ~} &~ & 
~ & ~ & {\bf ~} & ~ & 
~ & ~ & {\bf ~} & ~ & 
~ & ~ & {\bf ~} \\
$\lambda$6461.95 & M17.04 & 
0.68 & 6.47$^{(3.2)}_{(3.4)}$ & ~ & 
0.25 & ~ & ~ & 
1.05 & 1.00$^{(2.3)}_{(2.4)}$& ~ & 
~ & ~ & ~ & 
2.35 & 2.23$^{(3.5)}_{(3.6)}$& ~ & 
0.74 & 3.21 \\ [0.9ex]
{\bf M17.04 6g $^{2}$G -- 4f $^{2}$F$^{{\rm o}}$} & 
~ & ~ & {\bf ~} &~ & 
~ & ~ & {\bf ~} \\
$\lambda$5342.40 & M17.06 & 
1.65 & 3.02$^{(2.4)}_{(2.4)}$ & ~ & 
0.10 & ~ & ~ &
1.27 & 2.32$^{(4.9)}_{(4.9)}$& ~ & 
~ & ~ & ~ & 
4.49 & 2.74$^{(3.3)}_{(3.3)}$& ~ & 
0.31 & 2.98 \\ [0.9ex]
{\bf M17.06 7g $^{2}$G -- 4f $^{2}$F$^{{\rm o}}$} & 
~ & ~ & {\bf ~} &~ & 
~ & ~ & {\bf ~} \\
Average & 
~ & ~ & {\bf 2.66~(6.9)} & 
~ & ~ & {\bf 1.55} & 
~ & ~ & {\bf 2.53~ (5.8)}& 
~ & ~ & {\bf 1.62} & 
~ & ~ & {\bf 3.34~(4.1)}& 
~ & ~ & {\bf 3.29} \\
\hline
\end{tabular}
\begin{description}
\item[$^{\rm a}$] \citet{LSB00}, observations of the entire nebula.
\item[$^{\rm b}$] \citet{LLB01}.
\item[$^{\rm c}$] \citet{LBZ06}, 2~arcsec~wide-slit spectroscopy.
\item[$^{\rm *}$] Values in parentheses are the propagated uncertainties based on observational error (in per cent). The super- and subscripts are the errors added to and subtracted from the observed fluxes, respectively, and then corrected for extinction.
\end{description}
\label{tab:cabund}
\end{table*}

\subsection{Plasma diagnostics based on the H~{\sc i} recombination 
spectrum}

Table~\ref{tab:plasmadiag} presents plasma diagnostic results 
based on the hydrogen recombination spectrum.  These temperatures 
were derived from the intensity ratio of the nebular continuum at 
Balmer jump (3646\,{\AA}) to the H11 $\lambda$3770 line, defined as 
[$I_{\rm c}$($\lambda$3643) $-$ $I_{\rm c}$($\lambda$3681)]/$I$(H11), 
where $I_{\rm c}$($\lambda$3643) and $I_{\rm c}$($\lambda$3681) are 
intensities of the nebular continua at 3643\,{\AA} and 3681\,{\AA}, 
respectively. 
We derived the Balmer jump temperature using the fitting formula 
given by \citet[][Equation~3 therein]{LLB01}.  Our Balmer jump 
diagnostics are generally consistent with those of \citet{ZLW04}.

The intensity ratios of high-order $I$($n$\,$\rightarrow$2, 
$n\geq$10) Balmer lines relative to H$\beta$ are sensitive to the 
electron density but insensitive the electron temperature 
\citep{ZLW04}.  This diagnostic works well for the high-density 
plasmas ($N_\mathrm{e}\geq$10$^{6}$~cm$^{-3}$; \citealt{LSB00}). 
The Balmer lines in our deep UVES spectra can be resolved up to 
$n$=24. 
For even higher-$n$ Balmer lines, measurements of line fluxes become 
unreliable due to possible line blending.  Fig.~\ref{fig:balmerdec} 
shows the Balmer line intensities relative to H$\beta$ as a function 
of $n$ (10$\leq$\,$n$\,$\leq$24) for NGC\,6153, M\,1-42 and Hf\,2-2. 
For M\,1-42, there were fewer Balmer lines detected in the spectra 
due to the low S/N ratio.  Not only the H14 and H16 are overestimated 
due to line blending, but some other high-order Balmer lines might 
also be affected, as shown in Fig.~\ref{fig:balmerdec}.  Also 
over-plotted in the Fig.~\ref{fig:balmerdec} are the theoretical 
Balmer decrements at different electron densities at the electron 
temperature deduced from the Balmer jump.  The theoretical relative 
intensities of Balmer lines were calculated from the Case~B 
emissivities of H~{\sc i} lines \citep{SH95}.  The electron densities 
yielded by a least-squares optimization are presented in 
Table~\ref{tab:plasmadiag}.

Electron temperature can also be diagnosed from the Paschen jump 
around 8204\,{\AA} \citep[e.g.,][Equation~7 therein]{FL11}.  The 
fitting errors introduced by this formula are less than 5 per cent 
in the range 3500$\leq\,T_\mathrm{e}\leq$14\,000~K, and are less 
than 16 per cent at 3200$\leq\,T_\mathrm{e}\leq$25\,000~K 
\citep{FL11}.  The derived Paschen jump temperatures are presented 
in Table~\ref{tab:plasmadiag} for NGC\,6153, M\,1-42 and Hf\,2-2. 
Fig.~\ref{fig:paschendec} shows the relatively intensities of Paschen 
lines as a function of $n$ for 11$\leq$\,$n$\,$\leq$23 measured from 
the deep spectra of NGC\,6153, M\,1-42 and Hf\,2-2.  Also over-plotted 
in Fig.~\ref{fig:paschendec} are the theoretical relative intensities 
of the Paschen lines as a function of the principle quantum number $n$ 
at different electron densities.  For M\,1-42, same as the Balmer 
lines, there are fewer high-order Paschen lines that can be used to 
constrain the electron density.  For Hf\,2-2, measurements of the 
high-order Paschen lines have relatively large uncertainties due to 
faintness of this PN which produce relatively low S/N of the spectrum, 
yielding unreliable electron densities.  Some high-order Paschen lines 
are stronger than the lower-order lines probably due to line blending.  
The electron temperatures derived from the Paschen jump, as given 
in Table \ref{tab:plasmadiag}, was assumed when calculating the 
theoretical curves of the Paschen decrements.

\subsection{Temperature diagnostics based on the He~{\sc i} recombination 
spectrum}

The temperature diagnostics based on the He~{\sc i} recombination 
line ratios have been developed by \citet{ZLL05b}, who adopted the 
theoretical He~{\sc i} line emissivities of \citet{BSS99}. 
Table~\ref{tab:plasmadiag} presents the electron temperatures 
diagnosed from the intensity ratios of the He~{\sc i} 
$\lambda\lambda$4471,5876,6678,7281 optical recombination lines for 
NGC\,6153, M\,1-42 and Hf\,2-2.  Here the analytical formula of 
\citep[][Equation~1 therein]{ZLL05b} was used to derived the 
temperatures.  Equation~$1$ in \citet{ZLL05b} is valid from 5000~K 
to 20\,000~K. 
For the range $T_\mathrm{e}\,<$5000~K, \citet{ZLL05b} made 
analytical fits based on the theoretical He~{\sc i} line emissivities 
of \citet{S96} and the collisional data among low-$n$ levels of 
\citet{SB93}. 
We derived electron temperatures from the He~{\sc i} 
$\lambda$7281/$\lambda$6678, $\lambda$7281/$\lambda$5876, 
$\lambda$6678/$\lambda$4471 and $\lambda$5876/$\lambda$4471 line ratios, 
as listed in Table~\ref{tab:plasmadiag}. 
Since $\lambda$7281/$\lambda$6678 is the best diagnostic ratio of 
He~{\sc i} for temperature determination \citep{ZLL05b}, we adopted 
the temperatures derived from this line ratio for all three PNe.

%\clearpage
%\begin{portrait}
\begin{table*}
\centering
\setlength{\tabcolsep}{0.005in}
\caption{Recombination line N$^{2+}$/H$^{+}$ abundances for NGC\,6153, 
M\,1-42 and Hf\,2-2. Intensities are normalized such that H$\beta$ = 100. 
The N~{\sc ii} effective recombination coefficients were adopted 
from \citet{FSL13}.}
\begin{tabular}{llrrcrrcrrcrrcrrcrr}
\hline
\hline
~ & ~ & \multicolumn{5}{c}{NGC\,6153} & ~ & 
              \multicolumn{5}{c}{M\,1-42} & ~ & 
              \multicolumn{5}{c}{Hf\,2-2} \\
\cline{3-7}
\cline{9-13}
\cline{15-19}
Line & Mult.  & \multicolumn{2}{c}{20 minutes} & ~ & \multicolumn{2}{c}{Literature$^{\rm a}$} & 
                        & \multicolumn{2}{c}{30 minutes} & ~ & \multicolumn{2}{c}{Literature$^{\rm b}$} & 
                        & \multicolumn{2}{c}{30 minutes} & ~ & \multicolumn{2}{c}{Literature$^{\rm c}$} \\
\cline{3-4}
\cline{6-7}
\cline{9-10}
\cline{12-13}
\cline{15-16}
\cline{18-19}
(\AA) & ~ & \multicolumn{1}{c}{\underline{$I_{\rm obs}$}} & \multicolumn{1}{c}{N$^{2+}$/H$^{+}$$^{*}$} & ~ & \multicolumn{1}{c}{\underline{$I_{\rm obs}$}} & \multicolumn{1}{c}{N$^{2+}$/H$^{+}$} 
          & ~ & \multicolumn{1}{c}{\underline{$I_{\rm obs}$}} & \multicolumn{1}{c}{N$^{2+}$/H$^{+}$$^{*}$} & ~ & \multicolumn{1}{c}{\underline{$I_{\rm obs}$}} & \multicolumn{1}{c}{N$^{2+}$/H$^{+}$} 
          & ~ & \multicolumn{1}{c}{\underline{$I_{\rm obs}$}} & \multicolumn{1}{c}{N$^{2+}$/H$^{+}$$^{*}$} & ~ & \multicolumn{1}{c}{\underline{$I_{\rm obs}$}} & \multicolumn{1}{c}{N$^{2+}$/H$^{+}$}\\
~ & ~ & $I ({\rm H}\beta)$ & ($\times$10$^{-3}$) & ~ & $I ({\rm H}\beta)$ & ($\times$10$^{-3}$) 
    & ~ & $I ({\rm H}\beta)$ & ($\times$10$^{-3}$) & ~ & $I ({\rm H}\beta)$ & ($\times$10$^{-3}$)
    & ~ & $I ({\rm H}\beta)$ & ($\times$10$^{-3}$) & ~ & $I ({\rm H}\beta)$ & ($\times$10$^{-3}$) \\
\hline
%~ & ~ & \multicolumn{5}{c}{\bf 3--3 transitions} \\
~ & ~ & \multicolumn{17}{c}{\bf 3--3 transitions} \\
$\lambda$5666.63 & M3 & 
1.45 & 1.57$^{(2.5)}_{(1.9)}$ & ~ & 
0.22 & 1.68 & ~ & 
1.89 & 5.14$^{(1.4)}_{(1.4)}$& ~ & 
0.67 & 5.67 & ~ & 
4.97 & 2.70$^{(7.0)}_{(7.0)}$& ~ & 
0.58 & 2.48 \\ [0.9ex]
$\lambda$5676.02 & M3 & 
1.23 & 2.88$^{(2.8)}_{(3.1)}$ & ~ & 
0.13 & 2.17 & ~ & 
1.40 & 6.57$^{(1.8)}_{(1.8)}$& ~ & 
0.32 & 6.06 & ~ & 
4.53 & 2.66$^{(3.0)}_{(3.8)}$& ~ & 
0.27 & 2.60 \\ [0.9ex]
$\lambda$5679.56 & M3 & 
2.61 & 1.62$^{(1.9)}_{(1.9)}$ & ~ & 
0.43 & 1.78 & ~ & 
3.18 & 5.93$^{(5.0)}_{(5.0)}$& ~ & 
1.27 & 5.77 & ~ & 
6.58 & 2.04$^{(5.0)}_{(5.0)}$& ~ & 
1.05 & 2.41 \\ [0.9ex]
$\lambda$5686.21 & M3 & 
1.55 & 2.04$^{(3.9)}_{(3.9)}$ & ~ & 
0.09 & 2.09 & ~ & 
1.43 & 5.66$^{(1.6)}_{(1.4)}$& ~ & 
0.20 & 5.20 & ~ & 
4.33 & 3.42$^{(12.6)}_{(13.2)}$& ~ & 
0.26 & 3.34 \\ [0.9ex]
$\lambda$5710.77 & M3 & 
1.32 & 2.00$^{(2.5)}_{(3.0)}$ & ~ & 
0.06 & 1.97 & ~ & 
1.57 & 7.14$^{(3.1)}_{(3.1)}$& ~ & 
0.20 & 7.53 & ~ & 
3.67 & 4.45$^{(2.2)}_{(2.2)}$& ~ & 
0.24 & 4.64 \\ [0.9ex]
{\bf M3 3p $^{3}$D -- 3s $^{3}$P$^{{\rm o}}$} & 
~ & ~ & {\bf 2.02} {\bf (2.3)}& 
~ & ~ & {\bf 1.94} &  
~ & ~ & {\bf 6.09}  {\bf (11.5)} & 
~ & ~ & {\bf 6.05} & 
~ &  ~ & {\bf 3.05} {\bf (2.7)}& 
~ & ~ & {\bf 3.09} \\
$\lambda$4630.54 & M5 & 
1.28 & 1.57$^{(8.0)}_{(8.1)}$ & ~ & 
0.21 & 1.78 & ~ & 
1.04 & 3.83$^{(1.6)}_{(1.6)}$& ~ & 
0.44 & 3.91 & ~ & 
2.46 & 2.01$^{(6.0)}_{(4.0)}$& ~ &
0.46 & 2.17 \\ [0.9ex]
{\bf M5 3p $^{3}$P -- 3s $^{3}$P$^{{\rm o}}$} \\
$\lambda$3994.99 & M12 & 
1.50 & 3.19$^{(9.1)}_{(9.1)}$ & ~ & 
~ & ~ & ~ & 
0.66 & 2.10$^{(2.4)}_{(2.9)}$& ~ & 
~ & ~ & ~ & 
2.32 & 3.70$^{(3.0)}_{(5.0)}$& ~ & 
~ & ~ \\ [0.9ex]
{\bf M12 3p $^{1}$D -- 3s $^{1}$P$^{{\rm o}}$} \\
$\lambda$4803.29 & M20 & 
1.00 & 1.76$^{(6.0)}_{(11.0)}$ & ~ & 
0.10 & 1.50 & ~ & 
0.70 & 3.69$^{(3.8)}_{(4.1)}$& ~ & 
~ & ~ & ~ & 
2.13 & 2.50$^{(4.0)}_{(5.0)}$& ~ & 
~ & ~ \\ [0.9ex]
{\bf M20 3p $^{3}$D$^{{\rm o}}$ -- 3s $^{3}$D} \\
Average & 
~ & ~ & {\bf 2.17} {\bf (3.3)} & 
~ & ~ & {\bf 1.64} & 
~ & ~ & {\bf 3.20}  {\bf (2.5)} & 
~ & ~ & {\bf 3.91} & 
~ & ~ & {\bf 2.74} {\bf (2.6)}& 
~ & ~ & {\bf 2.17} \\ [0.9ex]
~ & ~ & \multicolumn{17}{c}{\bf 4f--3d transitions} \\
$\lambda$4035.08 & M39a & 
1.52 & 1.30$^{(8.5)}_{(9.2)}$ & ~ & 
0.09 & 1.04 & ~ & 
0.71 & 2.43$^{(2.9)}_{(2.5)}$& ~ & 
0.30 & 2.38 & ~ & 
2.58 & 2.93$^{(6.5)}_{(6.5)}$& ~ & 
0.53 & 3.19 \\ [0.9ex]
$\lambda$4043.53 & M39a & 
1.59 & 1.45$^{(7.6)}_{(8.6)}$& ~ & 
0.15 & 1.30 & ~ & 
0.79 & 2.86$^{(1.4)}_{(1.4)}$& ~ & 
0.68 & 3.36 & ~ & 
2.79 & 2.54$^{(7.9)}_{(8.7)}$& ~ & 
0.49 & 2.11 \\ [0.9ex]
Average & 
~ & ~ & {\bf 1.38} {\bf (5.5)}& 
~ & ~ & {\bf 1.29} & 
~ & ~ & {\bf 2.64}  {\bf (8.1)} & 
~ & ~ & {\bf 2.49} & 
~ & ~ & {\bf 2.69} {\bf (3.9)}& 
~ & ~ & {\bf 2.69} \\
\hline
\end{tabular}
\begin{description}
\item[$^{\rm a}$] \citet{LSB00}, observations of the entire nebula.
\item[$^{\rm b}$] \citet{LLB01}.
\item[$^{\rm c}$] \citet{LBZ06}, 2~arcsec~wide-slit spectroscopy.
\item[$^{\rm *}$] Values in parentheses are the propagated uncertainties based on observational error (in per cent). The super- and subscripts are the errors added to and subtracted from the observed fluxes, respectively, and then corrected for extinction.
\end{description}
\label{tab:nabund}
\end{table*}
%\end{portrait}

\subsection{Plasma diagnostics based on the heavy element ORLs}

In a low-density nebula, the relative populations of the lowest 
fine-structure levels of a recombining ion (e.g., N$^{2+}$ 
$^{2}$P$_{1/2}^{{\rm o}}$ and $^{2}$P$_{3/2}^{{\rm o}}$ in the case 
of N~{\sc ii}) vary with the electron density, resulting a density 
dependence of the relative line emissivities of the fine-structure 
lines belonging to a given multiplet of the recombined ion 
\citep{FSL11}.  The intensity ratio of two ORLs of the same multiplet 
but coming from different parent levels (of the recombining ion) can be 
used to determine the electron density \citep{FSL11}.  Under typical 
nebular conditions, the emissivity of an ORL has a weak, power-law 
dependence on the electron temperature.  This weak temperature 
dependence differs for the ORLs that come from levels of different 
orbital angular momentum quantum numbers $l$.  This difference is 
obvious if the two lines decay from levels with very different $l$. 
Therefore, using the intensity ratio of two ORLs with different $l$, 
we can determine the electron temperature, given that accurate 
measurements of ORLs are secured.

\begin{figure}
\begin{center}
\epsfig{file=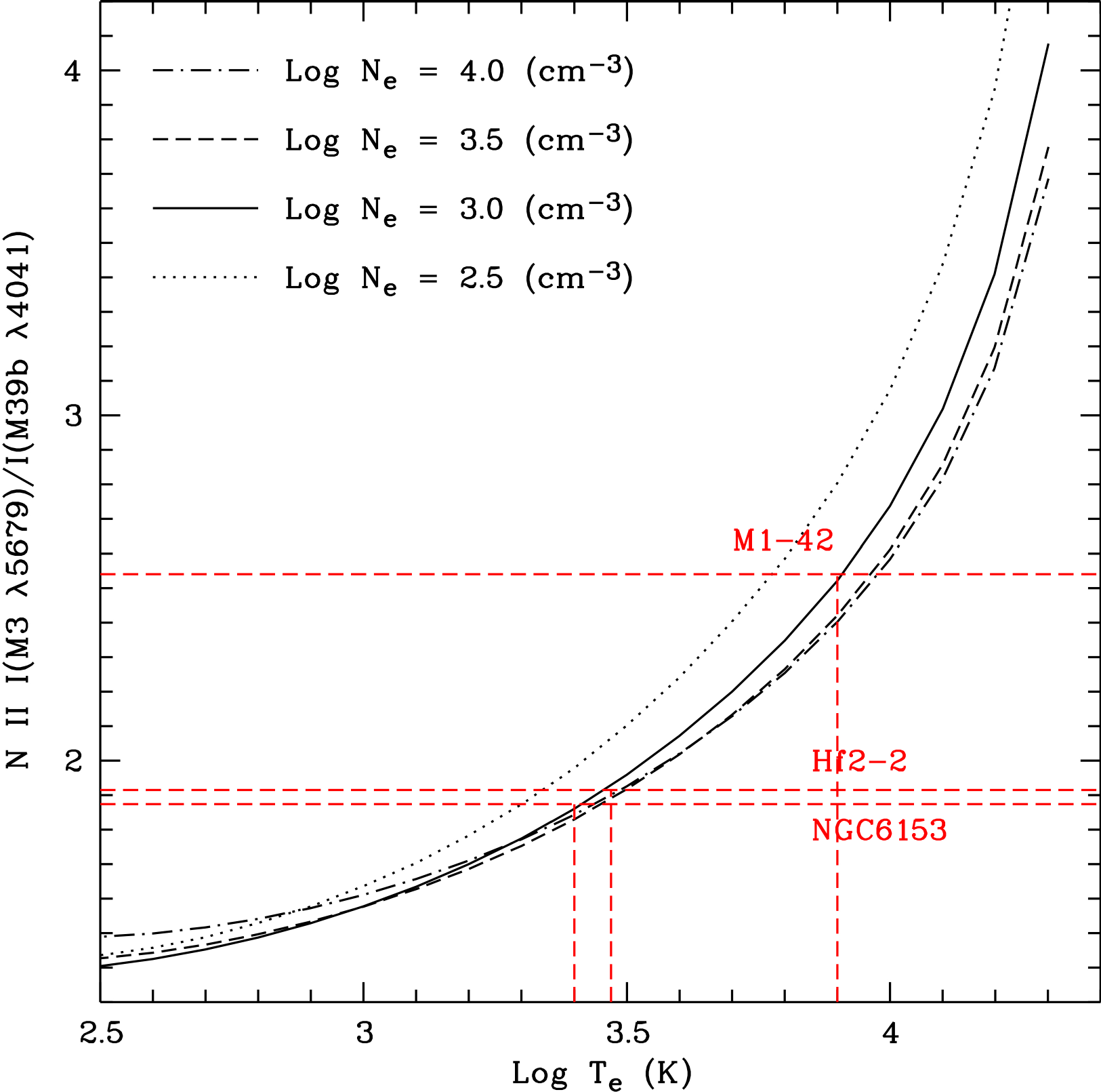,width=8.3cm,angle=0}
\caption{The N~{\sc ii} $\lambda$5679/$\lambda$4041 intensity ratio as 
a function of the electron temperature. 
Different density cases are represented by different line types. 
The curves were created using the N~{\sc ii} effective recombination 
coefficients of \citet{FSL11,FSL13}. 
The observed N~{\sc ii} line ratios of NGC\,6153, M\,1-42 and Hf\,2-2 
were indicated by horizontal red dashed lines.} 
\label{nii_te}
\end{center}
\end{figure}

\begin{figure}
\begin{center}
\epsfig{file=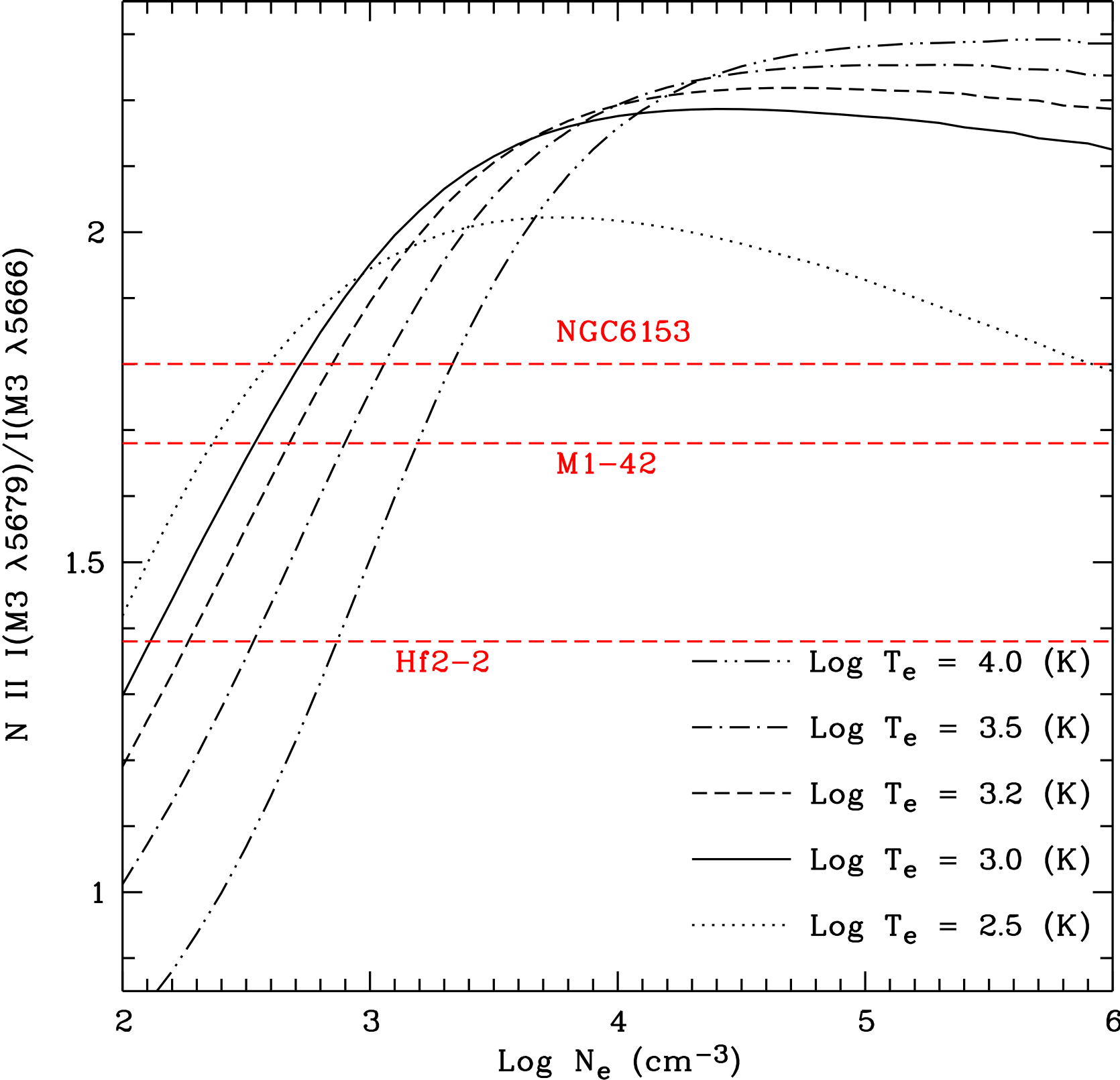,width=8.3cm,angle=0}
\caption{Same as Fig.~\ref{nii_te} but for the N~{\sc ii} 
$\lambda$5679/$\lambda$5666 intensity ratio as a function of the electron 
density.  Different temperature cases are represented by different line 
types.} 
\label{nii_ne}
\end{center}
\end{figure}

The electron temperatures and densities diagnosed from the N~{\sc ii} 
and O~{\sc ii} ORLs for the three PNe have been presented in 
\citet[][Table~1 and Figs\,6 and 7]{MFL13}.  They utilized a 
least-squares fitting method that compares the theoretical intensities 
computed from the emissivities of the N~{\sc ii} and O~{\sc ii} ORLs 
over a broad range of $T_\mathrm{e}$ and $N_\mathrm{e}$ with the 
intensities found in the literature.  Using a Monte Carlo-like method 
for generating simulated intensities within the observational errors, 
they were able to propagate those errors to the best-fit temperatures 
and densities for the N~{\sc ii} and O~{\sc ii} ORLs.  
Plasma diagnostics for NGC\,6153, M\,1-42 and Hf\,2-2 using the N~{\sc 
ii} and O~{\sc ii} line ratios are presented in Figs~\ref{nii_te}, 
\ref{nii_ne}, \ref{oii_te} and \ref{oii_ne}.  Generally the 
temperatures of the three PNe derived from the O~{\sc ii} 
$\lambda$4649/$\lambda$4089 intensity ratio are all below 2500~K 
(Fig.~\ref{oii_te}), and the temperatures yielded by the N~{\sc ii} 
$\lambda$5679/$\lambda$4041 line ratio are below 3000~K, except that of 
M\,1-42 whose N~{\sc ii} temperature is $\sim$8000~K (Fig.~\ref{nii_te}), 
much higher than the temperature diagnosed by \citet{MFL13}.  The electron 
densities derived from the N~{\sc ii} and O~{\sc ii} line ratios have 
large uncertainties mainly due to the very weak density-dependence 
of the ORL line ratios.

\section{Abundances}

\subsection{Ionic abundances from CELs}

We have observed numerous forbidden lines of heavy element CELs, 
including [N~{\sc i}], [\nii], [\oi], [\oii], [\oiii], [\neiii], 
[\neiv], [\sii], [\siii], [\clii], [\cliii], [\cliv], [\ariii] and 
[\ariv].  Ionic abundances were derived by solving population equations 
of multi-level atomic models. 
%The Fortran code {\sc equib} \citep{ha81} was used in ionic abundance 
%calculations. 
The CEL temperatures and densities in Table~\ref{tab:plasmadiag} were 
assumed in the abundance calculations.

\tel[\nii] is typically assumed for low-ionization species and 
\tel[\oiii] for ions of higher ionization degrees.  As discussed 
earlier, although electron temperatures derived from the [\nii] and 
[\oii] nebular-to-auroral line ratios can be overestimated due to 
recombination excitation (\citealt{R86}; \citealt{LSB00}) this effect 
is negligible in our spectra.  Abundances of N$^{0}$, N$^{+}$, 
O$^{+}$, O$^{++}$, Ne$^{++}$, S$^{+}$, S$^{++}$, Cl$^{++}$, Ar$^{++}$ 
and Ar$^{3+}$ relative to hydrogen have been determined from CELs.
We have adopted the average electron temperatures and densities derived 
from CELs.  Ionic abundances are listed in Table \ref{tab:celabund} for 
NGC\,6153, M\,1-42 and Hf\,2-2.  For each ion that have multiple CELs 
detected, the adopted ionic abundance was averaged from those lines. 
The ionic abundances were derived by solving population equations of 
multi-level atomic models.  Fortran code {\sc equib} was used in ionic 
abundance calculations.  The CEL temperatures and densities in Table 
\ref{tab:plasmadiag} were assumed in the calculations. 
%and correspond to the mean value of the abundances derived from all 
%the individual lines of each ion observed (weighted by their relative 
%strengths).
The CEL ionic abundances are compared with the literature.
Differences between our values and those in the literature can be 
explained by either the different instrumentation used or by the 
different analysis methodology implemented.

%\clearpage
%!TEX encoding = UTF-8 Unicode\begin{portrait}
\begin{table*}
\centering
\setlength{\tabcolsep}{0.005in}
\caption{Recombination line O$^{2+}$/H$^{+}$ abundances for NGC\,6153, 
M\,1-42 and Hf\,2-2. Intensities are normalized such that H$\beta$ = 100. 
The O~{\sc ii} effective recombination coefficients used for the 
abundance calculations were adopted from the unpublished calculations 
P.~J. Storey.}
\begin{tabular}{llrrcrrcrrcrrcrrcrr}
\hline
\hline
~ & ~ & \multicolumn{5}{c}{NGC\,6153} & ~ & 
              \multicolumn{5}{c}{M\,1-42} & ~ & 
              \multicolumn{5}{c}{Hf\,2-2} \\
\cline{3-7}
\cline{9-13}
\cline{15-19}
Line & Mult.  & \multicolumn{2}{c}{20 minutes} & ~ & \multicolumn{2}{c}{Literature$^{\rm a}$} & 
                        & \multicolumn{2}{c}{30 minutes} & ~ & \multicolumn{2}{c}{Literature$^{\rm b}$} & 
                        & \multicolumn{2}{c}{30 minutes} & ~ & \multicolumn{2}{c}{Literature$^{\rm c}$} \\
\cline{3-4}
\cline{6-7}
\cline{9-10}
\cline{12-13}
\cline{15-16}
\cline{18-19}
(\AA) & ~ & \multicolumn{1}{c}{\underline{$I_{\rm obs}$}} & \multicolumn{1}{c}{O$^{2+}$/H$^{+}$$^{*}$} & ~ & \multicolumn{1}{c}{\underline{$I_{\rm obs}$}} & \multicolumn{1}{c}{O$^{2+}$/H$^{+}$} 
          & ~ & \multicolumn{1}{c}{\underline{$I_{\rm obs}$}} & \multicolumn{1}{c}{O$^{2+}$/H$^{+}$$^{*}$} & ~ & \multicolumn{1}{c}{\underline{$I_{\rm obs}$}} & \multicolumn{1}{c}{O$^{2+}$/H$^{+}$} 
          & ~ & \multicolumn{1}{c}{\underline{$I_{\rm obs}$}} & \multicolumn{1}{c}{O$^{2+}$/H$^{+}$$^{*}$} & ~ & \multicolumn{1}{c}{\underline{$I_{\rm obs}$}} & \multicolumn{1}{c}{O$^{2+}$/H$^{+}$}\\
~ & ~ & $I ({\rm H}\beta)$ & ($\times$10$^{-3}$) & ~ & $I ({\rm H}\beta)$ & ($\times$10$^{-3}$) 
    & ~ & $I ({\rm H}\beta)$ & ($\times$10$^{-3}$) & ~ & $I ({\rm H}\beta)$ & ($\times$10$^{-3}$)
    & ~ & $I ({\rm H}\beta)$ & ($\times$10$^{-3}$) & ~ & $I ({\rm H}\beta)$ & ($\times$10$^{-3}$) \\
\hline
~ & ~ & \multicolumn{17}{c}{\bf 3--3 transitions} \\
$\lambda$4638.86 & M1 & 
1.61 & 5.22$^{(2.5)}_{(7.0)}$ & ~ & 
0.54 & 5.16 & ~ & 
1.16 & 6.57$^{(1.7)}_{(8.3)}$ & ~ & 
0.69 & 6.79 & ~ & 
3.02 & 0.92$^{(6.8)}_{(7.0)}$ & ~ & 
1.05 & 9.38 \\ [0.6ex]
$\lambda$4641.81 & M1 & 
6.88 & 3.92$^{(2.3)}_{(7.2)}$ & ~ & 
0.86 & 3.29 & ~ & 
4.46 & 3.81$^{(1.8)}_{(1.5)}$ & ~ & 
1.15 & 4.45 & ~ & 
2.09 & 7.13$^{(7.1)}_{(7.1)}$ & ~ & 
2.20 & 7.77 \\ [0.6ex]
$\lambda$4649.13 & M1 & 
2.83 & 2.35$^{(3.0)}_{(1.7)}$ & ~ & 
1.37 & 2.76 & ~ & 
2.93 & 3.91$^{(5.0)}_{(2.6)}$ & ~ & 
1.76 & 3.60 & ~ & 
7.02 & 5.20$^{(3.8)}_{(3.2)}$ & ~ & 
2.75 & 5.11 \\ [0.6ex]
$\lambda$4650.84 & M1 & 
1.56 & 3.08$^{(3.6)}_{(5.1)}$ & ~ & 
0.33 & 3.19 & ~ & 
~ & ~ & ~ & ~ & 
~ & ~ & ~ & 
~ & ~ & 
1.26 & 11.20 \\ [0.6ex]
$\lambda$4661.63 & M1 & 
1.72 & 3.33$^{(1.8)}_{(4.9)}$ & ~ & 
0.43 & 3.24 & ~ & 
1.37 & 4.98$^{(1.2)}_{(7.3)}$ & ~ & 
0.62 & 4.76 & ~ & 
3.37 & 7.82$^{(5.7)}_{(5.7)}$ & ~ & 
1.06 & 7.41 \\ [0.6ex]
$\lambda$4673.73 & M1 & 
1.11 & 3.75$^{(6.7)}_{(7.3)}$ & ~ & 
0.08 & 3.93 & ~ & 
0.61 & 3.42$^{(5.3)}_{(8.3)}$ & ~ & 
~ & ~ & ~ & 
2.13 & 3.61$^{(4.1)}_{(4.1)}$ & ~ & 
~ & ~ \\ [0.6ex]
$\lambda$4676.24 & M1 & 
1.43 & 2.34$^{(2.1)}_{(4.0)}$ & ~ & 
0.30 & 2.69 & ~ & 
1.07 & 4.21$^{(1.7)}_{(7.5)}$ & ~ & 
0.44 & 3.99 & ~ & 
2.67 & 3.28$^{(1.9)}_{(1.8)}$ & ~ & 
0.43 & 3.58 \\ [0.6ex]
$\lambda$4696.35 & M1 & 
0.96 & 5.25$^{(8.0)}_{(8.3)}$ & ~ & 
0.06 & 5.02 & ~ & 
~ & ~ & ~ & 
0.07 & 5.53 & ~ & 
1.79 & 6.90$^{(7.4}_{(7.4)}$ & ~ & 
~ & ~ \\ [0.6ex]
{\bf M1 3p $^{4}$D$^{\rm o}$ -- 3s $^{4}$P} & 
~ & ~ & {\bf 3.65} {\bf (2.9)} &
~ & ~ & {\bf 3.53} & 
~ & ~ & {\bf 4.48} {\bf (2.3)} & 
~ & ~ & {\bf 4.85} & 
 ~ & ~ & {\bf 4.98} {\bf (4.7)}& 
  ~ & ~ & {\bf 7.41} \\
$\lambda$4317.14 & M2 & 
1.33 & 2.06$^{(6.3)}_{(4.0)}$ & ~ & 
0.13 & 1.72 & ~ & 
0.85 & 3.28$^{(2.1)}_{(7.5)}$ & ~ & 
0.23 & 3.09 & ~ & 
2.57 & 3.33$^{(1.6)}_{(3.0)}$ & ~ &
0.59 & ~ \\ [0.6ex]
$\lambda$4319.63 & M2 & 
1.47 & 2.01$^{(4.0)}_{(3.0)}$ & ~ & 
0.16 & 1.95 & ~ & 
0.64 & 2.88$^{(1.4)}_{(7.8)}$ & ~ & 
0.21 & 3.09 & ~ & 
2.62 & 3.58$^{(2.7)}_{(2.7)}$ & ~ & 
0.27 & ~ \\ [0.6ex]
$\lambda$4325.76 & M2 & 
1.38 & 5.44$^{(16.2)}_{(7.9)}$ & ~ & 
0.08 & 5.23 & ~ & 
0.44 & 2.46$^{(8.9)}_{(8.4)}$ & ~ & 
0.04 & 2.39 & ~ & 
2.36 & 3.10$^{(2.4)}_{(2.4)}$ & ~ & 
~ & ~ \\ [0.6ex]
$\lambda$4336.86 & M2 & 
1.21 & 2.27$^{(14.5)}_{(5.5)}$ & ~ & 
~ & ~ & ~ & 
0.74 & 3.47$^{(3.0)}_{(7.9)}$ & ~ & 
~ & ~ & ~ & 
1.89 & 3.56$^{(4.7)}_{(4.7)}$ & ~ & 
~ & ~ \\ [0.6ex]
$\lambda$4345.56 & M2 & 
1.60 & 3.79$^{(7.1)}_{(6.1)}$ & ~ & 
0.26 & 3.40 & ~ & 
~ & ~ & ~ & 
0.44 & 5.60 & ~ & 
~ & ~ & ~ & 
0.79 & 8.91 \\ [0.6ex]
$\lambda$4349.43 & M2 & 
1.59 & 2.63$^{(2.7)}_{(4.1)}$ & ~ & 
0.41 & 2.18 & ~ & 
1.10 &2.74$^{(11.0)}_{(6.1)}$ & ~ & 
0.56 & 2.95 & ~ & 
2.77 & 3.46$^{(1.6)}_{(2,4)}$ & ~ & 
0.65 & 3.13 \\ [0.6ex]
$\lambda$4366.89 & M2 & 
1.39 & 2.58$^{(3.1)}_{(4.8)}$ & ~ & 
0.22 & 2.69 & ~ & 
0.88 & 4.34$^{(12.0)}_{(8.0)}$ & ~ & 
0.36 & 4.47 & ~ & 
2.50 & 5.30$^{(5.3)}_{(5.3)}$ & ~ & 
0.48 & 5.01 \\ [0.6ex]
{\bf M2 3p $^{4}$P$^{\rm o}$ -- 3s $^{4}$P} & 
~ & ~ & {\bf 2.97} {\bf (3.9)} &
~ & ~ & {\bf 2.86} & 
~ & ~ & {\bf 3.20} {\bf (19.1)} & 
~ & ~ & {\bf 3.14} & 
~ & ~ & {\bf 3.72}{\bf (19.5)}& 
~ & ~ & {\bf 5.68} \\
$\lambda$4414.90 & M5 & 
0.81 & 4.18$^{(4.1)}_{()}$ & ~ & 
0.18 & 3.60 & ~ & 
0.69 & 3.56$^{(1.0)}_{()}$ & ~ & 
0.22 & 5.37 & ~ & 
1.75 & 9.06$^{(11.5)}_{()}$ & ~ &
0.27 & 7.44 \\  [0.6ex]
{\bf M5 3p $^{2}$D$^{\rm o}$ -- 3s $^{2}$P} \\
$\lambda$3973.26 & M6 & 
1.50 & 4.92$^{(13.2)}_{(7.4)}$ & ~ & 
~ & ~ & ~ & 
0.62 & 10.25$^{(5.0)}_{(9.4)}$ & ~ & 
~ & ~ & ~ & 
2.21 & 10.35$^{(7.9)}_{(7.9)}$ & ~ & 
~ & ~ \\ [0.6ex]
{\bf M6 3p $^{2}$P$^{\rm o}$ -- 3s $^{2}$P} \\
$\lambda$4072.16 & M10 & 
2.70 & 4.66$^{(2.8)}_{(4.4)}$ & ~ & 
1.02 & 4.25 & ~ & 
2.23 & 5.76$^{(12.0)}_{(6.2)}$ & ~ & 
1.30 & 5.24 & ~ & 
4.07 & 7.01$^{(4.0)}_{(4.4)}$ & ~ & 
2.05 & 7.43 \\ [0.6ex]
$\lambda$4078.84 & M10 & 
1.30 & 4.77$^{(5.2)}_{(7.4)}$ & ~ & 
0.17 & 4.75 & ~ & 
0.68 & 4.98$^{(2.0)}_{(8.7)}$ & ~ & 
0.22 & 5.71 & ~ & 
2.25 & 11.05$^{(8.0)}_{(8.0)}$ & ~ & 
0.45 & 10.70 \\ [0.6ex]
$\lambda$4085.11 & M10 & 
1.41 & 4.99$^{(4.0)}_{(7.3)}$ & ~ & 
0.21 & 4.74 & ~ & 
0.83 & 4.69$^{(13.0)}_{(8.3)}$ & ~ & 
0.23 & 4.93 & ~ & 
2.34 & 8.30$^{(7.0)}_{(7.3)}$ & ~ & 
0.42 & 8.14 \\ [0.6ex]
$\lambda$4092.93 & M10 & 
1.41 & 5.98$^{(6.5)}_{(7.8)}$ & ~ & 
0.18 & 5.60 & ~ & 
0.64 & 5.42$^{(2.2)}_{(8.8)}$ & ~ & 
0.20 & 5.90 & ~ & 
2.29 & 5.83$^{(5.9)}_{(6.3)}$ & ~ & 
~ & ~ \\ [0.6ex]
{\bf M10 3d $^{4}$F -- 3p $^{4}$D$^{\rm o}$} & 
~ & ~ & {\bf 2.93} {\bf (12.9)} &
~ & ~ & {\bf 4.84} & 
~ & ~ & {\bf 5.78} {\bf (3.7)} & 
~ & ~ & {\bf 5.43} & 
~ & ~ & {\bf 8.60}{\bf (10.1)}& 
~ & ~ & {\bf 8.43} \\ [0.6ex]
$\lambda$4132.80 & M19 & 
1.51 & 2.44$^{(7.4)}_{(4.3)}$ & ~ & 
0.11 & 2.02 & ~ & 
0.76 & 4.91$^{(1.8)}_{(8.5)}$ & ~ & 
0.23 & 4.03 & ~ & 
2.63 & 9.66$^{(7.1)}_{(7.5)}$ & ~ & 
0.61 & 9.24 \\ [0.6ex]
$\lambda$4153.30 & M19 & 
1.64 & 3.00$^{(4.0)}_{(4.8)}$ & ~ & 
0.29 & 3.69 & ~ & 
0.94 & 4.33$^{(12.0)}_{(7.9)}$ & ~ & 
0.41 & 4.99 & ~ & 
2.98 & 8.74$^{(6.4)}_{(6.8)}$ & ~ & 
0.80 & 8.48 \\ [0.6ex]
$\lambda$4169.22 & M19 & 
1.44 & 3.36$^{(6.0)}_{(6.0)}$ & ~ & 
0.10 & 3.64 & ~ & 
0.73 & 4.54$^{(4.6)}_{(8.5)}$ & ~ & 
0.15 & 4.29 & ~ & 
2.62 & 12.23$^{(7.9)}_{(7.9)}$ & ~ & 
0.33 & 10.36 \\ [0.6ex]
{\bf M19 3d $^{4}$P -- 3p $^{4}$P$^{\rm o}$} & 
~ & ~ & {\bf 4.16} {\bf (3.5)} &
~ & ~ & {\bf 3.12} & 
~ & ~ & {\bf 4.59} {\bf (5.2)} & 
~ & ~ & {\bf 4.44} & 
~ & ~ & {\bf 10.21} {\bf (14.5)}& 
~ & ~ & {\bf 9.34} \\ [0.6ex]
Average & 
~ & ~ & {\bf 3.62} {\bf (22.2)} &
~ & ~ & {\bf 3.59} & 
~ & ~ & {\bf 4.67} {\bf (18.8)} & 
~ & ~ & {\bf 4.46} & 
~ & ~ & {\bf 6.88} {\bf (19.1)}& 
~ & ~ & {\bf 7.71} \\ [0.6ex]
~ & ~ &  \multicolumn{17}{c}{\bf 4f--3d transitions} \\
$\lambda$4089.29 & M48a & 
2.22 & 4.45$^{(2.5)}_{(5.1)}$ & ~ & 
0.54 & 4.80 & ~ & 
1.34 & 5.37$^{(9.0)}_{(7.5)}$ & ~ & 
0.80 & 5.76 & ~ & 
3.54 & 5.90$^{(3.9)}_{(4.2)}$ & ~ & 
1.01 & 5.10 \\ [0.6ex]
{\bf M48a 4f G[5]$^{\rm o}$--3d $^{4}$F} & 
~ & ~ & {\bf ~} & {\bf ~} &
~ & ~ & {\bf ~} \\
$\lambda$4083.90 & M48b & 
1.39 & 6.15$^{(15.1)}_{(8.1)}$ & ~ & 
0.20 & 6.05 & ~ & 
0.81 & 7.13$^{(3.9)}_{(8.9)}$ & ~ & 
0.30 & 7.54 & ~ & 
2.19 & 7.76$^{(6.7)}_{(7.7)}$ & ~ & 
0.45 & 7.82 \\ [0.6ex]
{\bf M48b 4f G[4]$^{\rm o}$--3d $^{4}$F} & 
~ & ~ & {\bf ~} & {\bf ~} &
~ & ~ & {\bf ~} & 
~ & ~ & {\bf ~} & {\bf ~} & 
~ & ~ & {\bf ~} \\
Average & 
~ & ~ & {\bf 5.30} {\bf (16.0)} &
~ & ~ & {\bf 5.42} & 
~ & ~ & {\bf 6.25} {\bf (14.1)} & 
~ & ~ & {\bf 5.92} & 
 ~ & ~ & {\bf 6.83} {\bf (13.6)}& 
  ~ & ~ & {\bf 6.46} \\ [0.6ex]
\hline
\end{tabular}
\begin{description}
\item[$^{\rm a}$] \citet{LSB00}, observations of the entire nebula.
\item[$^{\rm b}$] \citet{LLB01}.
\item[$^{\rm c}$] \citet{LBZ06}, 2~arcsec~wide-slit spectroscopy.
\item[$^{\rm *}$] Values in parentheses are the propagated uncertainties based on observational error (in per cent). The super- and subscripts are the errors added to and subtracted from the observed fluxes, respectively, and then corrected for extinction.
\end{description}
\label{tab:oabund}
\end{table*}
%\end{portrait}

\subsection{Ionic abundances from ORLs}

We identified more than 200 permitted lines in the deep spectrum for 
each of the three PNe.  These lines mainly come from the C, N, 
O and Ne ions and are mostly excited by recombination.  The most 
numerous ionic species of C, N, O and Ne in the three PNe are in the 
doubly ionized stage, and the recombination lines emitted by them are 
prominent.  Accurate measurements of these recombination lines allow 
us to derive reliable ionic abundances of the heavy elements.  Setting 
$I$($\lambda$) as the observed intensity of a recombination line with 
wavelength $\lambda$ emitted by an ion X$^{i}$, we derive the ionic 
abundance X$^{i+1}$/H$^{+}$ (i.e., the recombining ion) using 

\begin{equation}
\frac{{\rm X}^{i+1}}{{\rm H}^{+}} = \frac{\lambda}{4861} 
\frac{\alpha_{\rm eff}({\rm H}\beta)}{\alpha_{\rm eff}(\lambda)} 
\frac{I(\lambda)}{I({\rm H}\beta)}, 
\end{equation}
\noindent
where $\alpha_{\rm eff}(\lambda)$ and $\alpha_{\rm eff}({\rm H}\beta)$ 
are the effective recombination coefficients for $\lambda$ and 
H$\beta$, respectively.  
The $\alpha_{\rm eff}$($\lambda$)/$\alpha_{\rm eff}$(H$\beta$) 
coefficient ratio is insensitive to the electron temperature and 
almost independent of the density.
Here the effective recombination coefficients of helium and the heavy 
element ions, $\alpha_{\rm eff}$($\lambda$), were adopted from the 
literature, and $\alpha_{\rm eff}$(H$\beta$) was adopted from 
\citet{SH95}.  Calculations and discussion of the ORL ionic 
abundances are presented below.

\begin{figure}
\begin{center}
\epsfig{file=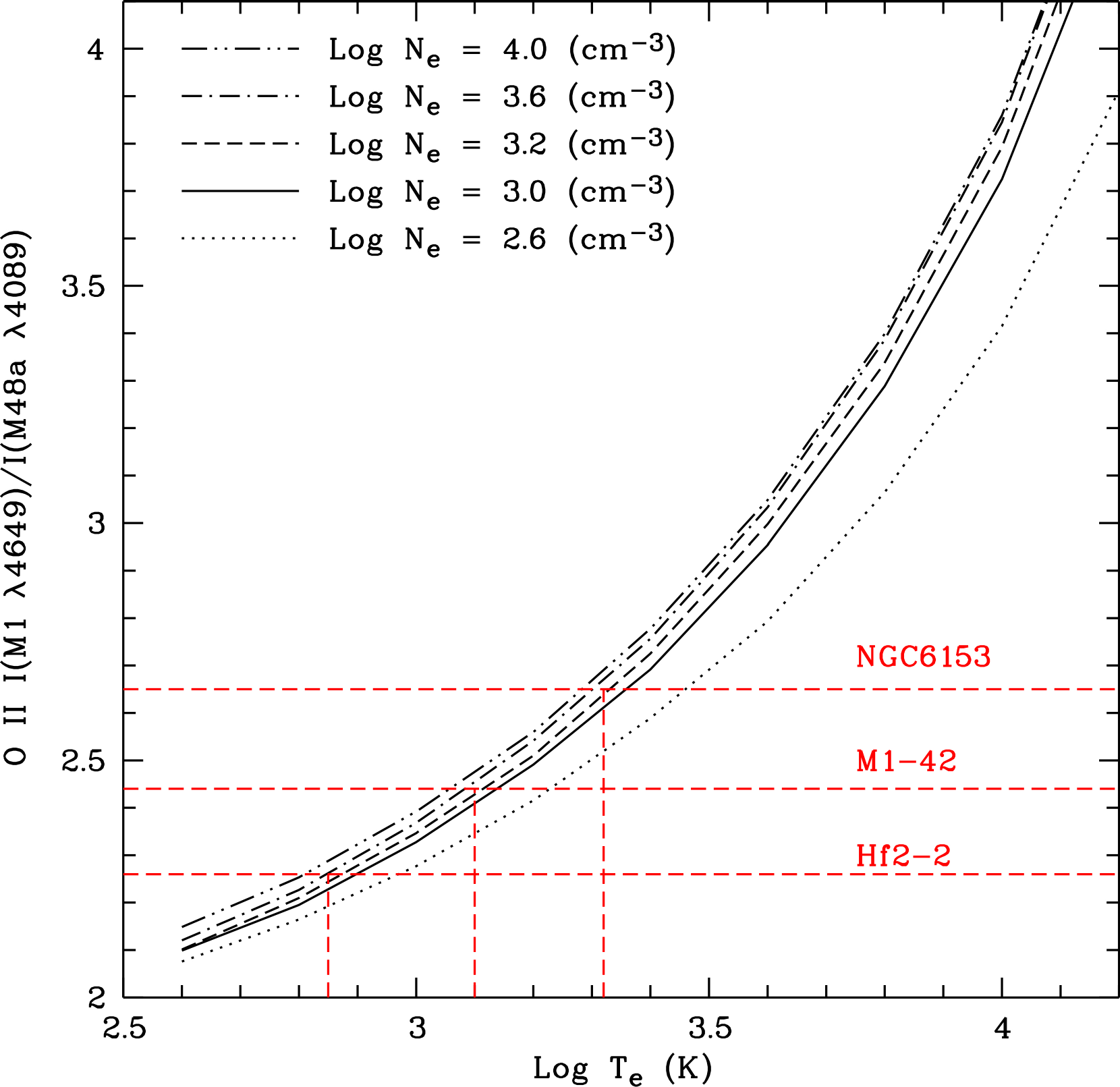,width=8.3cm,angle=0}
\caption{The O~{\sc ii} $\lambda$4649/$\lambda$4089 intensity ratio 
as a function of the electron temperature.  Different density cases 
are represented by different line types.  The curves were created 
using the O~{\sc ii} effective recombination coefficients of P.~J. 
Storey (unpublished).  The observed O~{\sc ii} line ratios of 
NGC\,6153, M\,1-42 and Hf\,2-2 were indicated by horizontal red 
dashed lines.} 
\label{oii_te}
\end{center}
\end{figure}

\begin{figure}
\begin{center}
\epsfig{file=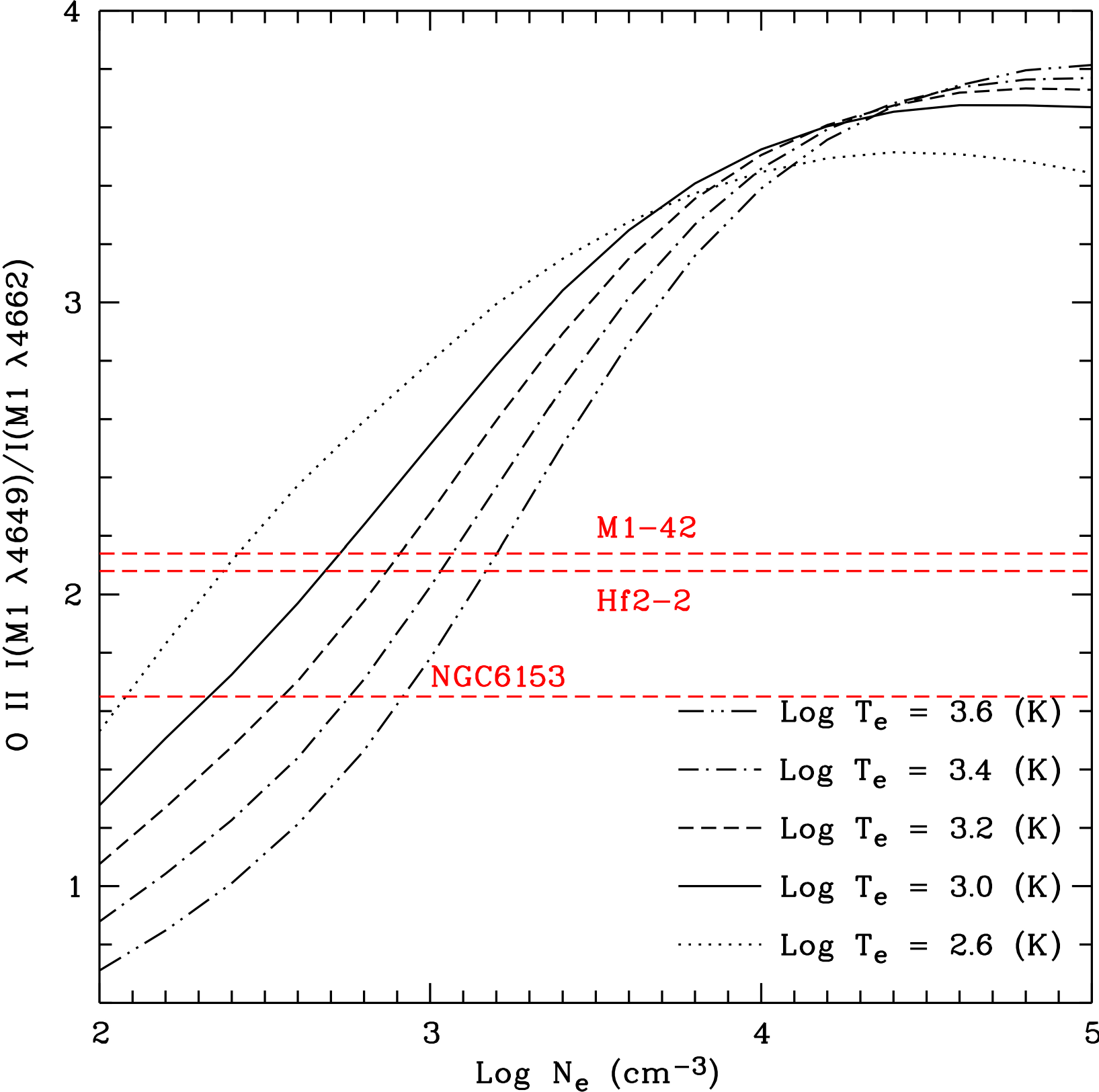,width=8.3cm,angle=0}
\caption{Same as Fig.~\ref{oii_te} but for the O~{\sc ii}
$\lambda$4649/$\lambda$4662 intensity ratio as a function of the 
electron density.  Different temperature cases are represented by 
different line types.} 
\label{oii_ne}
\end{center}
\end{figure}

\subsubsection{He$^{+}$/H$^{+}$ and He$^{2+}$/H$^{+}$}

In this section, we derived the He$^{+}$/H$^{+}$ ionic abundances 
using the theoretical emissivities of \citet{PFS12}, using the best 
observed He~{\sc i} lines, $\lambda$4471, $\lambda$5786 and 
$\lambda$6678.  The electron temperature diagnosed from the 
He~{\sc i} $\lambda$7281/$\lambda$6678 line ratio was assumed in the 
abundance calculations.  The He$^{2+}$/H$^{+}$ ionic abundance ratio 
was derived using the He~{\sc ii} $\lambda$4686 line, and the 
hydrogenic effective recombination coefficients of \citet{SH95} was 
adopted.  The He$^{+}$/H$^{+}$ and He$^{2+}$/H$^{+}$ ionic abundance 
ratios derived for NGC\,6153, M\,1-42 and Hf\,2-2 are presented in 
Table~\ref{tab:heabund}.
The ionic and elemental abundances from the literature are also 
presented in Table~\ref{tab:heabund} for comparison.  For NGC\,6153 
and M\,1-42, our He$^{+}$/H$^{+}$ abundance ratios generally agree 
with the literature, while for Hf\,2-2, our abundance value is higher 
than that given by \citet{LBZ06}.  Difference in the He$^{2+}$/H$^{+}$ 
ratios is very small due to its low abundance compared with He$^{+}$.

There are about 40 He~{\sc i} emission lines identified in our spectra.
While these lines mainly arise from recombination, the emissivities of 
some the He~{\sc i} lines may be contributed by collisional excitation 
and/or affected by self-absorption \citep[e.g.,][]{BSS02}. 
%On the other hand, singlets are generally more suitable for deriving 
%an accurate He$^{+}$/H$^{+}$ ratio, because they are not affected by 
%self-absorption effects.
In general the singlet lines are affected by Cases A or B, while the 
triplet lines are not.  However, triplet lines could be affected by 
self-absorption from the 2s $^{3}$S meta-stable state.  Because of the 
large number of singlet lines detected with good signal-to-noise ratios, 
we have decided to derive the He$^{+}$/H$^{+}$ ratio making use of 
these lines.  The He~{\sc i} lines detected in the UVES spectra are 
little affected by line blending.  Thus both singlet and triplet lines 
can be used to derive He$^{+}$/H$^{+}$.  We used the average 
He$^{+}$/H$^{+}$ ratio from the best detected He~{\sc i} lines 
(Table~\ref{tab:heabund}).

\subsubsection{C$^{2+}$/H$^{+}$}

Most of the C~{\sc ii} optical recombination lines observed in 
our UVES spectra are excited by recombination, and a few are probably 
dominated by dielectronic recombination. 
Processes of dielectronic recombination, capture and radiative
cascade, and radiative recombination of the C$^{+}$ ion are described 
in \citet[][Fig.~$10$ therein]{FL13}.  A free electron with energies 
equivalent to the energy of a Rydberg state of C~{\sc ii} can be 
captured and then cascade to the ground levels, emitting a series of 
lines.  The captured electron on an autoionizing state can also return 
to a continuum level through autoionization and thus leaves an ion 
(i.e., the recombining ion) in its ground state.  In the physical 
conditions of PNe, autoionization usually dominates and 
the population of autoionizing states can be defined by the Saha 
and Boltzmann equations (i.e., the local thermodynamical equilibrium 
-- LTE).  The emissivity of a dielectronic recombination line is 
proportional to $\exp{-E/kT_\mathrm{e}}$ ($E$ is the excitation energy 
of this line).  The intensity ratio of a dielectronic recombination 
line and a line excited by pure radiative recombination, can used to 
derive the electron temperature \citep{FL13}.

%\clearpage
%\begin{portrait}
\begin{table*}
\centering
\setlength{\tabcolsep}{0.001in}
\caption{Recombination line Ne$^{2+}$/H$^{+}$ abundances for NGC\,6153, 
M\,1-42 and Hf\,2-2. Intensities are normalized such that H$\beta$ = 100. 
The effective recombination coefficients of the 3--3 transitions of 
Ne~{\sc ii} were adopted from \citet{KSD98}, while the coefficient data 
of the 4f--3d lines were from the unpublished calculations of P.~J. 
Storey.} 
\begin{tabular}{llrrcrrcrrcrrcrrcrr}
\hline
\hline
~ & ~ & \multicolumn{5}{c}{NGC\,6153} & ~ & 
              \multicolumn{5}{c}{M\,1-42} & ~ & 
              \multicolumn{5}{c}{Hf\,2-2} \\
\cline{3-7}
\cline{9-13}
\cline{15-19}
Line & Mult.  & \multicolumn{2}{c}{20 minutes} & ~ & \multicolumn{2}{c}{Literature$^{\rm a}$} & 
                        & \multicolumn{2}{c}{30 minutes} & ~ & \multicolumn{2}{c}{Literature$^{\rm b}$} & 
                        & \multicolumn{2}{c}{30 minutes} & ~ & \multicolumn{2}{c}{Literature$^{\rm c}$} \\
\cline{3-4}
\cline{6-7}
\cline{9-10}
\cline{12-13}
\cline{15-16}
\cline{18-19}
(\AA) & ~ & \multicolumn{1}{c}{\underline{$I_{\rm obs}$}} & \multicolumn{1}{c}{Ne$^{2+}$/H$^{+}$$^{*}$} &  ~ & \multicolumn{1}{c}{\underline{$I_{\rm obs}$}} & \multicolumn{1}{c}{Ne$^{2+}$/H$^{+}$} 
          & ~ & \multicolumn{1}{c}{\underline{$I_{\rm obs}$}} & \multicolumn{1}{c}{Ne$^{2+}$/H$^{+}$$^{*}$} &  ~ & \multicolumn{1}{c}{\underline{$I_{\rm obs}$}} & \multicolumn{1}{c}{Ne$^{2+}$/H$^{+}$} 
          & ~ & \multicolumn{1}{c}{\underline{$I_{\rm obs}$}} & \multicolumn{1}{c}{Ne$^{2+}$/H$^{+}$$^{*}$} &  ~ & \multicolumn{1}{c}{\underline{$I_{\rm obs}$}} & \multicolumn{1}{c}{Ne$^{2+}$/H$^{+}$}\\
~ & ~ & $I ({\rm H}\beta)$ & ($\times$10$^{-3}$) & ~ & $I ({\rm H}\beta)$ & ($\times$10$^{-3}$) 
    & ~ & $I ({\rm H}\beta)$ & ($\times$10$^{-3}$) & ~ & $I ({\rm H}\beta)$ & ($\times$10$^{-3}$)
    & ~ & $I ({\rm H}\beta)$ & ($\times$10$^{-3}$) & ~ & $I ({\rm H}\beta)$ & ($\times$10$^{-3}$) \\
\hline
~ & ~ & \multicolumn{17}{c}{\bf 3--3 transitions} \\
$\lambda$3694.21 & M1 & 
3.17 & 0.84$^{(2.9)}_{(2.9)}$& ~ & 
0.45 & 1.36 & ~ & 
1.00 & 1.31$^{(1.4)}_{(1.4)}$& ~ & 
0.45 & 1.39 & ~ & 
5.97 & 1.58$^{(6.0)}_{(6.1)}$& ~ & 
~ & ~ \\ [0.9ex]
$\lambda$3709.62 & M1 & 
2.91 & 1.29$^{(11.5)}_{(11.6)}$ & ~ & 
0.15 & 1.16 & ~ & 
0.90 & 1.59$^{(12.1)}_{(12.2)}$& ~ & 
~ & ~ & ~ & 
6.18 & 1.09$^{(4.7)}_{(4.8)}$ & ~ & 
~ & ~ \\ [0.9ex]
{\bf M1 3p $^{4}$D -- 3s $^{4}$P$^{\rm o}$} & 
~ &  ~& {\bf 1.12} {\bf (15.7)}& 
~ & ~ & {\bf 1.26} & 
~ &  ~& {\bf 1.45} {\bf (9.7)}& 
~ & ~ & {\bf 1.39} & 
~ & ~&  {\bf 1.34} {\bf (18.4)}& 
~ & ~ & {\bf ~} \\ [0.9ex]
$\lambda$3334.84 & M2 & 
3.23 & 0.99$^{(4.3)}_{(4.3)}$& ~ & 
0.73 & 1.10 & ~ & 
1.99 & 1.83$^{(2.1)}_{(2.1)}$& ~ & 
1.15 & 1.67 & ~ & 
4.96 & 1.52$^{(9.6)}_{(9.9)}$& ~ & 
~ & ~ \\ [0.9ex]
$\lambda$3355.02 & M2 & 
3.37 & 0.99$^{(3.4)}_{(3.4)}$& ~ & 
0.46 & 1.31 & ~ & 
1.77 & 1.56$^{(7.2)}_{(7.2)}$& ~ & 
0.82 & 1.84 & ~ & 
5.68 & 1.67$^{(4.1)}_{(4.2)}$& ~ & 
~ & ~ \\ [0.9ex]
{\bf M2 3p $^{4}$P -- 3p $^{4}$D$^{\rm o}$} & 
~ &  ~& {\bf 0.99}  {\bf ~} & 
~ & ~ & {\bf 1.20} & 
~ & ~ & {\bf 1.70} {\bf (8.0)}& 
~ & ~ & {\bf 1.76} & 
~ & ~ & {\bf 1.60}  {\bf (4.7)} & 
~ & ~ & {\bf ~} \\ [0.9ex]
$\lambda$3218.19 & M13 & 
4.00 & 1.65$^{(8.7)}_{(8.7)}$& ~ & 
0.50 & 1.20 & ~ & 
~ & ~ & ~ & ~ & 
~ & ~ & 
7.24 & 1.49$^{(13.4)}_{(13.7)}$& ~ & 
~ & ~ \\ [0.9ex]
{\bf M13 3p $^{4}$D$^{\rm o}$ -- 3d $^{4}$F} & 
~ & ~ & {\bf ~} & ~ & 
~ & ~ & {\bf ~} \\
Average & 
~ & ~&  {\bf 1.17} {\bf (22.9)}& 
~ & ~ & {\bf 1.22} & 
~ & ~ & {\bf 1.57} {\bf (11.7)}& 
~ & ~ & {\bf 1.57} & 
~ & ~ & {\bf 1.47} {\bf (13.6)}& 
~ & ~ & {\bf ~} \\ [0.9ex]
~ & ~ & \multicolumn{17}{c}{{\bf 4f--3d transitions}$^{\rm d}$} \\
$\lambda$4391.99 & M55e & 
1.46 & 1.27$^{(4.8)}_{(4.9)}$& ~ & 
0.14 & 1.43 & ~ & 
0.74 & 1.71$^{(2.8)}_{(2.7)}$& ~ & 
0.18 & 1.85 & ~ & 
2.36 & 2.05$^{(5.5)}_{(5.9)}$& ~ & 
0.21 & 2.28 \\ [0.9ex]
$\lambda$4409.30 & M55e & 
1.36 & 2.41$^{(5.4)}_{(5.4)}$& ~ &  
0.13 & 2.07 & ~ & 
0.57 & 2.01$^{(3.9)}_{4.0)}$& ~ & 
0.16 & 2.43 & ~ & 
2.39 & 2.54$^{(8.8)}_{(9.2)}$& ~ &  
~ & ~ \\ [0.9ex]
{\bf M55 4f 2[5]$^{\rm o}$ -- 3d $^{4}$F} & 
~ & ~&  {\bf 1.84} {\bf (3.1)}& 
~ & ~ & {\bf 1.75} & 
~ & ~ & {\bf 1.86} {\bf (8.1)}& 
~ & ~ & {\bf 2.14} & 
~ & ~ & {\bf 2.30} {\bf (10.7)}& 
~ & ~ & {\bf ~} \\ [0.9ex]
$\lambda$4379.55 & M60b & 
2.36 & 2.35$^{(3.0)}_{(3.0)}$& ~ &  
~ & ~ & ~ & 
2.24 & 0.90$^{(1.2)}_{(1.2)}$& ~ & 
0.33 & 0.99 & ~ & 
2.77 & 0.69$^{(6.9)}_{(7.1)}$& ~ &  
~ & ~ \\ [0.9ex]
{\bf M60b 4f 1[4]$^{\rm o}$ -- 3d $^{2}$F} & 
~ &  {\bf ~} &{\bf ~}& 
~ & ~ & {\bf ~} \\
$\lambda$4428.64 & M60c & 
1.13 & 3.00$^{(7.7)}_{(7.9)}$& ~ &  
0.10 & 2.41 & ~ & 
~ & ~ & ~ & ~ & 
~ & ~ & 
1.77 & 1.41$^{(10.5)}_{(10.8)}$& ~ &  
~ & ~ \\ [0.9ex]
{\bf M60c 4f 1[3]$^{\rm o}$ -- 3d $^{2}$F} & 
~ & ~&  {\bf ~} & ~ & 
~ & ~ & {\bf ~} \\
$\lambda$4430.94 & M61a & 
1.07 & 2.63$^{(4.7)}_{(3.7)}$& ~ &  
0.06 & 2.16 & ~ & 
~ & ~ & ~ & ~ & 
~ & ~ & 
2.00 & 2.46$^{(14.0)}_{(15.0)}$& ~ &  
~ & ~ \\ [0.9ex]
{\bf M61a 4f 2[4]$^{\rm o}$ -- 3d $^{2}$D} & 
~ & ~ & {\bf ~} &~& 
~ & ~ & {\bf ~} \\
$\lambda$4457.05 & M61d & 
1.22 & 4.28$^{(2.5)}_{(2.5)}$& ~ &  
0.05 & 5.15 & ~ & 
~ & ~ & ~ & ~ & 
~ & ~ & ~ & 
~ & ~ & 
~ & ~ \\ [0.9ex]
{\bf M61d 4f 2[2]$^{\rm o}$ -- 3d $^{2}$D} & 
~ & ~ & {\bf ~} &~& 
~ & ~ & {\bf ~} \\
$\lambda$4413.22 & M65 & 
096 & 3.49$^{(20.6)}_{(2.08)}$& ~ &  
0.07 & 2.90 & ~ & 
0.64 & 4.65$^{(9.1)}_{(9.2)}$& ~ &  
0.11 & 4.63 & ~ & 
~ & ~ & ~ & 
~ & ~ \\ [0.9ex]
{\bf M65 4f 0[3]$^{\rm o}$ -- 3d $^{4}$P} & 
~ & ~ & {\bf ~} &~& 
~ & ~ & {\bf ~} \\
Average & 
~ & ~&  {\bf 3.15} {\bf (2.2)}& 
~ & ~ & {\bf 3.16} & 
~ & ~ & {\bf 2.32} {\bf (6.1)}&
~ & ~ & {\bf 3.16} & 
~ & ~ & {\bf 1.83} {\bf (3.8)}& 
~ & ~ & {\bf 2.28} \\ [0.9ex]
\hline
\end{tabular}
\\
\begin{description}
\item[$^{\rm a}$] \citet{LSB00}, observations of the entire nebula.
\item[$^{\rm b}$] \citet{LLB01}.
\item[$^{\rm c}$] \citet{LBZ06}, 2~arcsec~wide-slit spectroscopy.
\item[$^{\rm *}$] Values in parentheses are the propagated uncertainties based on observational error (in per cent). The super- and subscripts are the errors added to and subtracted from the observed fluxes, respectively, and then corrected for extinction.
\item[$^{\rm d}$] Effective recombination coefficients are only available at one electron temperature and density case with \tel~= 10,000 K and \nel~= 10,000 cm$^{-3}$.
\end{description}
\label{tab:neabund}
\end{table*}
%\end{portrait}

We have detected nearly a dozen C~{\sc ii} lines.  These lines come 
from the multiplets M3 and M6 as well as some $n$g--4f transitions. 
The best observed (and thus accurately measured) multiplets of C~{\sc 
ii} in our spectra are M3 $\lambda$7235 (3d $^{2}$D -- 
3p $^{2}$P$^{{\rm o}}$), M6 $\lambda$4267 (4f $^{2}$F$^{{\rm o}}$ -- 
3d $^{2}$D) and some transitions belonging to the $n$g--4f ($n\geq\,5$) 
array.  For these transitions, we have used effective recombination 
coefficients from \citet{DSK00}, whose calculations have only accounted 
for one electron density case at 10$^{4}$\,cm$^{-3}$.  They took great 
care in the relatively low temperatures (\tel~$<$ 5000 K).  These 
calculations were under the {\it LS}-coupling assumption and only for 
the transitions that come from the $^{1}$S state of the recombining 
ion C$^{2+}$.

%\clearpage
%\begin{landscape}
\begin{table*}
\centering
\setlength{\tabcolsep}{0.01in}
\caption{Total elemental abundances for NGC\,6153, M\,1-42 and Hf\,2-2.}
\begin{tabular}{lrrcrrcrrcrrcrrcrrcrrrr}
\hline
\hline
Element & \multicolumn{21}{c}{$\log~N({\rm X})/N({\rm H})$ + 12} \\
\cline{2-22}
~ & \multicolumn{5}{c}{NGC\,6153} & ~ & 
       \multicolumn{5}{c}{M\,1-42} & ~ & 
       \multicolumn{5}{c}{Hf\,2-2} & ~ & 
       \multicolumn{2}{c}{Avg PN$^{\rm a}$} & \multicolumn{1}{c}{Solar$^{\rm b}$} \\
\cline{2-6}
\cline{8-12}
\cline{14-18}
\cline{20-21}
~ & \multicolumn{2}{c}{20 minutes} & ~ & \multicolumn{2}{c}{Literature$^{\rm c}$} &  ~&
       \multicolumn{2}{c}{30 minutes} & ~ & \multicolumn{2}{c}{Literature$^{\rm d}$} &  ~&
       \multicolumn{2}{c}{30 minutes} & ~ & \multicolumn{2}{c}{Literature$^{\rm e}$} & ~ & 
       \multicolumn{1}{c}{Type 1} & \multicolumn{1}{c}{Type 2} & ~ \\
\cline{2-3}
\cline{5-6}
\cline{8-9}
\cline{11-12}
\cline{14-15}
\cline{17-18}
~ &  \multicolumn{1}{c}{ORL$^{*}$} &  \multicolumn{1}{c}{CEL$^{*}$} & ~ &  \multicolumn{1}{c}{ORL$^{*}$} &  \multicolumn{1}{c}{CEL$^{*}$} & ~ & 
        \multicolumn{1}{c}{ORL$^{*}$} &  \multicolumn{1}{c}{CEL$^{*}$} & ~ &  \multicolumn{1}{c}{ORL$^{*}$} &  \multicolumn{1}{c}{CEL$^{*}$} & ~ & 
        \multicolumn{1}{c}{ORL$^{*}$} &  \multicolumn{1}{c}{CEL$^{*}$} & ~ &  \multicolumn{1}{c}{ORL$^{*}$} &  \multicolumn{1}{c}{CEL$^{*}$} \\
        %ORL &  CEL & ~ &  ORL &  CEL & ~ &  ORL &  CEL & ~ &  ORL &  CEL & ~ & 
        %ORL &  CEL & ~ &  ORL &  CEL & ~ &  ORL &  CEL \\
\hline
He &  
11.12$^{(4.0)}_{(4.1)}$&   ~ & ~ &  
11.13 &   ~ &   ~&
11.16$^{(2.2)}_{(2.2)}$ &   ~ & ~ &  
11.17 &  ~ &  ~&
11.07$^{(4.3)}_{(4.5)}$ &   ~ & ~ &  
11.02 &   ~ &   ~&
11.11 &  11.05 &  10.93\\ [0.9ex]
C &  
9.46$^{(9.0)}_{(10.0)}$ &   ~ & ~ &   
9.42 &   8.44 &    ~&
9.40$^{(3.0)}_{(5.7)}$ &   ~ & ~ &   
9.35 &   7.80 &  ~&
9.62$^{(9.2)}_{(3.1)}$ &   ~ & ~ &  
9.63 &   ~ &    ~&
8.48 &   8.81 &   8.43\\ [0.9ex]
N &  
9.36$^{(12.7)}_{(13.5)}$ &   8.20$^{(3.4)}_{(2.4)}$ &   ~&
9.32 &   8.36 &   ~&
9.84$^{(7.6)}_{(7.6)}$ &   8.77$^{(2.8)}_{(1.7)}$ &  ~&
9.59 &   8.68 & ~&
9.54$^{(18.7)}_{(20.7)}$ &   8.00$^{(11.0)}_{(10.7)}$ &   ~&
9.52 &   7.77 &    ~&
8.72 &   8.14 &   7.83\\ [0.9ex]
O &   
9.51$^{(14.2)}_{(16.6)}$ &   8.51$^{(10.9)}_{(11.1)}$ &    ~&
9.66 &   8.69 &   ~&
9.56$^{(8.4)}_{(3.9)}$ &   8.75$^{(14.4)}_{(13.7)}$ &    ~&
9.79 &   8.63 & ~&
9.72$^{(13.1)}_{(13.6)}$ &   8.35$^{(14.1)}_{(16.2)}$ &   ~&
9.94 &   8.11 &    ~&
8.65 &   8.69 &   8.69\\ [0.9ex]
Ne &   
9.21$^{(5.5)}_{(4.9)}$ &   8.18$^{(17.1)}_{(22.5)}$ &   ~&
9.29 &   8.25 &     ~&
9.31$^{(5.4)}_{(4.8)}$ &   8.12$^{(11.3)}_{(10.1)}$ &    ~&
9.40& 8.12 & ~&
9.33$^{(5.3)}_{(4.9)}$ &   7.87$^{(13.2)}_{(13.4)}$ &       ~&
9.52& 7.62& ~&
8.09 &   8.10 &   7.93\\ [0.9ex]
S &   
~ &   7.00$^{(4.1)}_{(4.1)}$ &      ~&
~ &      7.23 &    ~&
~ &    6.90$^{(7.5)}_{(2.4)}$ &    ~&
 ~& 7.08 & ~&
~ &   6.69$^{(11.4)}_{(3.4)}$ &~&
~ &      6.36 &   ~&
6.91 &   6.91 &   7.12\\ [0.9ex]
Cl &   
~&~  5.77$^{(9.2)}_{(4.6)}$ &       ~&
~ &     5.62 &   ~& 
~ &    5.52$^{(5.2)}_{(10.8)}$ &    ~&
~ &   5.26 & ~&
 ~ &   5.64$^{(14.2)}_{(11.9)}$ &~&
   ~ &      ~ & ~&
~ &   ~ &   5.50\\ [0.9ex]
Ar &   
~&   6.20$^{(9.0)}_{(3.0)}$ &       ~&
~ &      6.40 &    ~&
~ &    6.10$^{(2.8)}_{(2.1)}$ &   ~&
 ~ &   6.56 & ~&
 ~ &   5.78$^{(18.1)}_{(12.3)}$ & ~&
 ~ &      6.13 &    ~\\
\hline
\end{tabular}
\\
\begin{description}
%\item $^{\rm a}$\citet{WL07}
\item[$^{\rm a}$] \citet{KB94}.
\item[$^{\rm b}$] \citet{AGS09}.
\item[$^{\rm c}$] \citet{LSB00}, observations of the entire nebula.
\item[$^{\rm d}$] \citet{LLB01}.
\item[$^{\rm e}$] \citet{LBZ06}, 2~arcsec~wide-slit spectroscopy.
\item[$^{\rm *}$] Values in parentheses are the propagated uncertainties based on observational error (in per cent). The super- and subscripts are the errors added to and subtracted from the observed fluxes, respectively, and then corrected for extinction.
\end{description}
\label{tab:totalabund}
\end{table*}
%\end{landscape}

In the abundance calculations of C$^{2+}$/H$^{+}$, an 
electron temperature of 1000~K, as derived from the N~{\sc ii} and 
O~{\sc ii} ORL ratios (Section~5.5; Figs~\ref{nii_te}--\ref{oii_ne}), 
was assumed.  We assumed that transitions between the doublet states 
of C~{\sc ii} were in Case~B, given that the ground state 
($^{2}$P$^{\rm o}$) of C$^{+}$ is a doublet. 
The C$^{2+}$/H$^{+}$ abundance ratios derived from the best observed 
lines are presented in Table~\ref{tab:cabund} for NGC\,6153, M\,1-42 
and Hf\,2-2.  The adopted C$^{2+}$/H$^{+}$ abundance ratios in the 
current work are averaged from the abundances derived from individual 
lines.  These values are also compared with those given in the 
literature. 
A review discussion of the current status of abundance 
determinations using the C~{\sc ii} ORLs is given by 
\citet{FL13}.

\subsubsection{N$^{2+}$/H$^{+}$}

Figs~$41$ and $42$ in \citet{FSL13} show the N$^{2+}$/H$^{+}$ 
and O$^{2+}$/H$^{+}$ abundance ratios calculated from the observed 
fine-structure lines belonging to different multiplets of the 3--3 
transitions as well as the 4f--3d transition array. 
The N$^{2+}$/H$^{+}$ abundance ratios derived using the old version 
of the N~{\sc ii} effective recombination coefficients (i.e., the 
{\it LS}-coupling calculations of \citealt{KS02}) were also presented 
in \citet{FL13} for purpose of comparison. 
They found that the ionic abundances derived from the 3--3 transitions 
generally consistent with those derived from the 4f--3d transitions 
of N~{\sc ii}, when the new effective recombination coefficients of 
\citet{FSL13} were used.

Several dozen N~{\sc ii} lines were identified in the spectra 
(Table~\ref{tab:orldist}).  The M3 multiplet is the strongest 
transition of N~{\sc ii} in the optical.  Located in the blue region 
of our spectra ($<$4500\,{\AA}), the 4f--3d lines of N~{\sc ii} have 
three lines that are best observed.  The best observed 3--3 transition 
lines of N~{\sc ii} in our spectra are 
M3 (3d $^{3}$D -- 3p $^{3}$P$^{{\rm o}}$), 
M5 (3p $^{3}$P -- 3s $^{3}$P$^{{\rm o}}$), 
M12 (3p $^{1}$D -- 3s $^{1}$P$^{{\rm o}}$) and 
M20 (3p $^{3}$D$^{{\rm o}}$ -- 3s $^{3}$D). 
The strongest transitions of the 4f--3d array include 
$\lambda$4035.08 (M39a 4f\,G[7/2]$_{3}$--3d\,$^{3}$F$_{2}^{\rm o}$), 
$\lambda$4041.31 (M39b 4f\,G[9/2]$_{5}$--3d\,$^{3}$F$_{4}^{\rm o}$), 
and $\lambda$4043.53 (M39a 4f\,G[7/2]$_{4}$--3d\,$^{3}$F$_{4}^{\rm 
o}$).  Some of these N~{\sc ii} lines have been adopted in plasma 
diagnostics \citep{MFL13} for Galactic PNe.  The effective 
recombination coefficients used in abundance calculations were 
adopted from \citet{FSL13}.

It has been known that the N~{\sc ii} triplet transitions among 
the 3s, 3p and 3d states, which are linked to the ground term 
2p$^{2}$~$^{3}$P resonance, can be excited by fluorescence excitation. 
\citet{G76} has shown that resonance fluorescence by the He~{\sc i} 
$\lambda$508.64 (1s8p~$^{1}$P$^{\rm o}_{1}$ -- 1s$^{2}$~$^{1}$S$_{0}$) 
line is a dominant mechanism to populate the 4s~$^{3}$P$^{\rm o}_{1}$ 
level of N~{\sc ii} in the Orion Nebula, and it should be responsible 
for the strengths of the M3 and M5 multiplets.  By comparing the 
N$^{2+}$/H$^{+}$ abundance ratios derived from the 3--3 and 4f--3d 
transitions, this effect is probably insignificant in NGC\,6153, 
M\,1-42 and Hf\,2-2.
In the physical conditions of PNe, the 2p4f levels of N~{\sc ii} are 
difficult to be enhanced by fluorescence (either by star light or by 
the resonance transitions) and they are only excited by radiative 
recombination.  The ionic abundances derived from the 4f--3d 
transitions are expected to be more reliable.  \citet{G76} also 
suggests that Multiplets 28 and 20 could be excited by a combination 
of starlight and recombination.  Abundances derived from best observed 
N~{\sc ii} lines are presented in Table~\ref{tab:nabund} for the three 
PNe.  The adopted N$^{2+}$/H$^{+}$ abundance ratio in the current 
paper is listed as the average values.  These values are also compared 
with those given in the literature.  \citet{FL13} provide a detailed 
review for deriving abundances using the N~{\sc ii} ORLs.

\subsubsection{O$^{2+}$/H$^{+}$}

The O$^{2+}$/H$^{+}$ ionic abundances using the unpublished new 
effective recombination coefficients of P.J. Storey (private 
communication; see also \citealt{FL13}).  More than 80 emission lines 
of O~{\sc ii} were detected in our spectra. 
The M1 3p\,$^{4}$D$^{\rm o}$ -- 3s\,$^{4}$P multiplet is the strongest 
among all O~{\sc ii} transitions, and the ORLs of this multiplet is 
also the best observed (Fig.~\ref{oii_lines}). 
The measurement uncertainties of the seven O~{\sc ii} lines 
($\lambda\lambda$4638.86, 4641.81, 4649.13, 4650.84, 4661.63, 4673.73, 
4676.24 and 4696.35) are all less than 10 per cent.  In our 
abundance calculations, we did not consider the following lines: 
(1) lines with errors higher than 40 per cent, (2) lines affected by 
line blending, and (3) the O~{\sc ii} $\lambda$4156.54 line of 
multiplet M19 because of possible line blending \citep{LSB00}. 
An electron temperature of 1000~K (Section~5.5) was assumed in 
the abundance calculations of O$^{2+}$/H$^{+}$.

%\clearpage 
%\begin{landscape}
%\begin{rotate}{180}
\begin{table*}
\centering
\setlength{\tabcolsep}{0.05in}
\caption{Abundance discrepancy factors. }
%\vskip-0.1truein
\begin{tabular}{lrcrcrcrcrcrcr}
\hline
\hline
Element & \multicolumn{11}{c}{ADF} \\
\cline{2-12}
~ & \multicolumn{3}{c}{NGC\,6153} & ~ & \multicolumn{3}{c}{M\,1-42} & ~ & \multicolumn{3}{c}{Hf\,2-2 } \\
\cline{2-4}
\cline{6-8}
\cline{10-12}
~ & \multicolumn{1}{c}{20 minutes} & ~ & \multicolumn{1}{c}{Literature$^{\rm a}$} & ~ & 
       \multicolumn{1}{c}{30 minutes} & ~& \multicolumn{1}{c}{Literature$^{\rm b}$} & ~ & 
       \multicolumn{1}{c}{30 minutes} & ~ & \multicolumn{1}{c}{Literature$^{\rm c}$}\\
\hline
N &  
11.41$^{(16.0)}_{(15.9)}$ &   ~& ~ & ~ & 
22.44$^{(10.5)}_{(9.3)}$ &   ~& ~ & ~ & 
83.46$^{(5.2)}_{(3.1)}$ & ~ \\
O &   
11.17$^{(12.6)}_{(17.7)}$ &   ~& 9.00& ~ & 
21.85$^{(5.9)}_{(4.0)}$ &   ~& 22.00 & ~ & 
81.51$^{(12.2)}_{(4.3)}$ &   ~& 84.00 \\
Ne &   
11.27$^{(7.3)}_{(4.8)}$ & ~ & ~ & ~ & 
22.91$^{(5.5)}_{(4.7)}$ & ~ &  ~ & ~ & 
83.01$^{(5.5)}_{(4.8)}$ & ~ \\
\hline
\end{tabular}
\begin{description}
\item[$^{\rm a}$] \citet{LSB00}, observations of the entire nebula.
\item[$^{\rm b}$] \citet{LLB01}.
\item[$^{\rm c}$] \citet{LBZ06}, 2~arcsec~wide-slit spectroscopy.
\item[$^{\rm *}$] Values in parentheses are the propagated uncertainties based on observational error (in per cent). The super- and subscripts are the errors added to and subtracted from the observed fluxes, respectively, and then corrected for extinction.
\end{description}
\label{tab:adf}
\end{table*}
%\end{landscape}
%\end{rotate}

The O~{\sc ii} optical recombination spectrum is prominent for 
all three PNe.  The best observed 3--3 lines of O~{\sc ii} belong to 
the multiplets 
M1 (3p $^{4}$D$^{\rm o}$ -- 3s $^{4}$P), 
M2 (3p $^{4}$P$^{\rm o}$ -- 3s $^{4}$P), 
M5 ($^{2}$D$^{\rm o}$ -- 3s $^{2}$P), 
M6 (3p $^{2}$P$^{\rm o}$ -- 3s $^{2}$P), 
M10 (3d $^{4}$F -- 3p $^{4}$D$^{\rm o}$) and 
M19 (3d $^{4}$P -- 3p $^{4}$P$^{\rm o}$).
The best observed multiplets of the 4f--3d transitions of O~{\sc ii} 
include M48a (4f G[5]$^{\rm o}$--3d $^{4}$F) and 
M48b (4f G[4]$^{\rm o}$--3d $^{4}$F). 
These lines have also been adopted in plasma diagnostics for 
PNe and H~{\sc ii} regions \citep{MFL13}. 
Abundances derived from the best observed lines are presented in 
Table~\ref{tab:oabund}, with the adopted O$^{2+}$/H$^{+}$ abundance 
ratio taken from the averaged M1 abundance in the current work. 
These values are also compared with those given in the literature.
The adopted O$^{2+}$/H$^{+}$ abundances of the three PNe were averaged 
from the different transitions.  Differences between our values and 
those in the literature can be explained by either the different 
instrumentation used or by the different analysis methodology 
implemented.

\subsubsection{Ne$^{2+}$/H$^{+}$}

We have detected about a dozen permitted lines of Ne~{\sc ii}. 
In our abundance calculations of Ne$^{2+}$/H$^{+}$ from the the 
3--3 transitions, we adopted the effective recombination coefficients 
of \citet{KSD98}, which were calculated in the $LS$ coupling. 
Analytical fits, which are reliable in 2000--20\,000~K, of the 
effective recombination coefficients as a function of the electron 
temperature were given in \citet{KSD98}. 
The Ne$^{2+}$/H$^{+}$ ionic abundances are presented in 
Table~\ref{tab:neabund} for NGC\,6153, M\,1-42 and Hf\,2-2.  For 
For the 4f--3d transitions of Ne~{\sc ii}, the effective recombination 
coefficients were adopted from the unpublished calculations of P.J. 
Storey (private communications).  However, these coefficients are only 
available for the case where $T_\mathrm{e}$ = 10\,000 K and 
$N_\mathrm{e}$ = 10$^{4}$~cm$^{-3}$. 
Difference between the Ne$^{2+}$/H$^{+}$ abundances derived from 
the 4f--3d transitions and those derived from the 3--3 transitions 
is noticeable.  This is mainly due to the inadequacy of the current 
versions of the Ne~{\sc ii} effective recombination coefficients 
\citep{FL13}.

The Ne$^{2+}$/H$^{+}$ abundances adopted for the three PNe are averaged 
from the 4f--3d transitions whose measurements are reliable.  Abundances 
derived from the best-detected lines are listed in Table~\ref{tab:neabund} 
for NGC\,6153, M\,1-42 and Hf\,2-2.  The averaged N$^{2+}$/H$^{+}$ 
abundances are adopted for the three PNe.  These values are also compared 
with those given in the literature.
\citet{FSL13} provide a detailed review for deriving abundances using 
\neii\ ORLs.

Other recombination lines, such as C~{\sc iii}, N~{\sc iii} and O~{\sc 
iii}, could also be used as potential temperature diagnostics 
\citep{FL13}.  However, given their lack of adequate atomic data, 
these lines can not be adopted in plasma diagnostics now.

\subsection{Ionic abundance discrepancies}

\begin{figure}
\centering
\epsfig{file=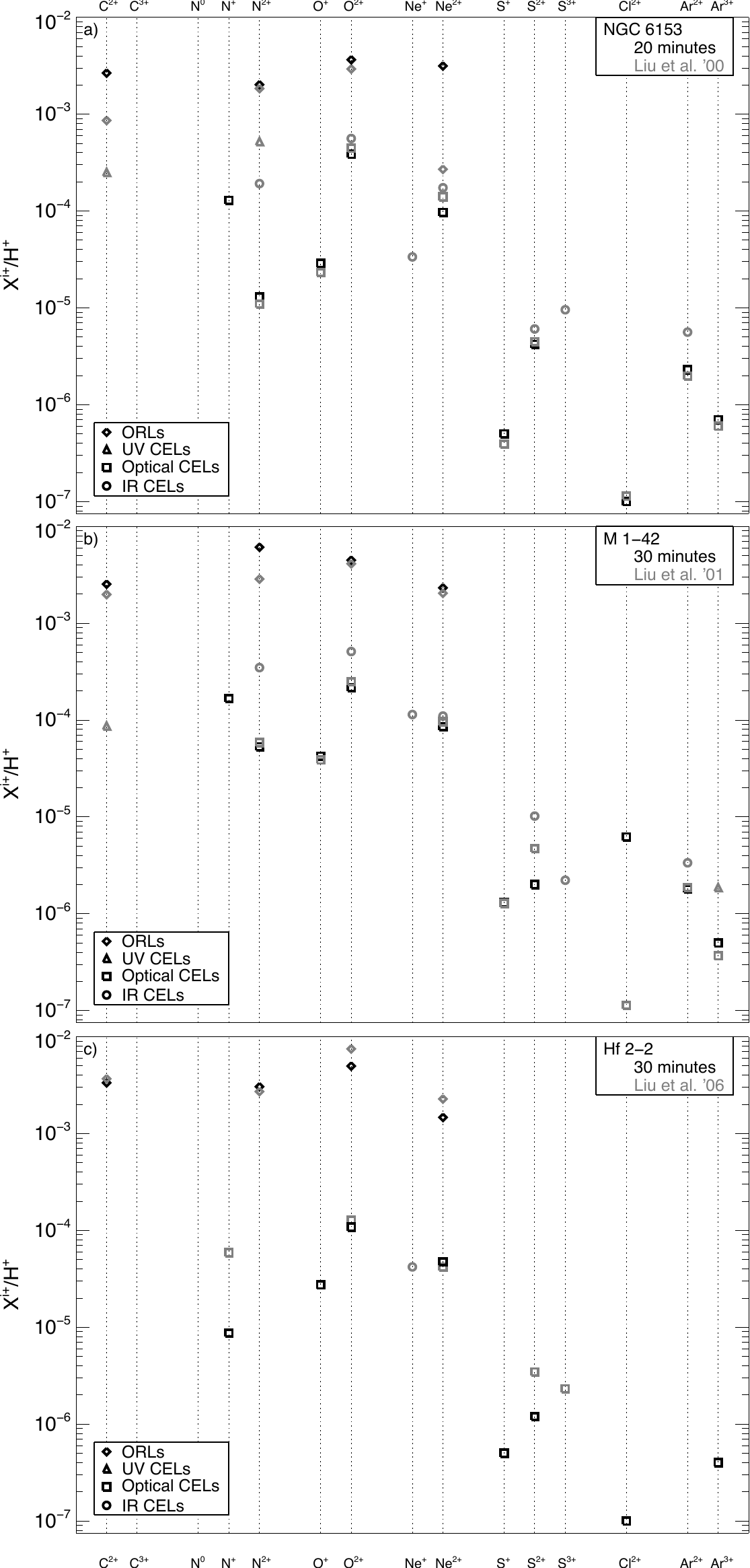,width=8.3cm,angle=0}
\caption{Comparison of the ionic abundances derived from ORLs and 
those from the UV, optical and IR CELs for NGC\,6153 (top), M\,1-42 
(middle), and Hf\,2-2 (bottom).  Different symbols indicate different 
types of transitions.  Data from the literature are presented for 
purpose of comparison: \citet{LSB00}--Liu00, \citet{LLB01}--Liu01, 
and \citet{LBZ06}--Liu06. }
\label{fig:abund}
\end{figure}

Comparison of the ionic abundances of heavy elements derived from 
ORLs and CELs is presented in Fig.~\ref{fig:abund}, where the ionic 
abundances of NGC\,6153, M\,1-42 and Hf\,2-2 from the literature are 
also overploted.  The ionic abundances derived from CELs are from 
Table~\ref{tab:celabund}, while the ORL abundances are from 
Tables~\ref{tab:cabund}--\ref{tab:neabund}.  The O$^{2+}$/H$^{+}$ 
ionic abundances derived from ORLs are higher than the CELs values 
by a factor of $\sim$10 for NGC\,6153, 25 for M\,42, and 63 for 
Hf\,2-2.  These discrepancies (ADFs) are consistent with the 
previous observations for the three PNe \citep{LSB00,LLB01,LBZ06}.

The ADF value in the Ne$^{2+}$/H$^{+}$ ionic abundance is 12 for 
NGC\,6153, 18 for M\,1-42, and 40 for Hf\,2-2, if we adopt the ionic 
abundances calculated from the 3--3 transitions of Ne~{\sc ii}. 
The ADFs of the former two PNe agree well with the literature, but 
differs in the case of Hf\,2-2.  If we adopt the abundances calculated 
from the 4f--3d transitions of Ne~{\sc ii}, the ADF(Ne$^{2+}$) will be 
$\sim$30 for NGC\,6153, 27 for M\,1-42 and $\sim$50 for Hf\,2-2.  The 
difference between our ADFs and those in the literature probably comes 
from the measurements of line fluxes.

It is also worth noting that the commonly used effective 
recombination coefficients for the 3--3 transitions of Ne~{\sc ii} 
were calculated by \citet{KSD98}, which are no longer adequate for 
nebular analysis and need to be thoroughly revised \citep{FL13}. 
The effective recombination coefficients for the 4f--3d lines 
of Ne~{\sc ii} have so far only been calculated at $T_\mathrm{e}$ 
= 10\,000~K and $N_\mathrm{e}$ = 10$^{4}$~cm$^{-3}$ (P.~J. Storey, 
private communications), and more temperature and density cases need 
to be considered.  New quantum-mechanical calculations of the 
Ne~{\sc ii} effective recombination coefficients are now in 
progress (X.~Fang et al., in preparation).  These calculations are 
in the intermediate-coupling scheme and will be extended to the 
electron temperatures well below 1000~K.

\subsection{Elemental abundances}

The elemental abundances derived from CELs and ORLs are presented 
in Table~\ref{tab:totalabund} for NGC\,6153, M\,1-42 and Hf\,2-2. 
We adopted the classical ionization correction factor (ICF) method 
from \citet{KB94}, who derived empirical ICFs of PNe based on 
observations of more than 60 southern Galactic PNe.  This method 
relies mainly on the similarity of ionization potentials between 
various ions.  The Equations~\ref{icf1}--\ref{icf8} were used to 
calculate total C, N, O and Ne abundances from the ionic abundances. 
The ADF values in N, O and Ne derived for the three PNe are 
presented in Table~\ref{tab:adf} along with the values from the 
literature.

\begin{equation}
\label{icf1}
\begin{split}
\frac{\rm O}{\rm H} = {\rm ICF}(\rm O) \times 
\left(\frac{{\rm O}^{+}}{{\rm H}^{+}} + 
\frac{{\rm O}^{2+}}{{\rm H}^{+}}\right) \\
= \left(\frac{{\rm He}^{+} + {\rm He}^{2+}}{{\rm He}^{+}}\right)^{2/3}
\times \left(\frac{{\rm O}^{+}}{{\rm H}^{+}} + 
\frac{{\rm O}^{2+}}{{\rm H}^{+}}\right).
\end{split}
\end{equation}
This equation was used to derive the CEL elemental abundance of 
oxygen.  The He$^{+}$/H$^{+}$ and He$^{2+}$/H$^{+}$ ionic abundances 
are given in Table~\ref{tab:heabund}.  For the ORL abundance of oxygen, 
we assumed that recombination line O$^{+}$/O$^{2+}$ number ratio is 
the same as that derived from the CELs.  Given that the very small 
fraction of oxygen in O$^{+}$, the errors introduced by this assumption 
are expected to be negligible.

For the carbon abundance, we adopted the equation
\begin{equation}
\label{icf2}
\frac{{\rm C}}{{\rm H}} = {\rm ICF}({\rm C}) \times 
\frac{{\rm C}^{2+}}{{\rm H}^{+}} \\
= \frac{{\rm O}}{{\rm O}^{2+}} \times \frac{{\rm C}^{2+}}{{\rm H}^{+}}.
\end{equation}
For the neon abundance, we used the equation
\begin{equation}
\label{icf3}
\frac{{\rm Ne}}{{\rm H}} = {\rm ICF}({\rm Ne}) \times 
\frac{{\rm Ne}^{2+}}{{\rm H}^{+}}\\
= \frac{{\rm O}}{{\rm O}^{2+}} \times \frac{{\rm Ne}^{2+}}{{\rm H}^{+}}.
\end{equation}
For nitrogen, we use the equation
\begin{equation}
\label{icf4}
\frac{{\rm N}}{{\rm H}} = {\rm ICF}({\rm N}) \times 
\frac{{\rm N}^{2+}}{{\rm H}^{+}}\\
= \frac{{\rm O}^{+}}{{\rm H}^{+}} \times \frac{{\rm N}^{2+}}{{\rm H}^{+}}.
\end{equation}

The elemental abundance of sulfur was derived using the equation
\begin{equation}
\label{icf5}
\frac{{\rm S}}{{\rm H}} = {\rm ICF}({\rm S}) \times 
\left(\frac{{\rm S}^{+}}{{\rm H}^{+}} + 
\frac{{\rm S}^{2+}}{{\rm H}^{+}}\right),
\end{equation}
where
\begin{equation}
\label{icf6}
{\rm ICF}({\rm S}) = \left[1 - \left(1 - \frac{{\rm O}^{+}}{{\rm O}}\right)^{3}\right]^{-1/3}.
\end{equation}
For argon, three stages of ionization (Ar$^{2+}$, Ar$^{3+}$ and Ar$^{4+}$) 
were observed, and its elemental abundance was calculated using the 
equation
\begin{equation}
\label{icf7}
\frac{{\rm Ar}}{{\rm H}} = {\rm ICF}({\rm Ar}) \times 
\left(\frac{{\rm Ar}^{2+}}{{\rm H}^{+}} + 
\frac{{\rm Ar}^{3+}}{{\rm H}^{+}} + 
\frac{{\rm Ar}^{4+}}{{\rm H}^{+}}\right),
\end{equation}
where
\begin{equation}
\label{icf8}
{\rm ICF}({\rm Ar}) = \frac{1}{1 - {\rm N}/{\rm N}^{+}}.
\end{equation}
The faint [Ar~{\sc v}] lines were detected in the spectra of our 
PNe, but their measurements are of large uncertainty.  However, the 
ionic concentration of Ar$^{4+}$ is very low (and thus Ar$^{4+}$/H$^{+}$ 
abundance ratio is not presented in Table~\ref{tab:celabund}) compared 
to the Ar$^{2+}$ and Ar$^{3+}$ ions. 
The ICF of the chlorine abundance was not discussed in \citet{KB94}. 
We estimated the Cl/H abundance ratio based on the similarities of the 
ionization potentials of the Cl and S ions.

%\clearpage
%\begin{portrait}
\begin{table}
\begin{minipage}{85mm}
\centering
\setlength{\tabcolsep}{0.025in}
\caption{Number of permitted lines identified for all ionic species.}
%\vskip-0.1truein
\begin{tabular}{lrcrcr}
\hline
\hline
Ion & \multicolumn{5}{c}{No. of lines} \\
\cline{2-6}
~ & \multicolumn{1}{c}{NGC\,6153} & ~ & 
       \multicolumn{1}{c}{M\,1-42} & ~ & 
       \multicolumn{1}{c}{Hf\,2-2} \\
\cline{2-2}
\cline{4-4}
\cline{6-6}
~ & \multicolumn{1}{c}{20 minutes} & ~ & 
\multicolumn{1}{c}{30 minutes} & ~ & 
\multicolumn{1}{c}{30 minutes} \\
\hline
\hi & 
61 & ~ & 
37 & ~ & 
55 \\
\hei & 
45 & ~ & 
25 & ~ & 
20 \\
\heii & 
29 & ~ & 
17 & ~ & 
10 \\
\ci & 
1 & ~ & 
~ & ~ & 
~  \\
\cii & 
8 & ~ &
6 & ~ &
6 \\
\ciii & 
5 & ~ & 
4 & ~ & 
2 \\
N~{\sc i} & 
2 & ~ &
~ & ~ &
~ \\
\nii & 
46 & ~ & 
29 & ~ & 
30 \\
\niii & 
15 & ~ & 
11 & ~ & 
7 \\
\oi & 
1 & ~ &
~ & ~ &
~ \\
\oii & 
77 & ~ & 
56 & ~ & 
57 \\
\oiii & 
26 & ~ & 
18 & ~ & 
3 \\
\neii & 
25 & ~ & 
16 & ~ & 
11 \\
\mgi$]$ & 
2 & ~ & 
2 & ~ & 
2 \\
Si~{\sc ii} & 
6 & ~ & 
6 & ~ & 
~ \\
Si~{\sc iii} & 
1 & ~ & 
2 & ~ & 
1 \\
\fei & 
1 & ~ & 
~ & ~ & 
~ \\
\feii & 
1 & ~ & 
~ & ~ & 
1 \\
\hline
\end{tabular}
\label{tab:orldist}
\end{minipage}
\end{table}
%\end{portrait}

The elemental He/H abundances derived from the current analysis agree 
well with the average value of the bulge (11.05) and disc sample 
(11.05) (\citealt{WL07}) as well as the Solar value (\citealt{AGS09}). 
The elemental N/H abundances derived from CELs for NGC\,6153, M\,1-42 
and Hf\,2-2 all agree well with the Solar value obtained by 
\citet{AGS09} as well as the average value of the bulge (8.61) and 
disc (9.05) samples compiled by \citet{WL07}.  The elemental Ne/H 
abundances derived from CELs for the three PNe are all slightly 
above the Sun as well as the average bulge (8.05) and disc 
(8.09) values compiled by \citet{WL07}.  The CEL O/H abundances 
derived for NGC\,6153, M\,1-42 and Hf\,2-2 also agree well with the 
Solar value obtained by \citet{AGS09} as well as the average value of 
the bulge (8.72) and disc (8.71) samples compiled by \citet{WL07}. 
The sulphur, chlorine and argon abundances of NGC\,6153, M\,1-42 and 
Hf\,2-2 also agree well with the average abundance of the bulge and 
disc samples of \citet{WL07} as well as the Solar values.

%\clearpage
%\begin{portrait}
\begin{table}
\begin{minipage}{80mm}
\centering
\setlength{\tabcolsep}{0.025in}
\caption{Number of forbidden lines identified for all ionic species.}
%\vskip-0.1truein
\begin{tabular}{lrcrcrcrcr}
\hline
\hline
Ion & \multicolumn{5}{c}{No. of lines} \\
\cline{2-6}
~ & \multicolumn{1}{c}{NGC\,6153} & ~ & 
       \multicolumn{1}{c}{M\,1-42} & ~ & 
       \multicolumn{1}{c}{Hf\,2-2} \\
\cline{2-2}
\cline{4-4}
\cline{6-6}
~ & \multicolumn{1}{c}{20 minutes} & ~ & 
\multicolumn{1}{c}{30 minutes} & ~ & 
\multicolumn{1}{c}{30 minutes} \\
\hline
$[$\ci$]$  & 
2 & ~ & 
1 & ~ & 
1 \\
$[$\nii$]$  & 
4 & ~ & 
4 & ~ & 
3 \\
$[$\oi$]$  & 
2 & ~ & 
2 & ~ & 
~ \\
$[$\oii$]$  & 
3 & ~ & 
3 & ~ & 
3 \\
$[$\oiii$]$  & 
5 & ~ & 
3 & ~ & 
3 \\
$[$\neiii$]$  & 
4 & ~ & 
4 & ~ & 
4 \\
$[$\neiv$]$  & 
1 & ~ & 
~ & ~ & 
1 \\
$[$\sii$]$  & 
3 & ~ & 
2 & ~ & 
3 \\
$[$\siii$]$  & 
3 & ~ & 
3 & ~ & 
3 \\
$[$\clii$]$  & 
2 & ~ & 
1 & ~ & 
1 \\
$[$\cliii$]$  & 
3 & ~ & 
2 & ~ & 
2 \\
$[$\cliv$]$  & 
2 & ~ & 
2 & ~ & 
1 \\
$[$\ariii$]$  & 
3 & ~ & 
3 & ~ & 
2 \\
$[$\ariv$]$  & 
4 & ~ & 
3 & ~ & 
3 \\
$[$\crii$]$  & 
1 & ~ & 
~ & ~ & 
~ \\
$[$\mnv$]$  & 
1 & ~ & 
~ & ~ & 
~ \\
$[$\feii$]$  & 
2 & ~ & 
~ & ~ & 
~ \\
$[$\feiii$]$  & 
2 & ~ & 
~ & ~ & 
1 \\
$[$\feiv$]$  & 
2 & ~ & 
1 & ~ & 
1 \\
$[$Ni~{\sc ii}$]$  & 
1 & ~ & 
~ & ~ & 
~ \\
\hline
\end{tabular}
\label{tab:celdist}
\end{minipage}
\end{table}

%\clearpage
\begin{figure*}
\centering
\begin{tabular}{cc}
\epsfig{file=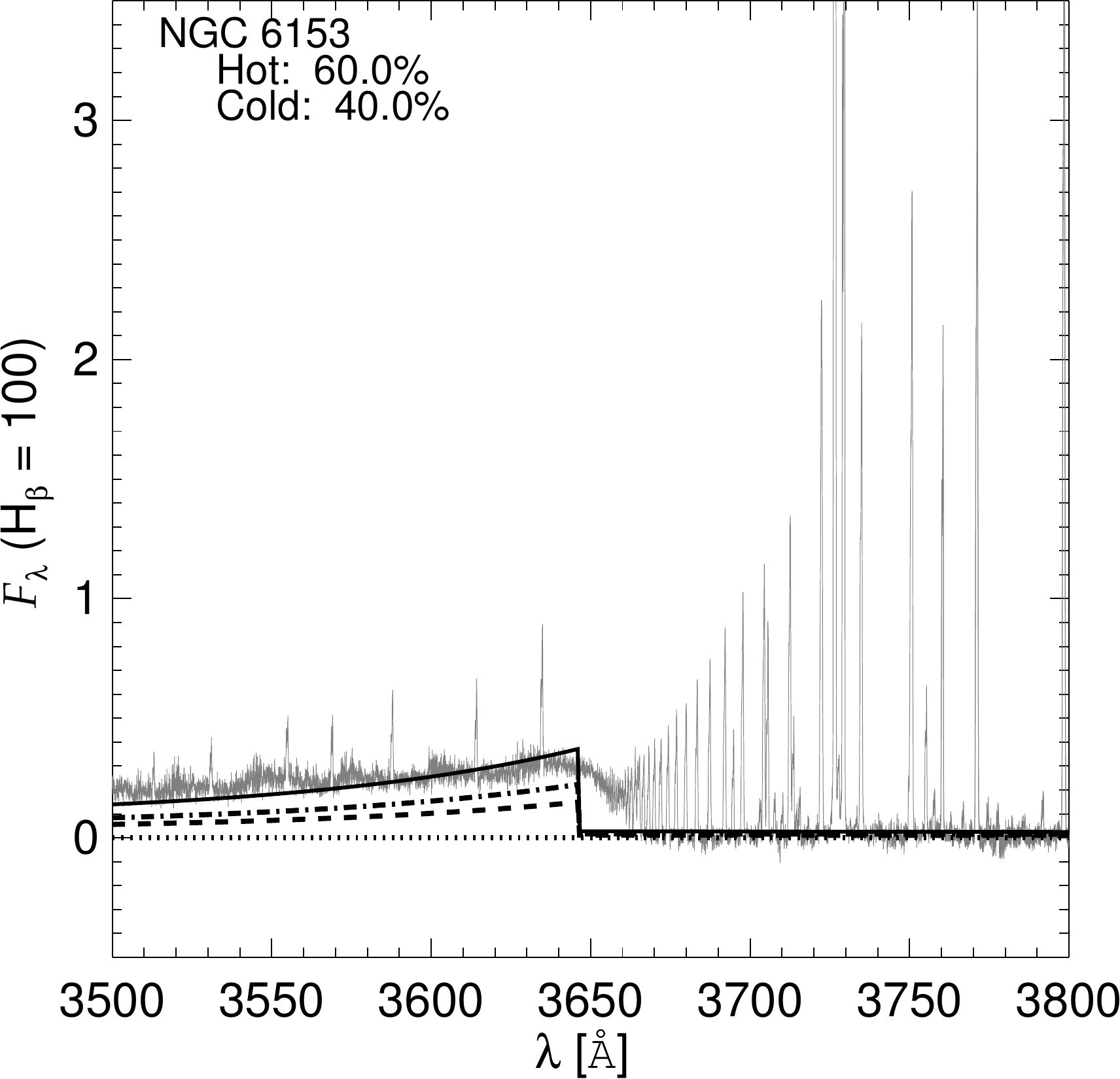,width=0.45\textwidth,angle=0}
\epsfig{file=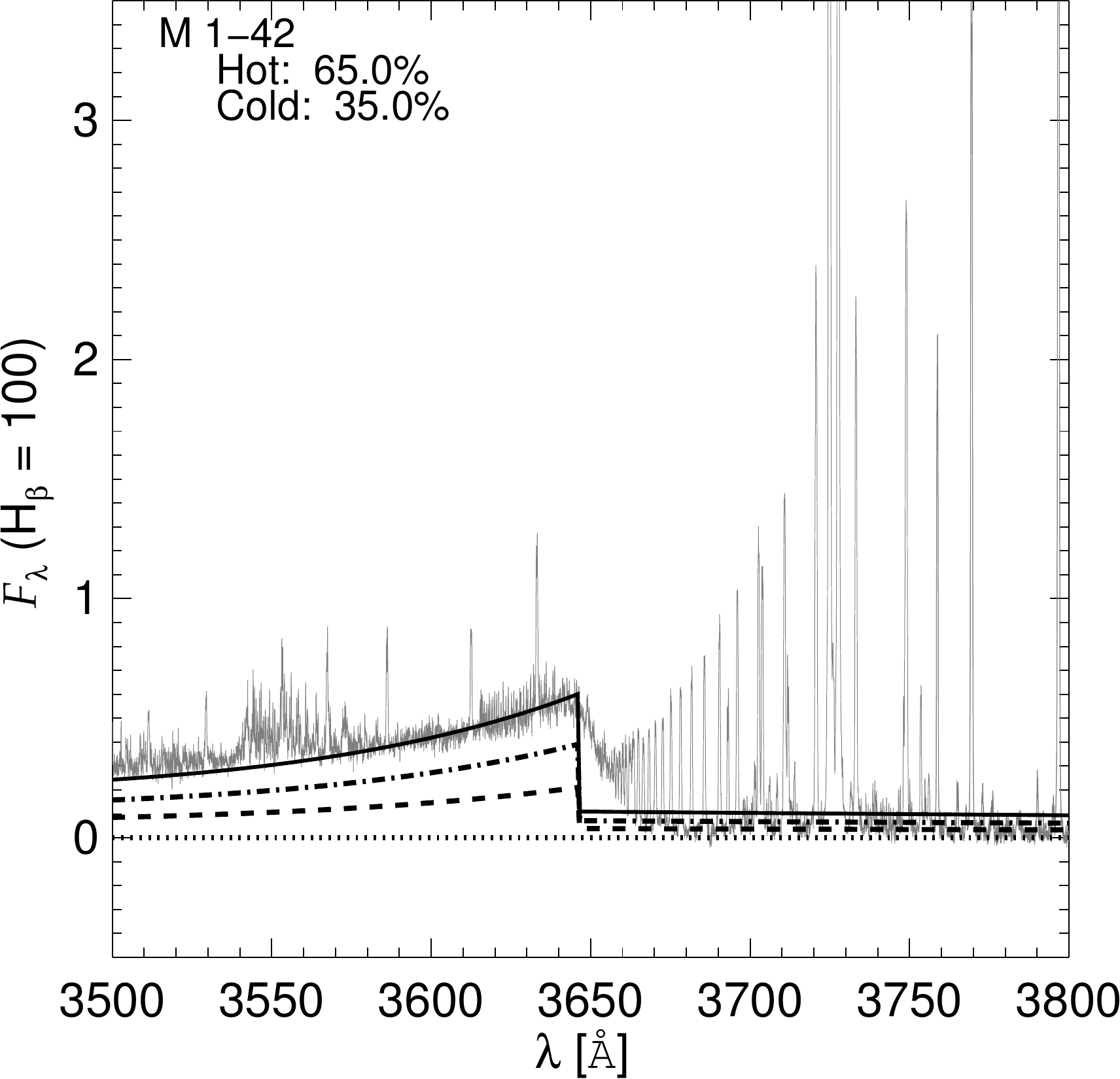,width=0.45\textwidth,angle=0}
\end{tabular}
\begin{tabular}{c}
\epsfig{file=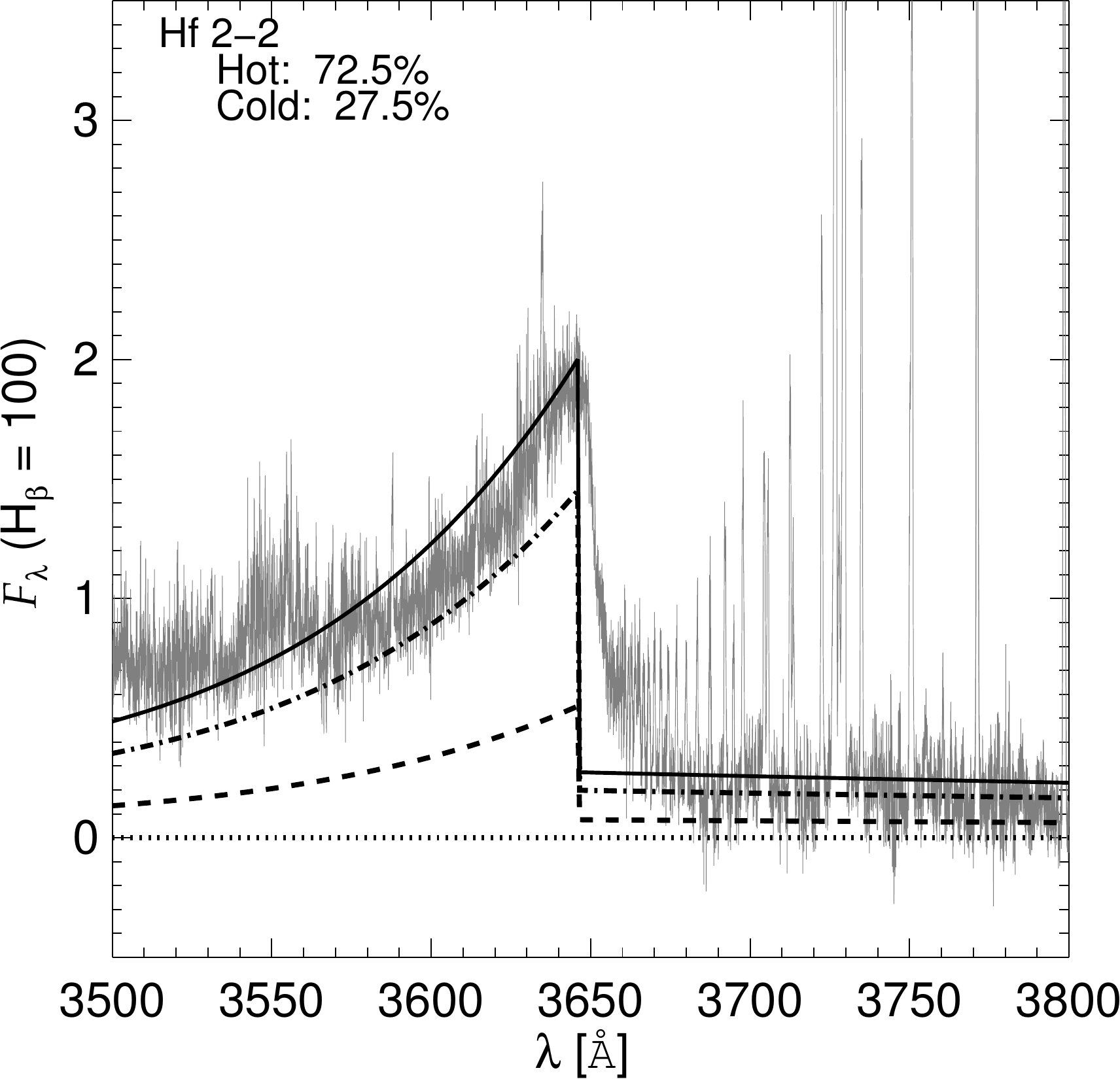,width=0.45\textwidth,angle=0}
\end{tabular}
\caption{The echelle spectra (in grey color) of NGC\,6153 (top-left 
panel), M\,1-42 (top-right panel) and Hf\,2-2 (bottom panel) showing 
the spectrum near the Balmer jump.  The black solid curve is the 
two-component fit to the continuum using the method of \citet{ZLZ14}. 
The continuum includes a very small contribution from the central 
star (the dotted line).  The dot-dashed and dashed lines are the 
modeled spectrum of the hot (10$^{4}$~K) and cold (10$^{3}$~K) 
components, respectively. 
Contributions in percentage from the hot and cold components are 
presented in each panel. 
The spectra have not been corrected for extinction and are normalized 
such that $F$(H$\beta$) = 100.} 
\label{fig:balmerjump}
\end{figure*}

\section{Discussion}

\subsection{The Balmer jump fit}

Our VLT UVES spectra of NGC\,6153, M\,1-42 and Hf\,2-2 
(Fig.~\ref{1d_spectra}) are very deep and of high spectral resolution. 
The line fluxes measured in the long-exposure spectra of NGC\,6153, 
M\,1-42 and Hf\,2-2 are presented in Table~\ref{lines} along with the 
measurement errors.

%Under the context of the bi-abundance nebular model \citep{LSB00}, 
We tried to determine the percentage of the hot and cold emission 
components in the three PNe by fitting the hydrogen recombination 
continuum around the Balmer jump using two analytical formulae. The 
method of \citet{ZLZ14} was used in the fitting. 
Fig.~\ref{fig:balmerjump} shows the continuum fitting around the 
Balmer jump for the long-exposure spectra of NGC\,6153, M\,1-42 and 
Hf\,2-2.  The fitting provides a rough estimate of how much ionized 
material comes from the cold ($\sim$1000~K) and hot ($\sim$10\,000~K) 
plasma.  Percentages of the emitting gas are presented in 
Fig.~\ref{fig:balmerjump}.  
Our analytical fits to the nebular continuum around the Balmer jump of 
the three PNe show that the majority of the observed continuum comes 
from the 10$^{4}$~K hot plasma (Fig.~\ref{fig:balmerjump}), indicating 
that a small amount of the cold component may account for the observed 
Balmer jump, which is generally consistent with the results of the 
\citet{ZLZ14}.  For Hf\,2-2, more than 70 per cent of the continuum 
emission comes from the hot plasma, and less than 30 per cent is 
emitted by the cold component.
Our continuum fits only provide a rough estimate, and more accurate 
fitting and comprehensive discussion are presented in \citet{ZLZ14}. 
%However, it is worth to try using the new UVES spectrum. 

\subsection{Physical conditions}

The density-sensitive CEL ratios [\oii] $\lambda$3726/$\lambda$3729, 
[\sii] $\lambda$6731/ $\lambda$6716 and [\cliii] 
$\lambda$5537/$\lambda$5517 were used to derive the electron densities 
for the three PNe.  The temperature-sensitive CEL ratios of [\oiii], 
[\oii], [\nii], [\neiii], [\ariii], [\sii] and [\siii] were used to 
derive the electron temperatures.  The electron temperatures and 
densities were calculated using the {\sc equib} code. 
The temperatures derived from the [\neiii] ($\lambda$3868 $+$ 
$\lambda$3967)/$\lambda$3342 
nebular-to-auroral line ratio for NGC\,6153, M\,1-42 and Hf\,2-2 were 
systematically higher than those derived from other diagnostic ratios. 
The [\neiii] temperature is probably unreliable due to two possible 
reasons: (1) The [\neiii] $\lambda\lambda$3868,3967 nebular lines are 
in a different order of the spectrum from the [\neiii] $\lambda$3342 
auroral line, and errors might be introduced when we scaled the two 
orders of the spectrum to the same level and joined them together; 
(2) like the auroral lines of [\nii], [\oii] and [\oiii], whose 
fluxes could be partially contributed by the recombination process 
\citep{LSB00}, the [Ne~{\sc iii}] $\lambda$3342 auroral line might 
also be enhanced by the recombination process, which will lead to an 
overestimated electron temperature.  Since our flux calibration is 
expected to be reliable, which can be inferred from the general 
consistency between the temperatures diagnosed from different CEL 
ratios (Table~\ref{tab:plasmadiag}), the second reason might be the 
main cause of the relatively high temperature derived from the 
[\neiii] line ratio.  However, detailed quantitative studies are 
needed to confirm this.

The electron densities are presented in Table~\ref{tab:plasmadiag}. 
For all three PNe studied, intensity ratio of the [S~{\sc ii}] 
$\lambda$6716/$\lambda$6731 doublet yields relatively lower electron 
densities than other diagnostic ratios.  The [S~{\sc ii}] densities 
might be problematic because the [S~{\sc ii}] lines could be 
suppressed by collisional de-excitation, given that the critical 
densities\footnote{At an electron temperature of 10\,000~K.} of the 
upper levels of $\lambda\lambda$6716,6731 are 1400 and 3600~cm$^{-3}$, 
respectively, which are comparable to the typical nebular densities. 
The same problem might happens to the [N~{\sc i}] nebular lines, 
whose intensity ratio also yields relatively low electron densities, 
compared with [O~{\sc ii}], [Cl~{\sc iii}], and [Ar~{\sc iv}].  The 
critical densities of the [N~{\sc i}] $\lambda\lambda$5198,5200 
nebular lines are 1800 and 780~cm$^{-3}$, respectively.  Therefore, 
the [N~{\sc i}] and [S~{\sc ii}] densities are not included in density 
average.

As discussed in Section~5.2, the auroral lines of N~{\sc ii}, 
O~{\sc ii} and O~{\sc iii} could be enhanced by recombination 
excitation\citep{LSB00}. 
It turns out that the contribution of recombination to the 
auroral line intensities of [N~{\sc ii}], [O~{\sc ii}], and 
[O~{\sc iii}] has very small effect on the derived electron 
temperatures, only reducing the temperatures by 1.55, 2.3 and 1.1 
per cent for NGC\,6153, M\,1-42 and Hf\,2-2, respectively.  These 
corrections are also presented in Fig.~\ref{fig:plasmadiag}. 
Conventionally, it is more appropriate to use the ionic abundances 
derived from the infrared forbidden lines instead of the 
recombination line abundances.

For the aforementioned cases, the CEL temperatures were adopted 
to derive the intensity contribution from the recombination process. 
This is suitable for a chemically homogeneous nebula.  If there is 
`cold', metal-rich component in the nebula, the recombination 
contribution to the auroral line fluxes of [N~{\sc ii}], [O~{\sc ii}] 
and [O~{\sc iii}] may need to be considered.  According to the 
previous discussion (see also \citealt{FL13}), although the metal-rich 
component in PNe may contribute to the recombination enhancement of 
the auroral lines, the majority of the emission still comes from the 
hot plasma.  Photoionization modeling is needed to estimate the 
relative contributions of these line fluxes from the two nebular 
components with largely different temperatures.

It is difficult to deduce a single electron density for the three 
PNe from the relative intensities of the high-$n$ Balmer/Paschen 
lines (Figs~\ref{fig:balmerdec} and \ref{fig:paschendec}) by simply 
using a least-squares optimization.  Fig.~\ref{fig:balmerdec} shows 
that the hydrogen Balmer lines with $n\,>$14 have relatively high 
intensities that yield electron densities 
$\sim$10$^{5}$--10$^{6}$\,cm$^{-3}$, 
while the lower-$n$ Balmer lines yield electron densities around 
10$^{4}$\,cm$^{-3}$.  Fig.~\ref{fig:paschendec} shows the same trend. 
Thus the electron densities deduced from the Balmer and Paschen 
decrements in Table~\ref{tab:plasmadiag} are only representative of 
the lower-$n$ hydrogen lines.  This density is higher than the 
CEL densities, indicating the condensation of ionized matter in 
these PNe, as suggested by \citet{LSB00}.  By fitting simultaneously 
the observed continuum near the Balmer jump and the high-order Balmer 
emission lines, \citet{ZLW04} derived an electron density of 
$\sim$400\,cm$^{-3}$ for Hf\,2-2, which is much lower than ours. 
Measurement errors of the Balmer lines are systematically smaller 
than those of the relatively weaker Paschen lines, and the low-order 
hydrogen lines generally have smaller errors than the high-$n$ lines.

\subsection{Possible origin of the abundance discrepancy}

The `cold', metal-rich inclusions proposed to exist in PNe 
generally well explain the observations \citep{L06b}. 
Our analyses of the UVES spectra of the three high-ADF PNe 
were carried out in a consistent manner:  the recombination-line 
temperature ($\sim$1000~K) was assumed in the ORL abundance analysis, 
and the CEL temperature ($\sim$10\,000~K) yielded was assumed in the 
forbidden-line abundance calculations.  Prior to the availability of 
the high-quality effective recombination coefficients of the ORLs 
of heavy elements, \citet{LSB00} adopted the [O~{\sc iii}] temperature 
of 9100~K to calculate the recombination-line abundances of heavy 
elements in NGC\,6153.  In the abundance analyses of M\,1-42, 
\citet{LLB01} adopted the H~{\sc i} Balmer jump temperature (3560~K) 
to derive the ORL abundances.  Given that these temperatures assumed 
were not based on the ORLs of heavy elements, the ORL 
abundances previously derived could be questionable \citep{FL13}.

The ORLs of N~{\sc ii} and O~{\sc ii} probe the regions where the 
heavy element recombination lines are emitted.  Thus the electron 
temperatures ($\sim$1000~K) derived from these ORL ratios reflect 
the physical conditions of these cold regions.  Such a huge difference 
between the ORL and the CEL temperatures, together with the high ADFs 
observed in the three PNe (NGC\,6153, ADF$\sim$10; M\,1-42, 
ADF$\sim$20; Hf\,2-2, ADF$\sim$70), can only be explained by the 
bi-abundance nebular model \citep[e.g.,][]{YLP11}, while temperature 
fluctuations and/or density inhomogeneities are difficult to interpret 
\citep{L06b}.  A temperature sequence $T_\mathrm{e}$(N~{\sc ii}, O~{\sc 
ii} ORLs) $\lesssim$ $T_\mathrm{e}$(He~{\sc i}) $\lesssim$ 
$T_\mathrm{e}$(H~{\sc i}) $\lesssim$ $T_\mathrm{e}$(CELs), which is 
in line with what is predicted by the bi-abundance nebular model 
\citep{L03,Peq03}, has been confirmed in our target PNe.  All these 
lead to our conclusion that our assumption of bi-abundance model for 
the PNe is probably reasonable.

The `cold', H-deficient plasma component in PNe is still not 
well understood and cannot be predicted by the current stellar 
evolution models.  \emph{Hubble Space Telescope} imaging reveals many 
[O~{\sc iii}] line-emitting, H-deficient knots close to the central 
stars of a rare group of PNe including Abell\,30 and Abell\,78 
\citep[e.g.,][]{BHT93,BHT95}.  These knots are embedded in a round, 
limb-brightened faint nebula with angular diameter as large as 2~arcmin. 
These particular objects are identified as the `born-again' PNe: 
PNe experience a final helium shell flash that brings the central star 
back to the AGB phase, subsequently repeating the PN evolution 
\citep{IKT83}. 
Spectroscopic observations of the Abell\,30 and Abell\,58 reveal that 
emission from these H-deficient knots are dominated by ORLs of C, N, 
O and Ne, and the ADF values of the knots are as large as several 
hundred \citep{WLB03,WBL08}.  Detailed 3D photoionization modeling of 
the H-deficient knots in Abell\,30 confirms that the ORLs are emitted 
from the electron temperature as low as several hundred K \citep{EBS03}. 
The above studies lead to the hypothesis that the `born-again' event 
might be an explanation of the H-deficient inclusions in PNe. 
However, detailed ORL abundance analyses show that the knots in 
Abell\,30 and Abell\,58 are both O-rich, not C-rich as expected from 
the theory of the `born-again' PNe \cite{IKT83}.

It is suggested that the abundance discrepancy may be related 
to the binary central stars of PNe, which is an intriguing topic as 
the central star of the extremely high-ADF PN Hf\,2-22 
(ADF$\sim$70; \citealt{LBZ06}) is a close binary system \citep{LAB98}. 
Spectroscopic observations of three high-ADF PNe (Abell\,46, Abell\,63 
and Ou\,5) with post common-envelope binary central stars seem to 
support the existence of the cold ($\sim$10$^{3}$~K), metal-rich 
ionized component \citep{Corradi15}.  The link between ADF and binarity 
of the central star has recently been further strengthened through 
studies of NGC\,6778 \citep{Jones16}, a PN with ADF$\sim$20.  In spite 
of these efforts, detailed investigations are still needed to clarify 
whether the abundance discrepancy is physically related with binarity 
and how.

Another postulation is that the H-deficient inclusions in PNe might 
have their origins in the debris planetary systems of the 
progenitor stars \citep[e.g.,][]{L03,L06a}.  In the future, 
high spatial-resolution imaging and high-dispersion spectroscopic 
observations, with an aid of detailed 3D photoionization modeling, 
will help to understand the origin and evolution of the H-deficient 
plasma inclusions in PNe.  Oxygen is not the only element that we 
found to have ADFs when comparing its abundances 
derived from both ORLs and CELs.  Nitrogen and Neon are also 
relatively abundant heavy elements in both ORLs and CELs, and their 
ADF values are above unity, indicating they could also arise 
from the cold, H-deficient secondary component.  We find that our 
results compare well with the previous literature.  Table \ref{tab:adf} 
lists the ADFs found for Nitrogen, Oxygen and Neon for all seven 
exposures of NGC\,6153, M\,1-42 and Hf\,2-2.

\subsection{Reliability of atomic data}

Calculations of the He~{\sc i} recombination spectrum started 
as early as 1950s and continued to improve until recently 
\citep[e.g.,][]{M57,B72,S96,BSS99,BSS02,BPF05,PBF05,PFS12}. 
Using the He~{\sc i} atomic model of \citet{S96} and the 
collisional data of \citet{SB93}, \citet{BSS99} calculated the 
He~{\sc i} line emissivities.  The data of \citet{BSS99} has been 
adopted in plasma diagnostics for PNe and the results are consistent 
with what is expected from the bi-abundance nebular model 
\citep{ZLL05b}. 
The more recent theoretical calculations of the He~{\sc i} line 
emissivities of \citep{PFS12} have incorporated ab initio 
photoionization cross-sections to the Case~B collisional-recombination 
calculations of the He~{\sc i} spectrum.

The reliability of the effective recombination coefficients (i.e.,
the collision-recombination theory) of heavy element ions can be 
assessed by comparing ionic abundances derived from ORLs within a 
given multiplet, or from different multiplets. 
\citet{LSB95,LSB00,LLB01} and \citet{FL13} provide critical analyses 
of ORLs of the heavy elements, and show improvement in the new 
recombination coefficients when compared with the old atomic data.

The new calculations of the N~{\sc ii} \citep{FSL11,FSL13} and O~{\sc 
ii} (P.~J. Storey, private communications) effective recombination 
coefficients are a great improvement over the previous work 
\citep[e.g.,][]{EV90,PPB91,S94,KS02,BS06}.  Discussion of the 
reliability of these new effective recombination coefficients is 
presented in \citet{Liu12b}.  \citet{FL13} critically analyzed the 
rich ORLs of the heavy elements observed in the very deep spectrum 
of NGC\,7009.  These new effective recombination coefficients 
were then utilized to construct diagnostic tools by \citet{MFL13}, 
who carried out plasma diagnostics for more than 100 Galactic PNe, 
including NGC\,6153, M\,1-42 and Hf\,2-2.  The electron temperature 
and density of each nebula were derived simultaneously by locating 
the minimal difference between the observed and theoretical 
line intensities \citep[][Sections~2.3 and 2.4 therein]{MFL13}. 
The theoretical line intensities were derived from the new 
effective recombination coefficients of N~{\sc ii} and O~{\sc ii}. 
%Once the electron temperatures and densities were derived for the 
%ionic abundances can be derived. 

\subsection{Emission line numbers}

Table~\ref{tab:orldist} summarizes the numbers of permitted 
lines identified for different ionic species in the three PNe. 
In total, 353 permitted lines were identified for the long exposure 
(20\,min) spectrum of NGC\,6153 and 168 for the short-exposure 
(2\,min) one with most of them excited mainly by recombination. 
For M\,1-42, 229 permitted lines are identified for the long-exposure 
(30\,min) spectrum, 205 for the medium-exposure (15\,min) spectrum 
and 116 for the short-exposure (1\,min) one.  For Hf\,2-2, 205 
permitted lines are identified in the long-exposure (30\,min) spectrum 
and 202 for the medium-exposure (15\,min) spectrum. 
Table~\ref{tab:celdist} presents the numbers of CELs.  In total 50 
lines were identified in the long-exposure (20\,min) spectrum of 
NGC\,6153 and 35 in the short-exposure (2\,min) spectrum, 229 in the 
long-exposure (30\,min) spectrum of M\,1-42, 34 for the medium-exposure 
(15\,min) one and 32 for the short-exposure (1\,min) spectrum and 32 
for the long-exposure (30\,min) spectrum of Hf\,2-2 and 34 for the 
medium-exposure (15\,min) one.

Nearly 80 O~{\sc ii} lines were identified in the deep spectrum of 
NGC\,6153.  Additionally, over 70 N~{\sc ii} and Ne~{\sc ii} lines 
were identified.  Only a handful of C~{\sc ii} lines were identified. 
Similar to the previous analysis of the spectrum of NGC\,7009 
\citep{FL11}, we noticed that line flux errors generally have a weak 
dependence on the wavelength, but sensitive to the line strengths. 
ORLs of C~{\sc ii}, N~{\sc ii}, O~{\sc ii} and Ne~{\sc ii} are the 
most numerous in our spectra, and have typical fluxes of 
$\sim$10$^{-4}$--10$^{-3}$ of H$\beta$, with most of the measurement 
errors $\sim$10--20 per cent.  Errors of the best observed ORLs, such 
as O~{\sc ii} M1 3p\,$^{4}$D$^{\rm o}_{7/2}$ -- 3s\,$^{4}$P$_{5/2}$ 
$\lambda$4649.13, N~{\sc ii} M3 3p\,$^{3}$D$_{3}$ -- 
3s\,$^{3}$P$_{2}^{\rm o}$ $\lambda$5679.56, and Ne~{\sc ii} M2 
3p\,$^{4}$D$_{7/2}^{\rm o}$ -- 3s\,$^{4}$P$_{5/2}$ $\lambda$3334.84, 
have errors much less than 10 per cent. 
Accurate measurements of these ORLs and others are absolutely 
essential for conclusive plasma diagnostics and abundance 
determinations using ORLs.

We have identified more than 200 emission lines in the deep 
echelle spectra of NGC\,6153, M\,1-42 and Hf\,2-2, with fluxes of 
the faintest lines measured down to the level $\leq$10$^{-5}$ of 
the H$\beta$.  Emission lines were identified using the code {\sc 
emili}.  We integrated over the emission line profiles across the 
whole UVES coverage (3040--10,940\,{\AA}) to obtain line fluxes. 
First, we manually identify all the emission lines in the spectra, 
utilizing the state-of-the-art atomic database, and constructed a 
preliminary emission line table that contains the information of 
wavelengths and fluxes.  Later on, all items in the line table 
were further identified using {\sc emili}, and the results were 
double-checked.  Table~\ref{lines} is the final line list of the 
three PNe.

\section{Summary and Conclusions}

We present deep echelle spectra of three Galactic high-ADF PNe 
NGC\,6153, M\,1-42 and Hf\,2-2.  The spectra were obtained with UVES 
on the ESO 8.2\,m VLT.  The high-resolution spectra cover a broad 
wavelength range ($\sim$3040--11\,000\,{\AA}).  Numerous ORLs of the 
C, N, O and Ne ions were well detected in our spectra.  We analyzed 
the heavy element ORLs using the newly calculated N~{\sc ii} and 
O~{\sc ii} effective recombination coefficients.  Results of plasma 
diagnostics and abundance determinations based on ORLs were compared 
with those derived from CELs.

We present critical analyses of the optical recombination lines 
of heavy elements observed in the UVES spectra of NGC\,6153, M\,1-42 
and Hf\,2-2. 
In our UVES spectra of the three PNe, we have detected several 
hundred emission lines, and 74 per cent of them are relatively 
faint permitted lines.  We carefully identified all emission lines 
using the line-identification code {\sc emili}.  Emission line 
fluxes of the heavy elements, mostly emitted by C~{\sc ii}, N~{\sc 
ii}, O~{\sc ii} and Ne~{\sc ii}, were measured.  Tables of line 
fluxes were compiled so that they can be readily used by the 
astrophysical communities. 
The electron temperatures derived from the ratios of the N~{\sc 
ii} and O~{\sc ii} ORLs are close to 1000~K, which is lower than 
the average CEL temperature by one order of magnitude.  The 
N$^{2+}$/H$^{+}$, O$^{2+}$/H$^{+}$ and Ne$^{2+}$/H$^{+}$ ionic 
abundance ratios derived from ORLs are higher than their CEL 
counterparts.  ADFs of the heavy elements in this paper are 
generally consistent with the literature.  The deep echelle spectra 
presented in this paper are not only critical for the basic nebular 
analysis, but also important for future detailed photoionization 
modeling of these high-ADF PNe, which will help to reveal the probable 
origin of the abundance and temperature discrepancies.  Although our 
current treatment of the recombination and collisional-radiative 
processes of N~{\sc ii} \citep{FSL11,FSL13} and O~{\sc ii} (P.~J. 
Storey, unpublished; private communications) in gaseous nebulae has 
been a great improvement over previous work, the atomic data for the 
Ne~{\sc ii} recombination lines that are adequate for the deep 
spectroscopy of nebulae are still missing. 
New quantum-mechanical calculations of the Ne~{\sc ii} effective 
recombination coefficients are now in progress.

Through the past decades, innovation of the observational techniques 
that enables accurate measurements of the faint ORLs of heavy elements 
and improvement in the calculations of atomic data, especially 
the recombination theories of the 2nd-row elements (C, N, O, and Ne) 
in nebular conditions, have help us to better understand the ORL 
versus CEL ``abundance and temperature discrepancies'' observed in 
PNe (also in H~{\sc ii} regions). 
Although several mechanisms have been raised to explain the observed 
discrepancies in the nebulae, including temperature fluctuations, 
density inhomogeneities, bi-abundance nebular model, and the 
$\kappa$-distributed electrons, considerable debate of these problems 
continues.  Future high-spatial and high-spectral resolution 
spectroscopy with the next-generation large telescopes may advance 
the study of these long-standing problems, and even stimulate new 
topics in nebular astrophysics.  In the meantime, sophisticated 
photoionization modeling as well as relativistic quantum-mechanical 
calculations of the atomic data for nebular physics, is still needed 
to reach further understanding in this area.

\section*{Acknowledgements}
We thank the anonymous referee, whose comments and suggestions have 
greatly improved this article.  Yong Zhang from the University of 
Hong Kong (HKU) is acknowledged for discussion.  We thank Peter J. 
Storey from University College London (UCL) for providing us 
the effective recombination coefficients of the O~{\sc ii} nebular 
lines as well as some of the Ne~{\sc ii} 4f\,--\,3d transitions 
before publication. 
Support from the Natural Science Foundation of China (No. 
10933001) is acknowledged.

%\clearpage

%\include{1d_spectra}
%\include{line_lists}

\appendix

\section{The spectra}

\begin{figure*}
\centering
\epsfig{file=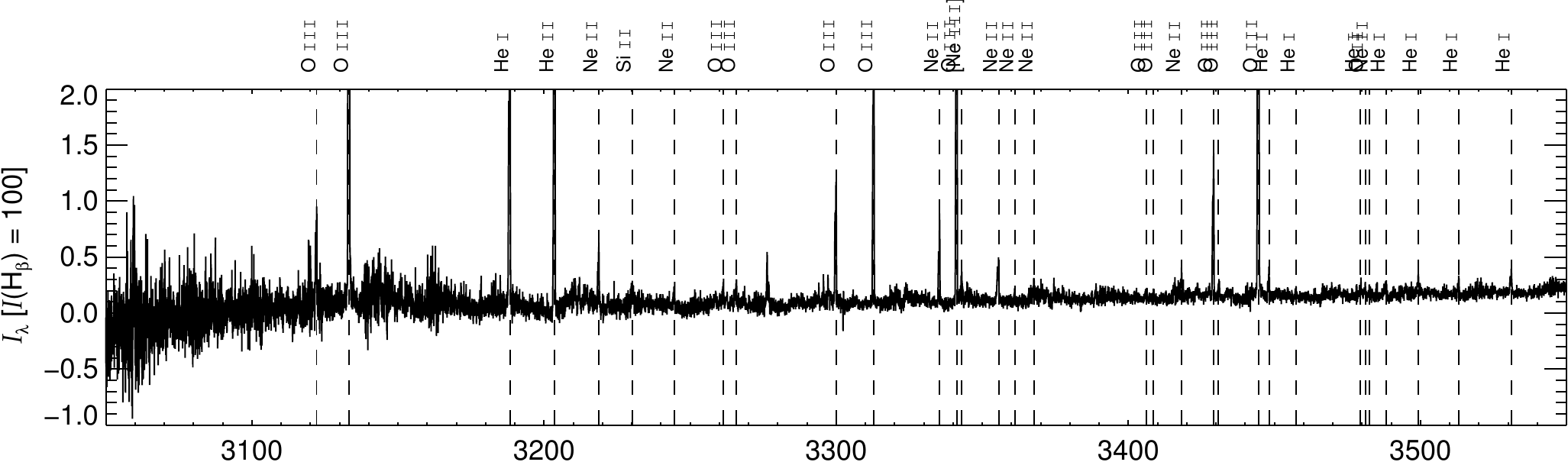,width=0.9\textwidth,angle=0}
\epsfig{file=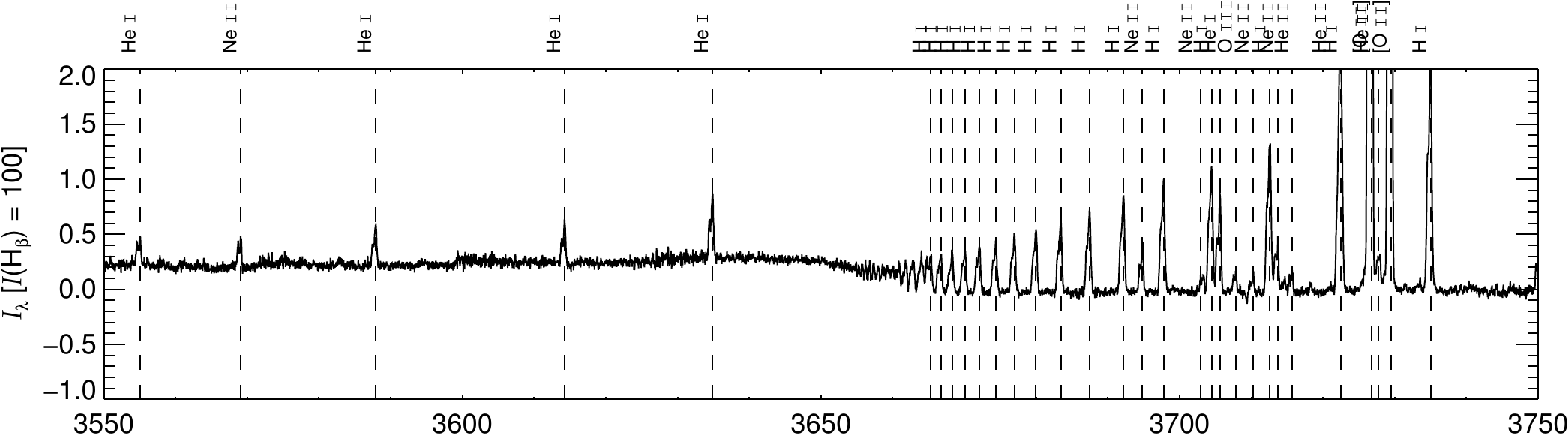,width=0.9\textwidth,angle=0}
\epsfig{file=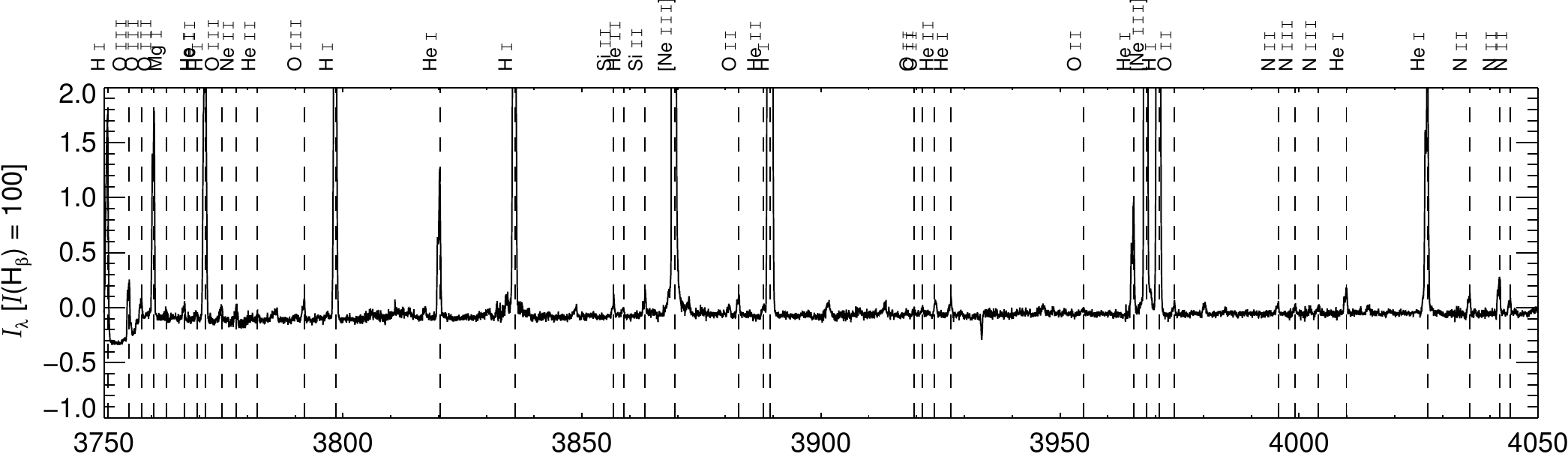,width=0.9\textwidth,angle=0}
\epsfig{file=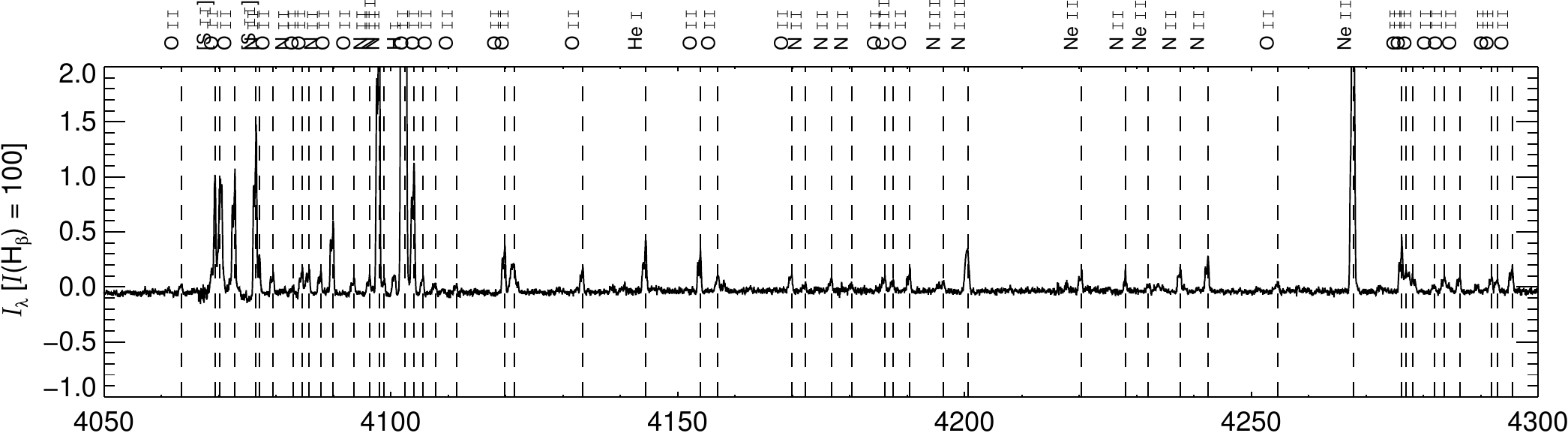,width=0.9\textwidth,angle=0}
\epsfig{file=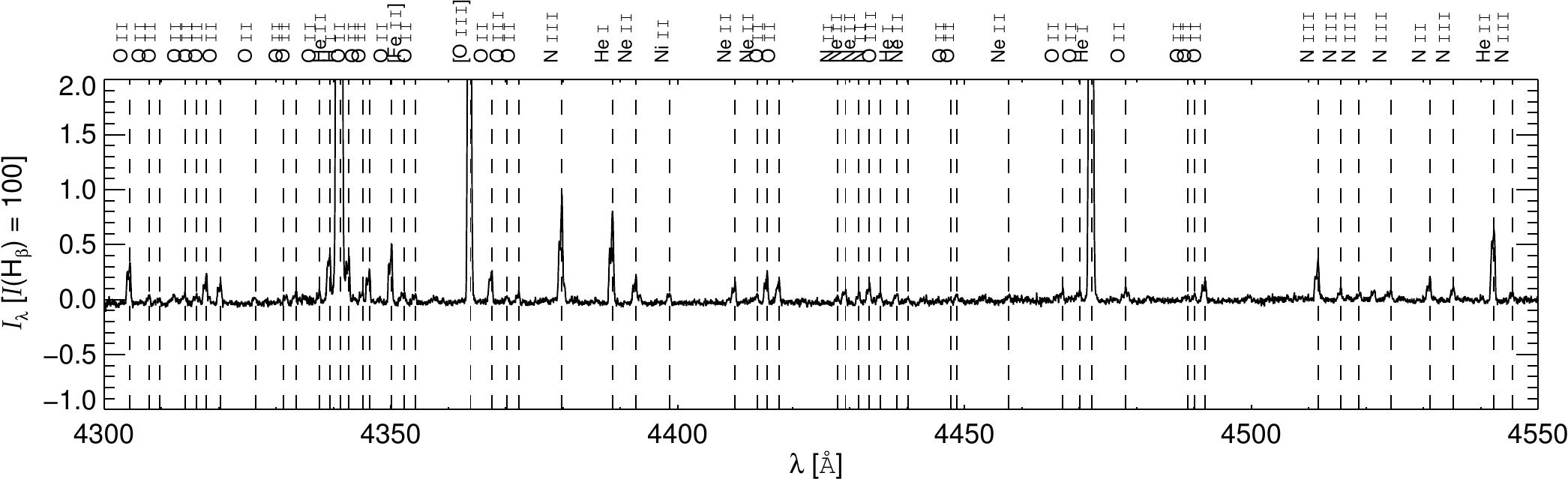,width=0.9\textwidth,angle=0}
\caption{The long-exposure spectra of NGC\,6153, M\,1-42 and Hf\,2-2 
from $\sim$3040 to 10,600\,{\AA}.  Extinction has not been corrected 
for.  Spectra are normalized such that H$\beta$ has an integrated flux 
of 100.  The spectrum of NGC\,6153 is presented here as an example. 
Readers may contact the corresponding author for the spectra of the 
other two PNe. }
\label{1d_spectra}
\end{figure*}

%\clearpage
\setcounter{figure}{0}
\begin{figure*}
\centering
\epsfig{file=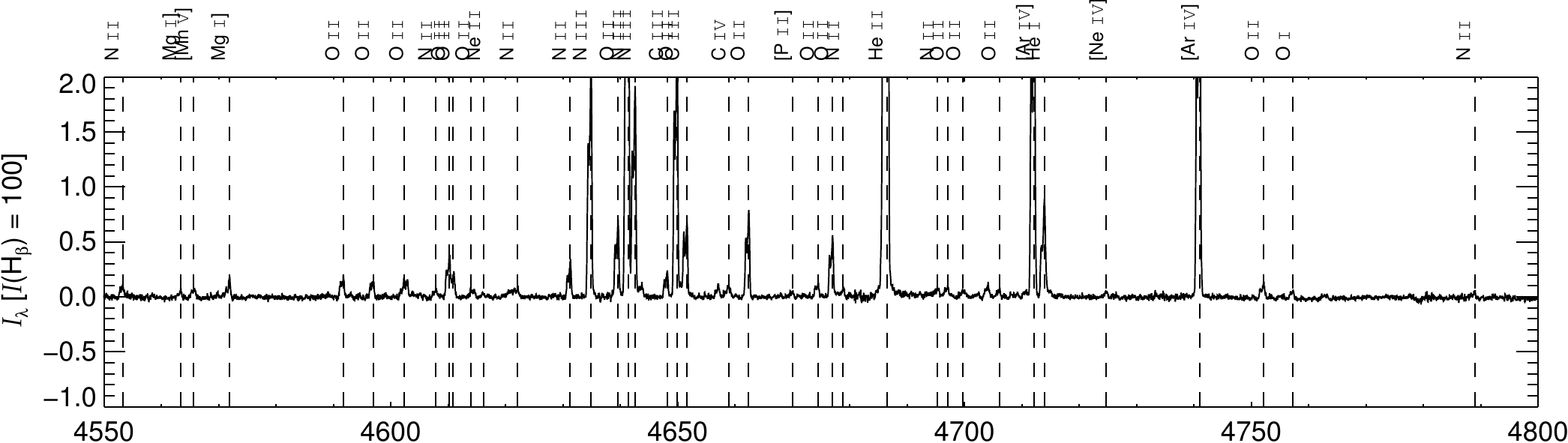,width=0.9\textwidth,angle=0}
\epsfig{file=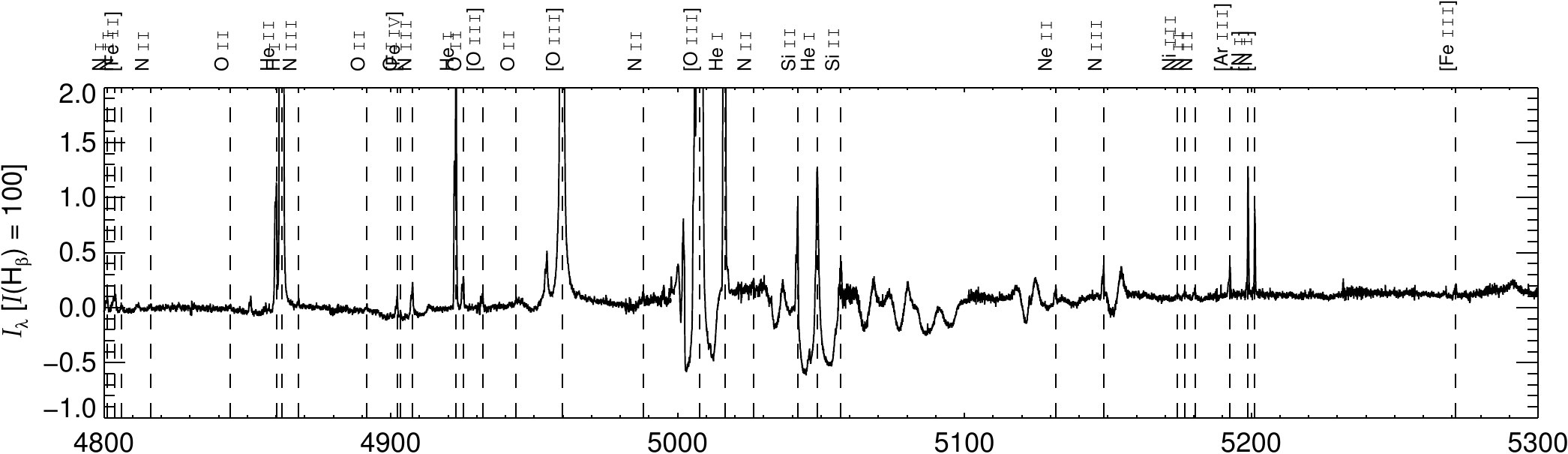,width=0.9\textwidth,angle=0}
\epsfig{file=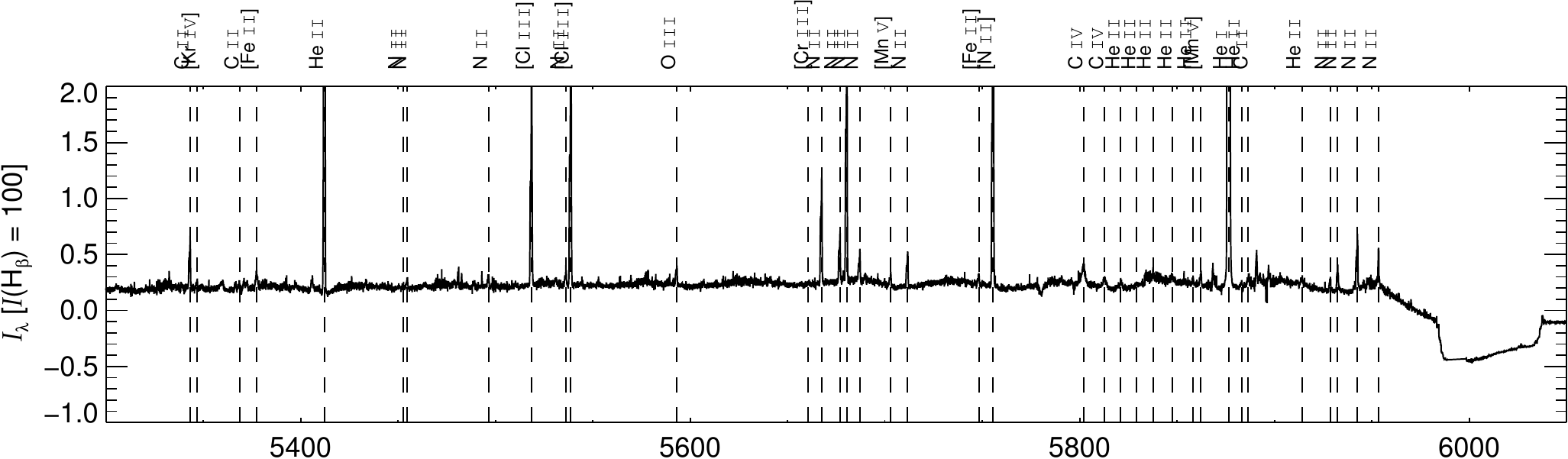,width=0.9\textwidth,angle=0}
\epsfig{file=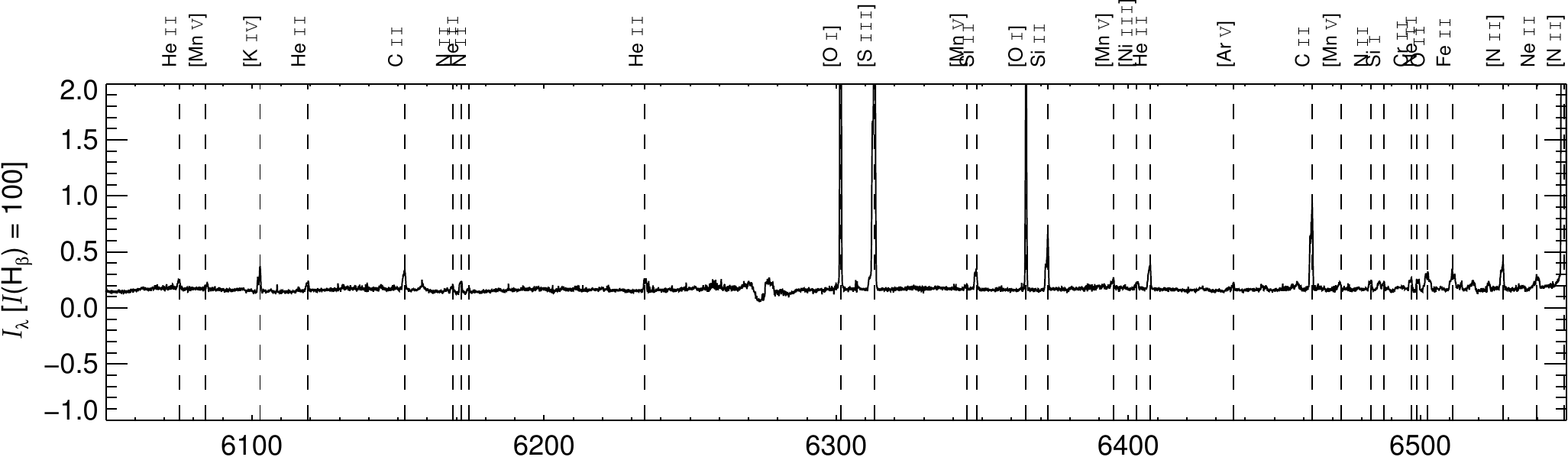,width=0.9\textwidth,angle=0}
\epsfig{file=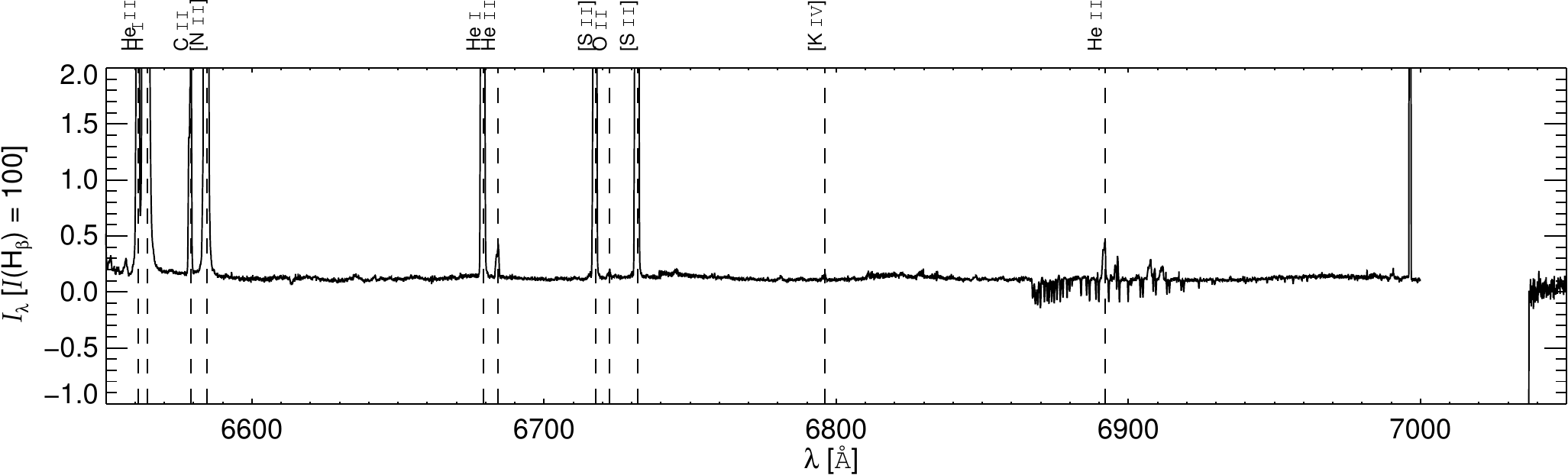,width=0.9\textwidth,angle=0}
\caption{(Continued)}
\end{figure*}

%\clearpage
\setcounter{figure}{0}
\begin{figure*}
\centering
\epsfig{file=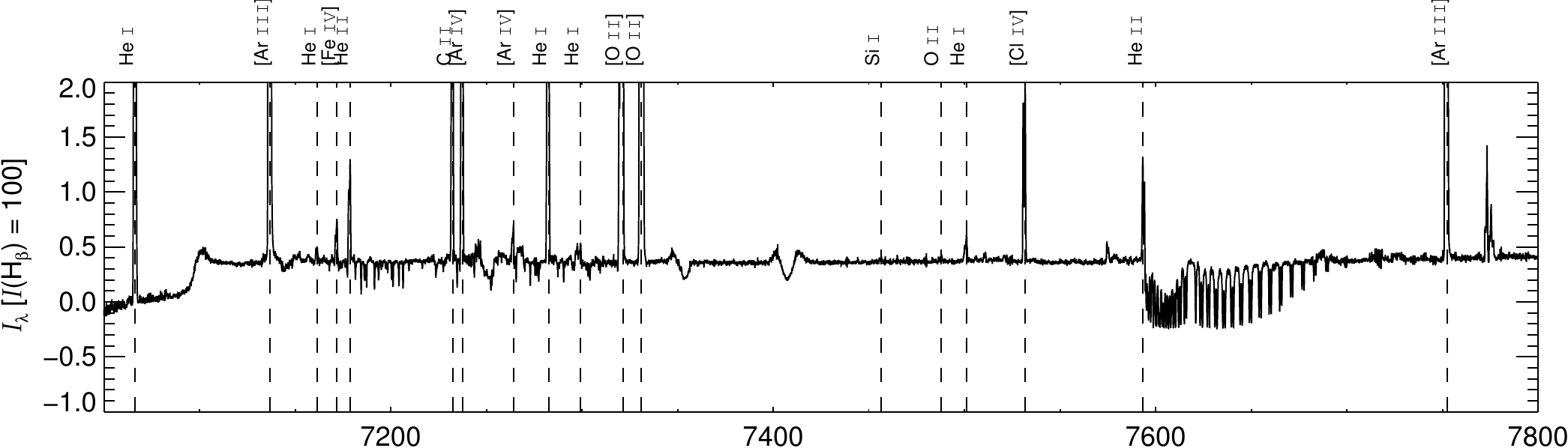,width=0.9\textwidth,angle=0}
\epsfig{file=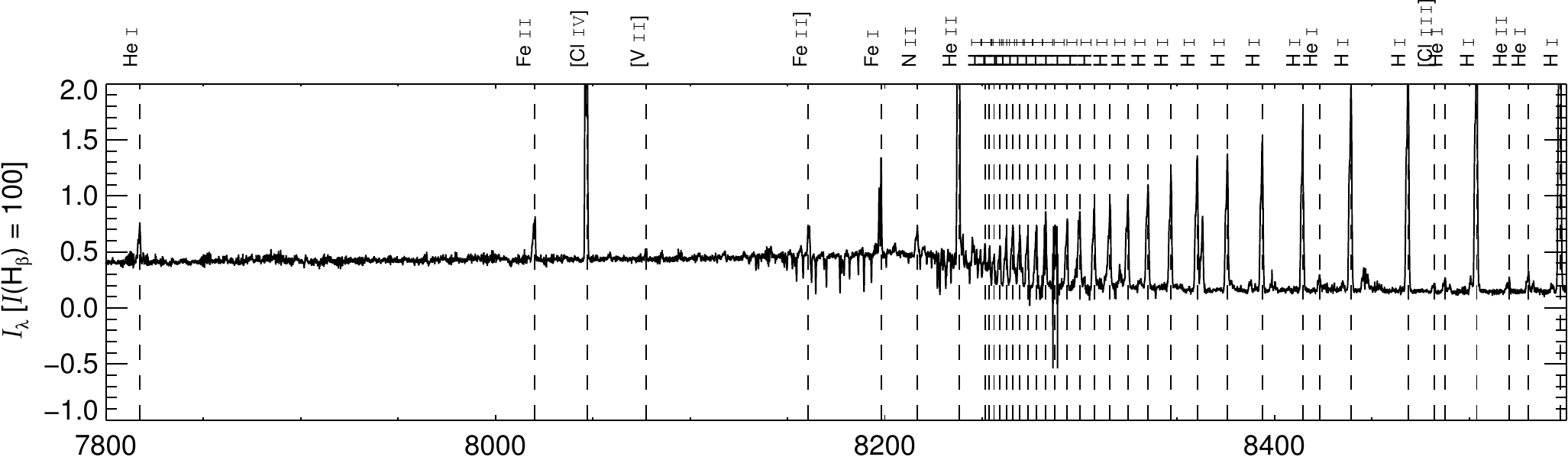,width=0.9\textwidth,angle=0}
\epsfig{file=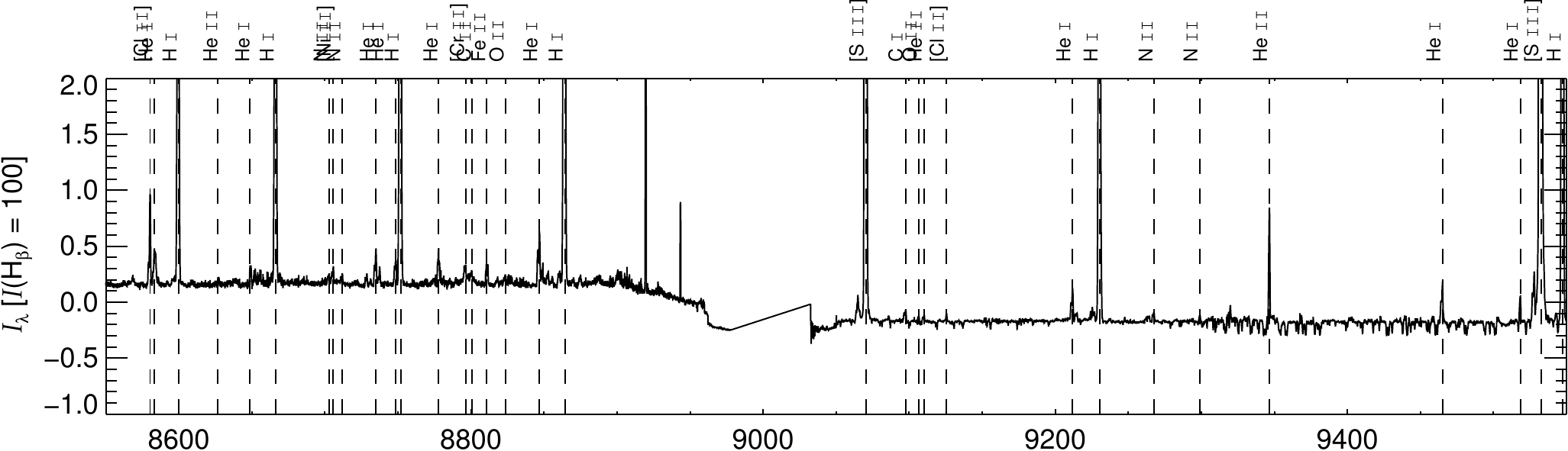,width=0.9\textwidth,angle=0}
\epsfig{file=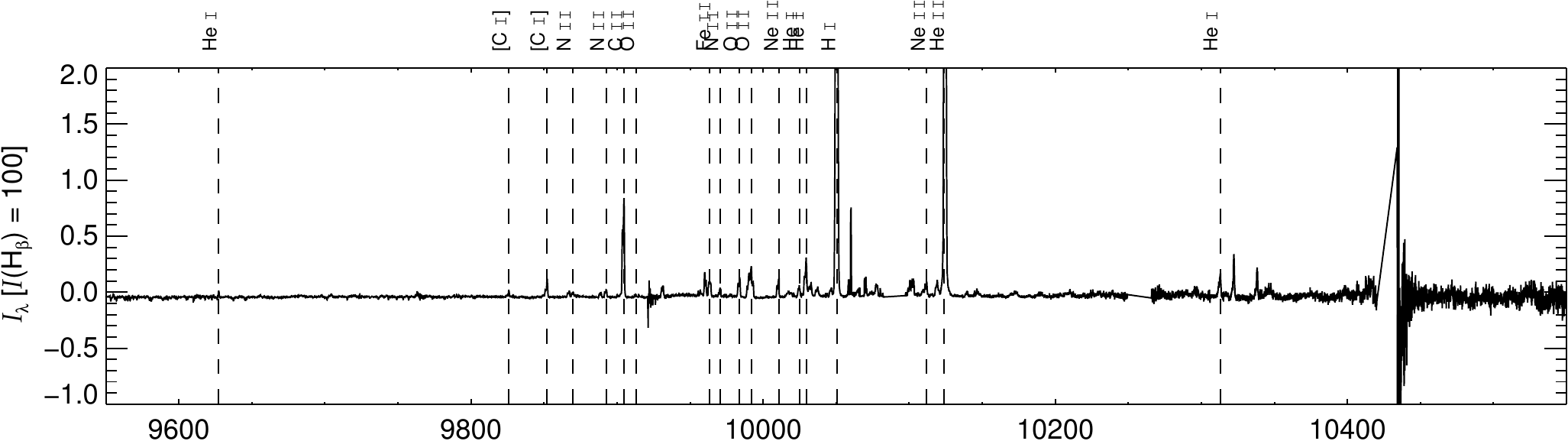,width=0.9\textwidth,angle=0}
\caption{(Continued)}
\end{figure*}

\end{document}